\documentclass[12pt]{iopart}
\usepackage{epsf}  
\usepackage{epsfig}
\begin{document}
\title[Report of Beyond the Standard Model Working Group]{Report of the Beyond the Standard
Model Working Group of the 1999 UK Phenomenology
Workshop on Collider Physics (Durham)\footnote[3]{To appear in Journal of
Physics G}}

\author{B~C~Allanach$^{1*}$, J~J~van~der~Bij$^2$, A~Dedes$^3$, A~Djouadi$^{4}$,
J~Grosse-Knetter$^5$, J~Hetherington$^{6}$, S~Heinemeyer$^7$, J~Holt$^5$, D~Hutchcroft$^8$, J~Kalinowski$^9$, G~Kane$^{10}$,
V~Kartvelishvili$^{11}$, S~F~King$^{12}$, S~Lola$^{13*}$, R~McNulty$^{14}$,
M~A~Parker$^{6}$, G~D~Patel$^{14}$, G~G~Ross$^{15}$, M~Spira$^{16}$,
P~Teixeira-Dias$^{17}$, G~Weiglein$^{13}$, G~Wilson$^{11}$,
J~Womersley$^{18}$, P~Walker$^{11}$, B~R~Webber$^{6}$, T~Wyatt$^{11*}$
}


\address{$^1$DAMTP, Silver Street, University of Cambridge, Cambridge, CB3
9EW, UK}
\address{$^2$Fakultaet fuer Physik,
Universitaet Freiburg,
H. Herderstr. 3,
79104 Freiburg i. B., Deutschland}
\address{$^3$Rutherford Appleton Laboratory, Chilton, Didcot, Oxon, OX11 OQX,
UK}
\address{$^{4}$Laboratoire de Physique Math\'ematique et Th\'eorique, UMR--CNRS 5825, 
Universit\'e de Montpellier II, F-34095 Montpellier Cedex 5, France}
\address{$^5$Nuclear Physics Laboratory, University of Oxford,
      Keble Road, Oxford, OX1 3RH, UK}
\address{$^6$Cavendish Laboratory, University of Cambridge, Madingley Road,
      Cambridge, CB3 OHE, UK}
\address{$^7$Deutsches Elektronen Synchrotron (DESY),
      Notkestrasse 85, D-22603 Hamburg, Germany}
\address{$^8$Physics Department, Royal Holloway University of London, 
      Egham Hill, Egham, Surrey, TW20 0EX, UK}
\address{$^9$Instytut Fizyki Teoretycznej UW, Hoza 69, PL-00681 Warsaw,
Poland}
\address{$^{10}$Randall Physics Laboratory, University of Michigan,
Ann Arbor, MI 48109-1120, USA}
\address{$^{11}$Department of Physics and Astronomy, Schuster Laboratory, The
University of Manchester, 
      Manchester, M13 9PL, UK}
\address{$^{12}$Department of Physics and Astronomy, University of Southampton,
      Highfield, Southampton, SO17 1BJ, UK}
\address{$^{13}$Theory Division, CERN, CH-1211 Geveva 23, Switzerland}
\address{$^{14}$Department of Physics, Oliver Lodge Laboratory University of
      Liverpool, PO Box 147, Oxford Street, Liverpool, L69 3BX, UK}
\address{$^{15}$Department of Theoretical Physics, University of Oxford,
      1 Keble Road, Oxford, OX1 3NP, UK}
\address{$^{16}$Institut fur Theoretische Physik, Universitat Hamburg,
      Luruper Chaussee 149 , D-22761 Hamburg, Germany}
\address{$^{17}$Department of Physics and Astronomy, University of Glasgow, 
      Kelvin Building, Glasgow G12 8QQ, UK}
\address{$^{18}$Fermilab,
P.O.Box 500,
Batavia, IL 60510-0500, USA}

\address{$^*$ convenors}

\begin{abstract}
The Beyond the Standard Model Working Group discussed a variety of topics
relating to exotic searches at current and future colliders, and the
phenomenology of current models beyond the Standard Model.
For example, various supersymmetric (SUSY) and extra dimensions search
possibilities and constraints are presented. Fine-tuning implications
of SUSY searches are derived. The implications of Higgs
(non)-discovery are discussed, as well as the program HDECAY\@. The
individual contributions are included seperately. Much of the enclosed
work is original, although some is reviewed. 
\end{abstract}

\maketitle

\newcommand{\lrpair}{\mbox{$\tilde{e}_L^+\tilde{e}_R^-$}}

\section{Introduction}

The `Beyond the Standard Model' working group addressed the prospects for
searches for supersymmetry, the phenomenology of large extra dimensions and
the phenomenological implications of lower bounds upon the Higgs boson mass.
The present status of large extra dimensions, SUSY breaking and searches for
SUSY and leptoquarks were well covered in the plenary talks by G Ross and J
Womersley. 

There were three broad subgroups: Higgs phenomenology, SUSY breaking/large
extra dimensions and the study of events containing isolated charged leptons and missing
$p_{t}$ at LEP2. These subgroups had
their own agendas of seminar presentations, discussions and reports.
Summaries from each subgroup were given to the rest of the Beyond the Standard
Model working group, and
indeed the other working groups of the workshop. Much of the following work is
original and
carried out at the workshop, whereas some is the result of literature reviews.

The minimal supergravity (mSUGRA) reach potential of the 
LHC is
readdressed in section~\ref{reach}. The exclusion limits are produced in terms
of a naturalness parameter.
In section~\ref{sec:steve}, the fine-tuning implications of  supersymmetric
particle masses are presented in various SUSY breaking scenarios.
The possibilities of detecting gluino-gluino bound states at run II of the
Tevatron are examined in section~\ref{vat}.
The experimental signatures and fits to current data of two models of extra
dimensions are presented in section~\ref{sec:us}.
In section~\ref{sec:terry}, events containing isolated leptons 
  and missing $p_{t}$ at LEP2 are discussed as a means of detecting,
  e.g., single chargino and \lrpair\ production.
Lower bounds upon
the Higgs mass from LEP2 have been steadily increasing in the last two years,
and the next sections address this empirical information.
We present a review of what the precision electroweak fits imply once one
retreats from the SM Higgs sector in
section~\ref{higgsphen}. 
Stealthy Higgs models that may be undetectable by the LHC are reviewed in
section~\ref{sec:stealth}.
State-of-the art upper limits 
upon the lightest Higgs boson mass 
in the general (R-parity conserving) MSSM and M-SUGRA are presented
in section~\ref{sec:msugra}. Within the MSSM, using the most recent and
prospective future 
LEP2 data, limits on $\tan \beta$ are derived.
Continuing work upon the Higgs decay program HDECAY was carried out at the
workshop and the program is reviewed in section~\ref{hdecay}. 

\ack
Many thanks to St.\ John's College, Durham where the workshop was held, to M.~Whalley, J.~Forshaw and E.W.N.~Glover for their
organisation.
This work is partially supported by PPARC.

\newpage

\section{Naturalness Reach of the Large Hadron Collider in Minimal SUGRA\label{reach}}

\author{B~C~Allanach, J~P~J~Hetherington, M~A~Parker, G~G~Ross, B~R~Webber}
\begin{abstract}
We re-analyse the best SUGRA discovery channel at the LHC, in order to
re-express coverage in terms of a fine-tuning parameter and to extend the
analysis to higher $m_0=3$~TeV. 
Such high values of $m_0$ have recently been found to have a focus point,
leading to relatively low fine-tuning.
It is found that even for $m_0$ as high as 3~TeV, mSUGRA can
still be discovered for $M_{\frac{1}{2}}<490 \pm 20$ GeV.
For $\mu<0, A_0=0$ GeV and $\tan \beta=10$ (corresponding to the focus point),
all points in mSUGRA with a fine
tuning measure up to 220 are covered by the search.
\end{abstract}

Recent work~\cite{focuspoints} has shown that MSSM scalar masses as large as 2
to 3 TeV can be consistent with naturalness. This occurs near a `focus point'
where the renormalisation group trajectories of the mass squared of a Higgs
doublet ($m_{H_2}^2$) cross close to the electroweak scale. 
As a result, 
the electroweak symmetry breaking is insensitive to ultraviolet
boundary conditions upon SUSY breaking parameters~\cite{focuspoints}.
Previous predictions of the discovery reach of the LHC into mSUGRA parameter
space went only as far as $m_0 < 2$~TeV~\cite{ATLASTDR2}. The purpose of this
work is to extend this reach to 3 TeV and to present it in terms of a
naturalness measure. While interpretation of a naturalness measure is
inevitably subjective, we advocate its use as a
single parameter for defining the search reach of a collider in the context of
a particular model. The naturalness coverage could then be used to compare
between different colliders/experiments/models etc.

At tree-level, the $Z$ boson mass is determined to be 
\begin{equation}
\frac{1}{2} M_Z^2 = \frac{m_{H_1}^2 - m_{H_2}^2 \tan^2 \beta}{\tan^2 \beta -
1} - \mu^2 \label{FTtree}
\end{equation}
by minimising the Higgs potential. $\tan \beta$ refers to the ratio of Higgs
vacuum 
expectation values (VEVs) $v_1/v_2$ and $\mu$ to the Higgs mass parameter in
the MSSM superpotential. 
In mSUGRA, $m_{H_2}$ has the same origin as the super-partner masses ($m_0$). Thus as
search limits put lower bounds upon super-partners' masses, the lower bound
upon $m_0$ rises, and consequently so does $|m_{H_2}|$. A cancellation is then
required between the first and second terms of equation~\ref{FTtree} in order
to provide the measured value of $M_Z \ll |m_{H_2}|$. Various measures have
been proposed in order to quantify this cancellation~\cite{measures}.

The definition of naturalness $c_a$ of a `fundamental' parameter $a$ employed
in reference~\cite{focuspoints} is
\begin{equation}
c_a \equiv \left| \frac{\partial \ln M_Z^2}{\partial \ln a} \right|.
\end{equation}
From a choice of a set of fundamental parameters $\{ a_i \}$, the
fine-tuning of a particular model is defined to be $c=\mbox{max}(c_a)$.
Our choice of free, continuously valued, independent and fundamental mSUGRA
parameters also follows ref.~\cite{focuspoints}:
\begin{equation}
\{ a_i \} = \{ m_0, M_{1/2}, \mu(M_{GUT}), A_0, B(M_{GUT}) \}
\end{equation}
where $M_{GUT} \sim 10^{16}$ GeV is the GUT scale.

\begin{figure}
\begin{center}
\leavevmode
\hbox{\epsfysize=10cm
\epsffile{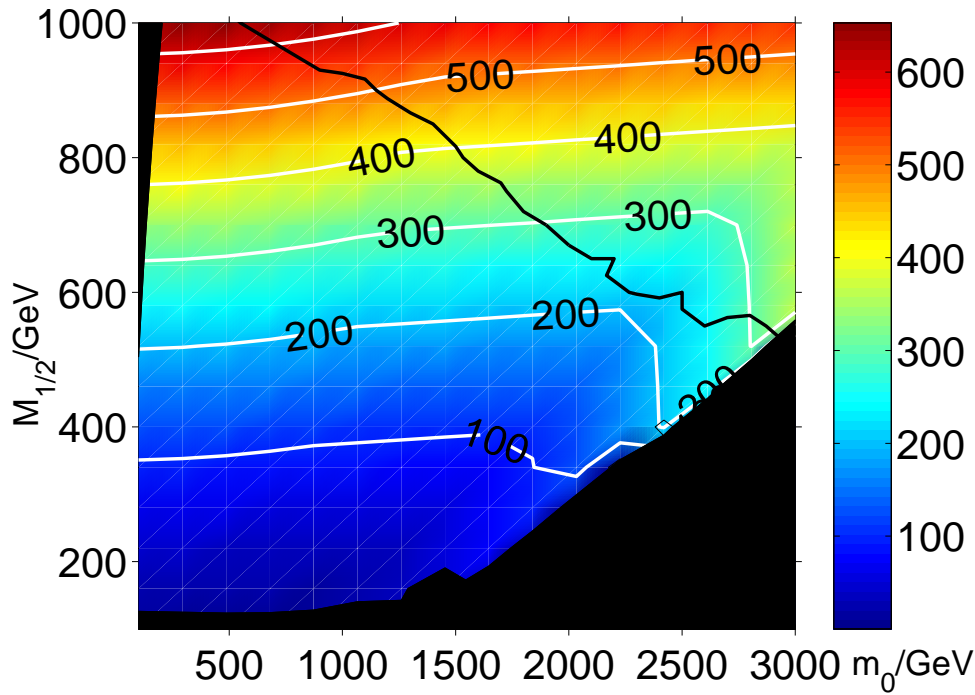}}
\end{center}
\caption{Naturalness reach at the LHC for $A_0=0$ GeV, $\tan \beta=10$,
$\mu<0$ in minimal SUGRA\@. The background represents the
degree of fine tuning $c$, as measured by the bar and white contours.
The black region to the top left hand corner is excluded by the requirement
that the LSP be neutral, and the black region at the bottom of the plot from
chargino exclusion limits and radiative symmetry breaking. The black line is
the LHC expectation contour for ten signal events in the 1 lepton 2 jets
channel described in the text for a luminosity of ${\mathcal L} =
10$~fb$^{-1}$.
}
\label{fig:ftreachpos}
\end{figure}
We now turn to the discussion of the LHC mSUGRA search.
In the ATLAS TDR~\cite{ATLASTDR2} the best reach
was found through a 1 lepton plus jets plus missing transverse momentum
signal, which looks mainly for chargino decays to
lepton and sneutrino. 
The following cuts were employed, which are the same as those
as in refs.~\cite{ATLASTDR2,hbaeretc} except where those original cuts are
included in parenthesis:
\begin{itemize}
\item
${/\!\!\!\!p_T}>400$ GeV
\item
At least 2 jets, with rapidity $\eta<2.0$, and $p_T>400$~GeV. 
\item
1 lepton, $\eta<2.0$, $p_T> 20$~GeV, 
lying further in $\eta, \phi$ space than 0.4 units from the centre
of any jet with cone-size 0.4 (0.7) (less than 5 GeV of energy within 0.3
units of the lepton).
\item
\begin{equation}
M_T =  \sqrt{2 (| {\bf p_l} | |{\bf /\!\!\!\!p_T}| - {\bf p_l} \cdot {\bf /\!\!\!\!p_T}) } >
100\mbox{~GeV},
\end{equation}
where ${\bf p_l}$, ${\bf /\!\!\!\!p_T}$ are transverse two-component lepton and
missing $p_T$ respectively.
\item
\begin{equation}
S_T = \frac{2 \lambda_2}{\lambda_1 + \lambda_2} > 0.2,
\end{equation}
where $\lambda_i$ are the eigenvalues of the sphericity matrix
$S_{ij} = \Sigma_{ij} p_i p_j $,
the sum being taken over all detectable final state particles, and
$p_i$ being the two-component transverse momentum attributed to the cell. 
\end{itemize}

mSUGRA events were simulated by employing the {\small ISASUSY} part of the
{\small ISAJET7.42}
package~\cite{ISASUSY} to calculate sparticle masses and branching ratios, and
{\small HERWIG6.1}~\cite{HERWIG} to simulate the events themselves.
Fig.~\ref{fig:ftreachpos} shows the contour for 10 signal events passing the
above cuts as a black
line in the $m_0/M_{1/2}$ plane for mSUGRA with $A_0=0$ GeV, $\tan \beta=10$,
$m_t(m_t)=160$ GeV, $\mu<0$ and for ${\mathcal L}=10$ fb$^{-1}$ of luminosity
(equivalent to a one year of running in the low luminosity
mode). Background at regions of parameter space with
such high energy cuts is estimated to be negligible.
The discovery contour is overlaid upon the density of naturalness $c$ 
(displayed as background) as
defined above. $c$ was calculated numerically to one-loop accuracy in soft
masses, with two-loop accuracy in supersymmetric parameters and step-function
decoupling of sparticles. Dominant one-loop top/stop corrections are added to
the Higgs potential and correct equation~\ref{FTtree}.
This approximation was also used to calculate the black (excluded) regions in
figure~\ref{fig:ftreachpos}. We note that the horizontal piece of the bottom
black region results from the limit $M_{\chi_1^\pm}>90$ GeV, whereas the
diagonal piece is from the constraint of electroweak symmetry breaking.
This last constraint is very sensitive to $m_t(m_t)$, and moves to the right
and off the plot for $m_t(m_t)=165$ GeV.
We note that while the naturalness contours displayed in
Figure~\ref{fig:ftreachpos} are of the same shape as those calculated
before~\cite{focuspoints}, the fine tuning is some 50\% 
higher. This is due~\cite{measures} to the approximation of using an
incomplete one-loop potential, and will be rectified~\cite{usnext} (as will
the approximation of using constant $m_t(m_t)$ over the $M_{1/2}/m_0$ plane).

\ack
Part of this work was produced using the Cambridge University High Performance
Computing Facility. BCA would like to thank K Matchev for valuable discussions
on checks of the numerical results. JH would like to thank C Lester for useful
discussions.

\renewcommand{\topfraction}{0.9}
\renewcommand{\bottomfraction}{0.9}
\renewcommand{\textfraction}{0.1}
\def\bea{\begin{eqnarray}}
\def\eea{\end{eqnarray}}
\section
{Fine-Tuning Constraints on Supergravity Models \label{sec:steve}}

\author
{M~Bastero-Gil, G~L~Kane and S~F~King}

\begin{abstract}
We discuss fine-tuning constraints
on supergravity models. The tightest constraints come from the
experimental mass limits on two key particles: the lightest CP
even Higgs boson and the gluino. We also include
the lightest chargino which is relevant when universal gaugino masses
are assumed.
For each of these particles we show how fine-tuning increases 
with the experimental mass limit, for four types of supergravity
model: minimal supergravity, no-scale supergravity (relaxing the 
universal gaugino mass assumption), D-brane models
and anomaly mediated supersymmetry
breaking models. Among these models, the D-brane model is less
fine tuned.The experimental prospects for an early discovery  
of Higgs and supersymmetry at LEP and the Tevatron are discussed
in this framework.
\end{abstract}

When should physicists give up on low energy supersymmetry?
The question revolves around the issue of how much fine-tuning one
is prepared to tolerate. Although fine-tuning is not a well defined concept,
the general notion of fine-tuning is unavoidable 
since it is the existence of fine-tuning in the
standard model which provides the strongest motivation for low energy 
supersymmetry, and the widespread belief that superpartners should
be found before or at the LHC\@. Although a precise measure of 
{\em absolute}\/ fine-tuning is impossible,
the idea of {\em relative fine-tuning}
\/ can be helpful in selecting certain models and regions
of parameter space over others\footnote{For a complete list of references and a fuller discussion
of these results see M. Bastero-Gil, G. L. Kane and S. F. King, 
hep-ph/9910506.}.

\begin{figure}
\begin{center}
\epsfysize=10cm
\epsfbox{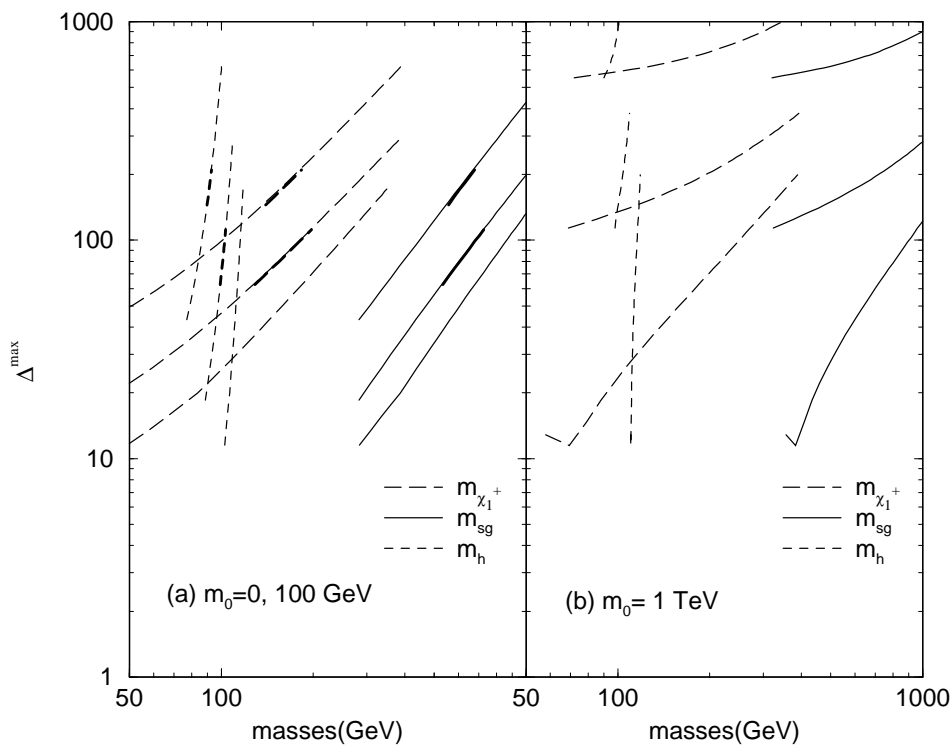}
\caption{Results for the minimal SUGRA model.
The maximum sensitivity parameter $\Delta^{max}$ is plotted 
as a function of the lightest CP even Higgs mass (short dashes),
gluino mass (solid line) and lightest chargino (long dashes).
For each particle type, the three sets of curves correspond to 
$\tan\beta$=2, 3, 10, from top left to bottom right, respectively. 
In panel (a) the shorter, thicker lines
correspond to $m_0=0$, while the longer lines are those for $m_0=100$
GeV. In panel (b) the results correspond to $m_0=1000$ GeV.}
\label{fig:stst}
\end{center}
\end{figure}
\begin{figure} \begin{center}
\epsfysize=10cm
\epsfbox{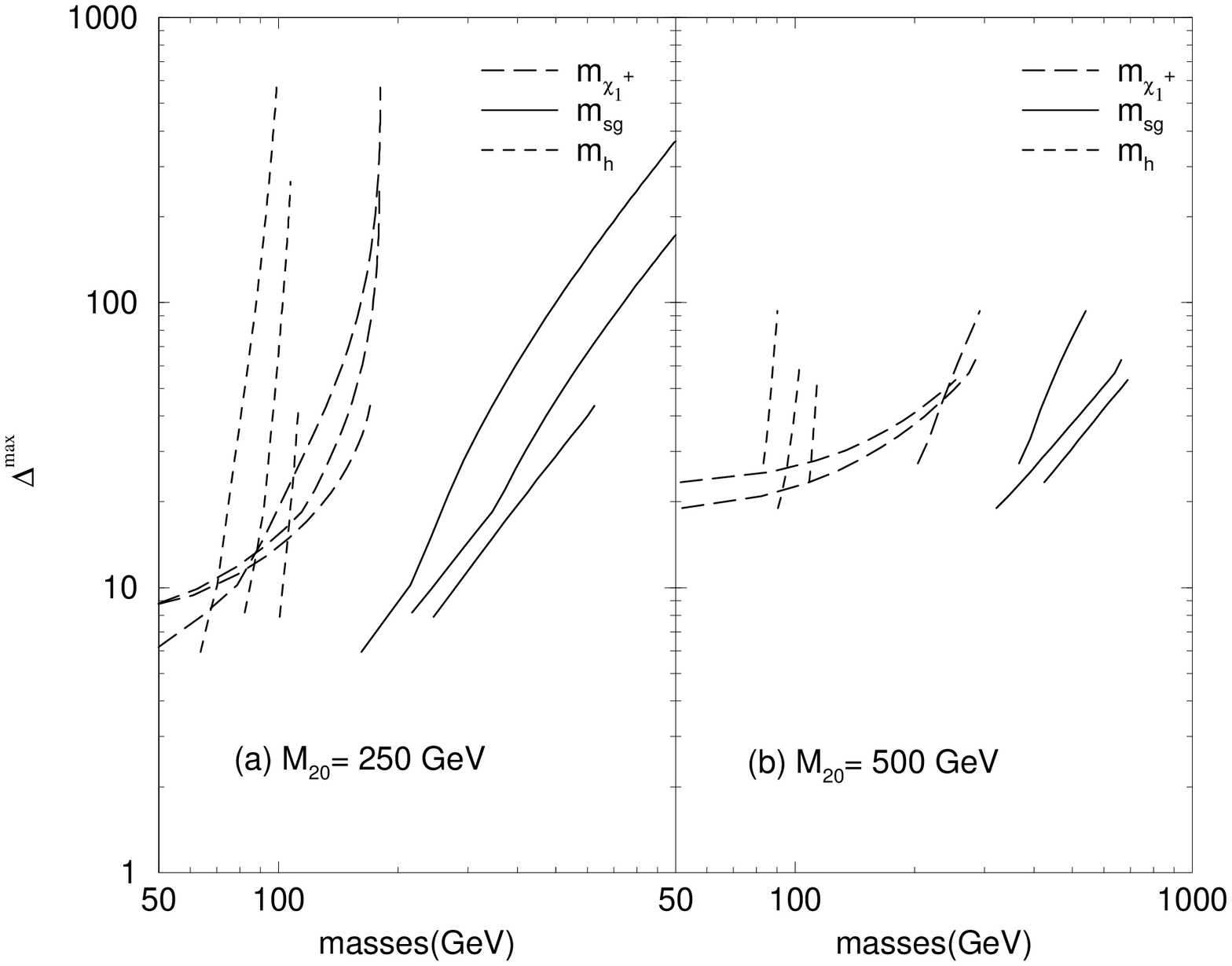} 
\caption{Results for the 
no-scale with non-universal gaugino masses. 
The maximum sensitivity parameter $\Delta^{max}$ is plotted 
as a function of the lightest CP even Higgs mass (short dashes),
gluino mass (solid line) and lightest chargino (long dashes).
For each particle type, the three sets of curves correspond to 
$\tan\beta$=2, 3, 10, from top left to bottom right, respectively. 
In panel (a) we fix $M_2(0)=250$ GeV, while in panel (b) 
$M_2(0)=500$ GeV.} \end{center}
\end{figure}
The models we consider, and the corresponding input parameters given
at the unification scale,  are listed below: 
\begin{enumerate}

\item Minimal supergravity 
\bea
a_{msugra}\in \{m_0^2, M_{1/2}, A(0), B(0), \mu(0)\}\,,
\label{msugra}
\eea 
where as usual $m_0$, $M_{1/2}$ and $A(0)$ are the universal scalar mass,
gaugino mass and trilinear coupling respectively, $B(0)$ is the
soft breaking bilinear coupling in the Higgs potential and $\mu(0)$ is
the Higgsino mass parameter. 

\item No-scale supergravity with non-universal gaugino
masses\footnote{This is in fact a new model not
previously considered in the literature, 
although the no-scale model with universal gaugino masses
is of course well known. As in the usual no-scale model,
this model has the attractive feature that 
flavour-changing neutral currents at low energies are very suppressed,
since all the scalar masses are generated by radiative corrections,
via the renormalisation group equations, which only depend on the
gauge couplings which are of course flavour-independent.}
\bea
a_{no-scale}\in \{M_1(0), M_2(0), M_3(0), B(0), \mu(0)\}
\label{noscalesugra}
\eea 

\item D-brane model
\bea
a_{D-brane}\in \{m_{3/2}, \theta , \Theta_1, \Theta_2, \Theta_3,
 B(0), \mu(0)\}\,,
\label{Dbrane}
\eea 
where $\theta$ and $\Theta_i$ are the goldstino angles, with
$\Theta_1^2+\Theta_2^2+\Theta_3^2=1$, and $m_{3/2}$
is the gravitino mass. The gaugino masses are given by
\bea
M_1(0)  =  M_3(0) & = & \sqrt{3}m_{3/2}\cos \theta \Theta_1 e^{-i\alpha_1}
\nonumber \,,\\
M_2(0) & = & \sqrt{3} m_{3/2}\cos \theta \Theta_2 e^{-i\alpha_2} \,,
\eea
and there are two types of soft scalar masses
\bea
m_{5152}^2 & = & m_{3/2}^2
[1-\frac{3}{2}(\sin^2 \theta +\cos^2 \theta \Theta_3^2) ]
\nonumber \,,\\
m_{51}^2 & = & m_{3/2}^2[1-3\sin^2 \theta] \,,
\eea

\item Anomaly mediated supersymmetry breaking
\bea
a_{AMSB}\in \{m_{3/2}, m_0^2, B(0), \mu(0)\}
\label{AMSB}
\eea 

\end{enumerate}

\begin{figure} \begin{center}
\epsfysize=10cm
\epsfbox{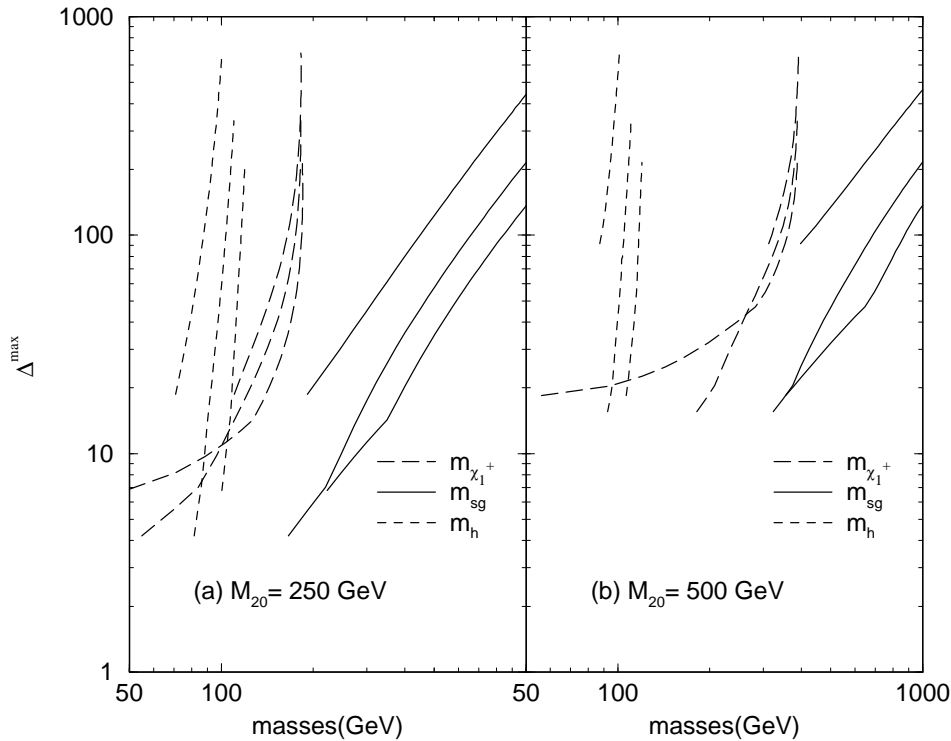}
\caption{Results for the D-brane model. 
The maximum sensitivity parameter $\Delta^{max}$ is plotted 
as a function of the lightest CP even Higgs mass (short dashes),
gluino mass (solid line) and lightest chargino (long dashes).
For each particle type, the three sets of curves correspond to 
$\tan\beta$=2, 3, 10, from top left to bottom right, respectively.
In panel (a) we fix $M_2(0)=250$ GeV, while in panel (b) 
$M_2(0)=500$ GeV.} \end{center}
\end{figure}
\begin{figure} \begin{center}
\epsfysize=10cm
\epsfbox{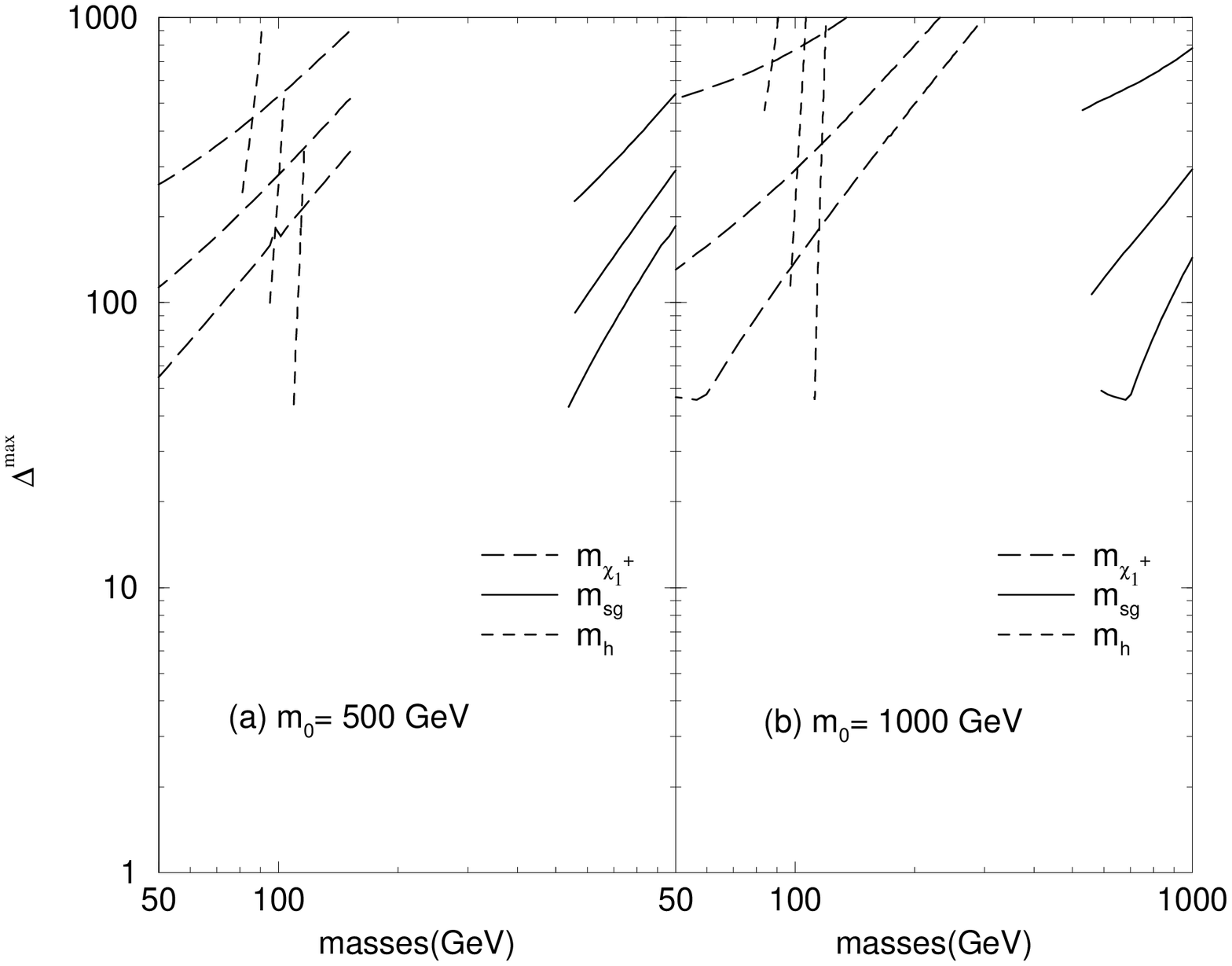}
\caption{Results for the
anomaly mediated supersymmetry breaking model.
The maximum sensitivity parameter $\Delta^{max}$ is plotted 
as a function of the lightest CP even Higgs mass (short dashes),
gluino mass (solid line) and lightest chargino (long dashes).
For each particle type, the three sets of curves correspond to 
$\tan\beta$=2, 3, 10, from top left to bottom right, respectively.
In panel (a) we fix $m_0=500$ GeV, while in panel (b) 
$m_0=1000$ GeV.} \label{fig:stend} \end{center}
\end{figure}
Our main results are shown in Figures~\ref{fig:stst}-\ref{fig:stend}, corresponding to SUGRA
models 1-4 above. In all models, 
fine-tuning is reduced as $\tan \beta$ is increased, with
$\tan \beta =10$ preferred over $\tan \beta =2,3$.
Nevertheless,  
the present LEP2 limit on the Higgs and chargino mass
of about 100 GeV and the gluino mass limit of about 250 GeV
implies that $\Delta^{max}$ is of order 10 or higher. 
The fine-tuning increases most sharply with the Higgs mass.
The Higgs fine-tuning curves are fairly model independent,
and as the Higgs mass limit rises above 100 GeV come to 
quickly dominate the fine-tuning. We conclude that the 
prospects for the discovery of the Higgs boson at LEP2 are good.
For each model there is a correlation between the
Higgs, chargino and gluino mass, for a given value of
fine-tuning. For example if the Higgs is discovered
at a particular mass value, then the corresponding
chargino and gluino mass for each $\tan \beta$ can be 
read off from Figures~\ref{fig:stst}-\ref{fig:stend}. 

The new general features of the results may then be summarised as follows:

\begin{itemize}
\item The gluino mass curves are
less model dependent than the chargino curves,
and this implies that in all models if the fine-tuning
is not too large then the prospects for the 
discovery of the gluino at the Tevatron are good.

\item The fine-tuning due to the chargino mass is model
dependent. For example in the no-scale model 
with non-universal gaugino masses and the
D-brane scenario the charginos may be relatively heavy compared
to mSUGRA.

\item Some models have less fine-tuning than others.
We may order the models on the basis of fine-tuning from
the lowest fine-tuning to the highest fine-tuning:
D-brane scenario $<$ generalised no-scale SUGRA $<$ mSUGRA $<$ AMSB. 

\item The D-brane model is less fine-tuned partly because the gaugino
masses are non-universal, and partly because there are large regions
where $\Delta_{m_{3/2}}$, 
$\Delta_{\mu(0)}$, and $\Delta_\theta$ are all close to zero
However in these regions the fine tuning is 
dominated by $\Delta_\Theta$, and this leads to an inescapable
fine-tuning constraint on the Higgs and gluino mass. 

\end{itemize}



\section{Gluino-gluino bound states \label{vat}}
\author{V Kartvelishvili and R McNulty}
\begin{abstract}
The properties of gluinonium are briefly reviewed. We then discuss
possibilities for detection at run II of the Tevatron via peaks in the di-jet
invariant mass spectrum.
\end{abstract}

If the decay of a gluino into a quark-squark pair is forbidden
kinematically and $R-$parity is conserved, 
the gluino can only decay   into a 
quark-antiquark pair an a neutralino, via a virtual squark, with a 
far longer lifetime. 
In this case the usual strategies for gluino
searches using high $P_T$ jets
and missing transverse energies are far less efficient --- the jets are 
more numerous and hence softer,
 while the missing energy is smaller. Consequently,
the reach for gluino searches is significantly reduced and it is quite
difficult to obtain a model-independent limit. 

In this case, however, there is a possibility
of observing the gluino indirectly, by detecting a gluino-gluino bound 
state{(see \cite{{Haber},{Chikovani96}} and references therein)}.
This has the advantage that the conclusions which can be drawn from a search 
for such states hold in a very wide class of supersymmetry models.
In addition, the detection of such a state would lead to a relatively
precise determination of the gluino mass, which could not be obtained easily
by observing the decay products of the gluino itself, as some of these
escape undetected. 

Gluino-gluino bound states (sometimes called gluinonium) can be
detected as narrow peaks in the di-jet
invariant mass distributions. The main problem is the high
background from QCD high $P_T$ jets, and thus it is vital to have 
two-jet invariant mass resolution as good as possible.

\subsection{Properties of gluinonium}

As strongly interacting fermions, gluinos have a lot in common with 
heavy quarks. There are important differences though:
\begin{itemize}
\item{gluino has no electroweak coupling, so its lifetime is defined
by its strong decays. For our case of interest, 
$m_{\tilde{g}}<m_{\tilde{q}}+m_{{q}}$, this means that gluino lives
long enough to form a bound state;}
\item{gluino is a colour octet; the potential between two gluinos is attractive 
not only if they are in a colour-singlet state, but also if they are in
colour octet states, both symmetric and antisymmetric;}
\item{gluino is a Majorana fermion (i.e.\ is its own antiparticle), and some
gluinonium states are forbidden due to the Pauli principle.} 
\end{itemize}

\bigskip
\begin{center}
\begin{tabular}{|c|c|c|c|}
\cline{1-4}
    & $ 1 $ & $ 8_S$ & $ 8_A $  \\
\cline{1-4}
\cline{1-4}
$ ^1S_0$ & $0^- \;  (\eta_{\tilde{g}}^0)$ &  $0^-  \; 
(\eta_{\tilde{g}}^8)$ &   \\
$ ^3S_1$ &  &  & $1^- \; (\psi_{\tilde{g}}^8)$   \\
\cline{1-4}
\end{tabular}
\noindent
\bigskip

{{\bf Table 1.} Spin-parities $J^P$ for the allowed low lying
states of
gluinonium with $L=0$. The three
columns correspond to the colour singlet state 1 and the  
symmetric and antisymmetric colour octet states $8_S$ and $8_A$ respectively.
}
\end{center}

The resulting spectra of low-lying gluinonium states 
\cite{Keung} 
are shown in Table 1. The allowed colour singlet 1 and 
symmetric octet $8_S$ states
have the same $J^P$ values as the charmonium states with $C = +1$, while
the allowed antisymmetric colour octet $8_A$ states
have the same $J^P$ values as the charmonium states with $C = -1$. 
In particular, the lowest lying  colour singlet and symmetric 
colour  octet states  are the  pseudoscalars  
$\eta_{\tilde{g}}^0$ 
and $\eta_{\tilde{g}}^8$ with $J^{P} = 0^{-}$, while
the lowest lying  antisymmetric colour octet state is 
vector gluinonium $\psi_{\tilde{g}}^8$ with 
$J^{P} = 1^{-}$.                                                         

All three $L=0$ states decay  via gluino-gluino annihilation.  
The pseudoscalars 
$\eta_{\tilde{g}}^{0,8}$ decay mainly to two gluons 
\cite{Keung} with decay widths $\sim 10^{-3}M$,
while  vector gluinonium $\psi_{\tilde{g}}^8$ decays predominantly 
into $q \bar{q}$ pairs \cite{Chikovani89}
with a decay width about $10^{-4}M$. 
Although  much larger than the free gluino decay width,
these widths are still very small compared to the gluinonium
mass $M \approx 2 m_{\tilde g}$. The size of all three states is of order 
$a^B \equiv 4(\alpha_s M)^{-1}$,
which is much smaller than the confinement length, thus justifying the
relative stability of  the  colour octet states (see
\cite{Haber,Chikovani89}). 

So, the vector gluinonium 
$\psi_{\tilde{g}}^8$ is a heavy compact object which behaves rather like a
heavy gluon, except that its coupling to quarks is much stronger than its
coupling to gluons. Hence it is most readily produced via
$q \bar{q}$ annihilation and  the Tevatron  is a 
promising place to look, being a source of  both
valence quarks and valence antiquarks. In contrast, the
pseudoscalar states $\eta_{\tilde{g}}^0$ and $\eta_{\tilde{g}}^8$  
couple predominantly to gluons, and can be produced equally well in both
$pp$ and $\bar{p}p$ collisions via the gluon-gluon fusion mechanism. 
Their production cross-section increases
more rapidly with energy than that for vector gluinonium,
and there is more chance of detecting them at
the LHC. 

\subsection{Vector gluinonium at the Tevatron}

The vector gluinonium is produced and decays
in $p \bar{p}$ collisions  via the subprocess
\begin{equation}\label{vecprod}
q + \bar{q} \rightarrow \psi_{\tilde{g}}^8\  \rightarrow q + \bar{q}, \; Q +
\bar{Q} 
\end{equation}
where  we use the symbols $q = u,d,s$  and $Q = c,b,t$ to distinguish
light and heavy quarks\footnote{Obviously, $t$-quarks contribute only if
the gluino is heavy enough, and even then for the range of gluino masses
accessible at the Tevatron this contribution is strongly
suppressed by the available phase space.}.

The nature of the background depends on $M/\sqrt{s}$.  
At  Tevatron energies, the range of interest lies 
mainly  in large values  $M/\sqrt{s}>0.2$, where the luminosity of 
colliding $q\bar{q}$ pairs prevails over that of gluon-gluon pairs.
In this region, the main sources of two-jet background are the subprocesses
\begin{equation}\label{bsub}
q + \bar{q} \stackrel{QCD}{\longrightarrow} 
g + g, \;\;\; q + \bar{q},\;\;\; Q + \bar{Q} 
\end{equation}
where the first two have the angular dependence 
$ \propto (1 - \cos^2 \theta^*)^{-2}$, peaking
sharply at $\cos \theta^* = \pm 1$, where $\theta^*$ 
is defined in c.m.\ frame of the two jets.
In contrast, the signal from the subprocess (\ref{vecprod})
has a much weaker dependence,  
$\sim 1 + \cos^2 \theta^*$.
Hence, a cut $|\cos \theta^*|<z$ should improve the signal-to-background ratio.

The usefulness of heavy quark tagging is clearly brought out by considering
the production ratios for the various final states in both the signal 
(\ref{vecprod}) and background (\ref{bsub}). The relative contribution of 
the three background subprocesses in (\ref{bsub}) 
at small $|\cos \theta^*|$ is given by \cite{Chikovani89}
\begin{equation}
g g : q \bar{q} : Q \bar{Q} = 14 : 65 : 6,
\end{equation}
while for the signal (\ref{vecprod}) one has 
\begin{equation}
g g : q \bar{q} : Q \bar{Q} = 0 : 3 : 2.
\end{equation}
Hence  by tagging the heavy quark jets one reduces the background
by a factor of $85/6\approx14$, while retaining 40\% of the signal.

At smaller gluinonium masses $M \approx 2 m_{\tilde{g}}< 200$ GeV, initial 
gluons  contribute much more significantly to the background, even with heavy 
quark jet tagging, through the subprocess
\begin{equation}\label{ggQQ}
g + g \stackrel{QCD}{\longrightarrow} Q + \bar{Q} \;.
\end{equation}
This makes the signal-to-background ratio hopelessly small for any
realistic di-jet invariant mass resolution. However, this region of
gluino masses is already covered by other methods.

\subsection{Simulation}

        \begin{figure}[top]
        \begin{center}
        \mbox{\epsfysize=10cm
        \epsfbox{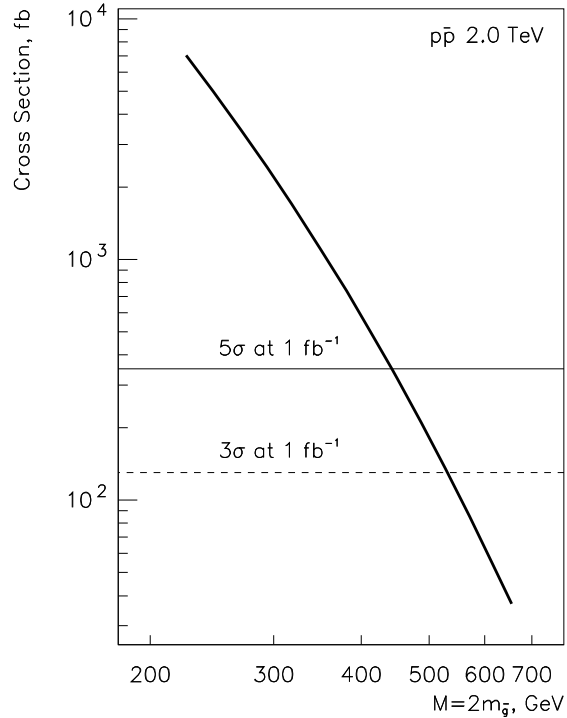}
        }
        \end{center}            
        \caption[1]{The calculated  production cross section of
        vector gluinonium in $p\bar p$ collisions at 2.0 TeV. The solid and 
        broken horizontal lines indicate the cross sections corresponding to a 
        statistical significance at the peak of 5 and 3 standard deviations 
        respectively, for a luminosity of 1 fb$^{-1}$. (See the text for the 
        cuts and resolutions used).}
        \end{figure}            

So, most of the two-jet QCD background at large invariant masses arises 
from  light quark and gluon jets, and the
signal-to-background ratio can be significantly enhanced by 
triggering on heavy quark jets \cite{Chikovani89}.
To check that this makes the detection of vector gluinonium a viable 
possibility at the upgraded Tevatron, 
we have simulated both the gluinonium signal and the 2-jet QCD background
using PYTHIA 5.7.
The vector gluinonium production and decay was simulated
by exploiting the fact that
$\psi_{\tilde{g}}^8$ behaves much like a heavy $Z^{\prime}$ with axial current
and lepton couplings set to zero and a known mass-dependent vector current
coupling to quarks, chosen to comply with the corresponding decay width
after
taking into account appropriate colour and flavour counting. This effective
coupling included the non-Coulomb corrections 
and an enhancement due to the fact that numerous radial excitations of the
$\psi_{\tilde{g}}^8$, which could not be separated from it for any
reasonable mass resolution, will also contribute. These yield an overall  
factor 
of between 1.8 and 1.6 depending on $M$, and the resulting  effective vector 
coupling $a_V$ falls exponentially from $a_V=0.225$ at 
$M=2m_{\tilde g}=225$ GeV to $a_V=0.172$ at $M=2m_{\tilde g}=450$ GeV. 
This signal sits on a much larger background, which has been simulated on 
the assumption
that it arises entirely from the leading order QCD subprocesses for
heavy quark pair production (\ref{bsub}) and (\ref{ggQQ}). A constant
$K$ factor $K=2.0$ has been used for both signal and background.

The cross section for vector gluinonium production at
the upgraded Tevatron with its energy increased to 2 TeV is
shown in Fig. 1. Only decays into heavy 
quark-antiquark pairs were taken into account, and the
tagging efficiency for at least one $c$- or $b$-quark jet
 was assumed to be $50\%$. The cut on the jet angle $\theta^*$
in the two-jet c.m\ frame was $|\cos \theta^*|<2/3$, and the cut on 
jet rapidity was $|y|<2.0$. 
The signal-to-background
ratio was found to be around $7-10\%$ at the peak for the assumed two-jet
invariant mass resolutions of 25 GeV, 30 GeV and 38 GeV at
$M=225$ GeV, 320 GeV and 450 GeV respectively.
One can hope to see
the gluinonium signal from
gluinos with masses up to 220 GeV as a 5 standard deviation peak,
and the signal from gluinos with masses up to 260 GeV as a 3 standard 
deviation peak.
Note that the statistical  significance of the peak is essentially
inversely proportional to the two-jet invariant mass resolution, so the
reach can be significantly extended if some way is found to improve the latter.

\subsection{Conclusion}

We
conclude that gluinonium states can be detected as narrow peaks in the
di-jet invariant mass spectra, effectively complementing more traditional
gluino searches, in the case when the gluino is lighter than the
squarks. 

In $p\bar p$ collisions one expects copious production of 
vector gluinonium, which decays predominantly to $q\bar q$ pairs.
The high efficiency of the heavy quark jet tagging
together with the boost of the Tevatron energy and luminosity should
allow one to reach
gluino masses of  220-260 GeV at $\sqrt{s}=2.0$ TeV and 1000 pb$^{-1}$, with
realistic efficiencies, resolutions and experimental cuts taken into
account. It is crucial, however, to improve tagging efficiency for both
$c-$ and $b-$quark jets, as well as the two-jet invariant mass resolution
for these jets.

\def\mpl{\ifmmode \overline M_{Pl}\else $\bar M_{Pl}$\fi}

\section{Experimental Signatures from Theories with
Extra Dimensions
\label{sec:us}}
\author{J Grosse-Knetter, J Holt and S Lola}

\begin{abstract}
We discuss possible experimental signatures 
and distinctions between two models
with extra dimensions.
In the first model a number $n$
of large extra dimensions is postulated, while
the second involves
the addition of only one extra dimension, but 
with a metric which is non-factorisable into 4+1 separate 
dimensions (Randall-Sundrum model).
\end{abstract}

An important issue in 
extending the Standard Model of Particle
Physics, is the hierarchy problem, arising from
the existence of two vastly different
fundamental scales ($M_{W}$ and $M_{Pl}$).
There are ways to evade this problem, such as
technicolour and supersymmetry.
A third solution which has recently received considerable attention,
is to {\em identify}\/ the Planck scale with the electroweak scale, by
introducing extra dimensions into which gravitons are able to propagate.
Here, we discuss some experimental
aspects of two classes of such models.

\subsection{Models with large Extra Dimensions} 
The first 
set of models considered here is the proposal of \cite{Dim}
where the Plank scale, $M_{Pl}$, is related to the scale
of gravitational interactions, $M_{D}$ in a space which includes
$n$ extra compact dimensions of radius $R$.
In this case, one finds that
$R^n M_D^{n+2} = M_{Pl}^2$ \cite{Dim}, where
$n$ is the number of the extra dimensions:
for $n=1, R \approx 10^{13}m $ which is obviously
excluded. However, already for $n=2$, $R \approx 1 mm$.
No effects  of the extra dimensions
on Standard Model
fields in accelerators have been observed, 
one therefore assumes
that our 4-dimensional world lies on a brane while the gravitons
(which feel the extra dimensions)
can propagate on the bulk. 
Since momentum in extra dimensions is seen as mass
in four dimensions, in computing graviton emission
one has to sum over a tower of massive 
Kaluza-Klein states, with
masses $m\approx \frac{2 \pi n}{R}$.
The coupling to any single mode has the normal gravitational 
strength ($\approx \frac{1}{\overline{M}_{Pl}}$,
where $\overline{M}_{Pl} = M_{Pl}/\sqrt{8 \pi}$), while the
mass of each mode is very small.
However the large multiplicity of modes, 
given approximately by
$\approx (ER)^n$, where $E$ denotes the energy that is
available to the graviton, increases the effective
coupling $1/M_s$ dramatically.

The Feynman rules for the new vertices \cite{GRW} are calculated
from ${\cal L} = -\frac{1}{\overline{M}_{Pl}} 
g_{\mu \nu}^j T^{\mu \nu}$,
where $j$ labels the Kaluza-Klein modes.
Some features for the interactions
that arise in this class of models, which are important
for accelerator searches, are the following:
(i) the interactions are flavour-independent. 
(ii) the individual modes are very light and couple
very weakly, thus may not be produced on resonance.
(iii) the spin-2 nature of the graviton can be determined via
angular distributions of the cross sections.
(iv) the effective coupling scales as
$\frac{1}{\overline{M}_{Pl}^2} (E R)^n \approx \frac{E^n}{M_D^{n+2}}$
and therefore a strong energy dependence 
(with increase of the cross sections as the energy increases)
should appear.

\begin{figure}[t]
\begin{center}
\mbox{\epsfig{file=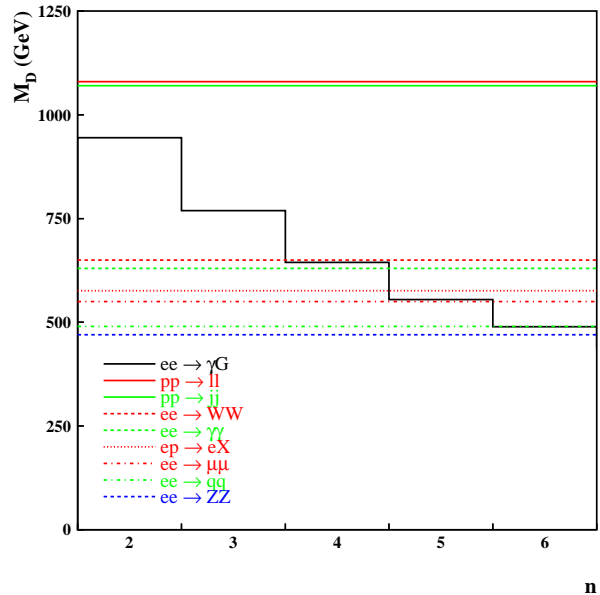,width=0.55\textwidth}}
\end{center}
\caption{\it Limits on the scale $M_{D}$ as a function of the number
of extra dimensions $n$ from different channels. 
References are given in the text.}
\label{fig:gravlim}
\end{figure}
\subsection{Limits on Models with Large Extra Dimensions}
The effects of gravity in models with large extra dimensions, have been 
searched for using the data from a number of experiments in different 
channels. No evidence for these effects has been found and lower
limits on the parameter $M_{D}$, as a function of the number of
extra dimensions, $n$, have been obtained from the different sets of data.
Some of these limits, taken from \cite{gravlim} together with the results 
presented below from HERA DIS, are shown in figure~\ref{fig:gravlim}.
The limits coming from $e^{+}e^{-} \rightarrow \gamma~{\mathrm{Graviton}}$
at LEP II show a strong dependence on the number of extra dimensions.
The cross-section for this process depends on the phase space available 
to the emitted gravitons which depends on $n$. The other limits 
are derived from processes which involve virtual exchange of gravitons.
The effective string scale $M_s$ 
has been taken to be equal to $M_{D}$.
The graviton exchange can
interfere constructively or destructively with the Standard Model processes,
set by a parameter $\lambda = \pm 1$; the above
limits  are for $\lambda = +1$

The best limits under these assumptions come from the TEVATRON
from di-lepton production using a combination of CDF and D0 data. Limits from
CDF alone on di-jet production are very competitive, suggesting that 
improved sensitivity could be obtained by including D0 di-jet data.
Combining all the channels studied by L3 at LEP II, gives a lower limit on
$M_{s}$ of 860 GeV~\cite{gravlim} from approximately 50 $pb^{-1}$ of 
data. The four LEP collaboration now have a total of more than 
1.6 $fb^{-1}$ worth of data collected at energies above $\sim 183$ GeV.
Combining all results sensitive to virtual graviton exchange, from all four 
experiments, could give results which would compete with those 
from the TEVATRON.

\subsection{Fits to HERA DIS data}
One of the processes with sensitivity to effects predicted
from Kaluza-Klein models with large extra dimensions
is the neutral-current (NC) 
deep-inelastic scattering (DIS) of positrons off protons.
Effects are expected through the exchange of gravitons coupling to
both $e^+ q$ and $e^+ g$ in addition to the SM-exchange of photons and 
$Z^0$ bosons~\cite{DIStheo}. These
additional contributions (expected 
at large $Q^2 \approx M_s$, 
lead to an enhancement in the
cross section $d\sigma/dQ^2$, where $Q^2$ is the squared four-momentum 
transferred between positron and proton. 

\begin{figure}
\begin{minipage}[h]{8in}
\hspace*{2 cm}
\epsfig{file=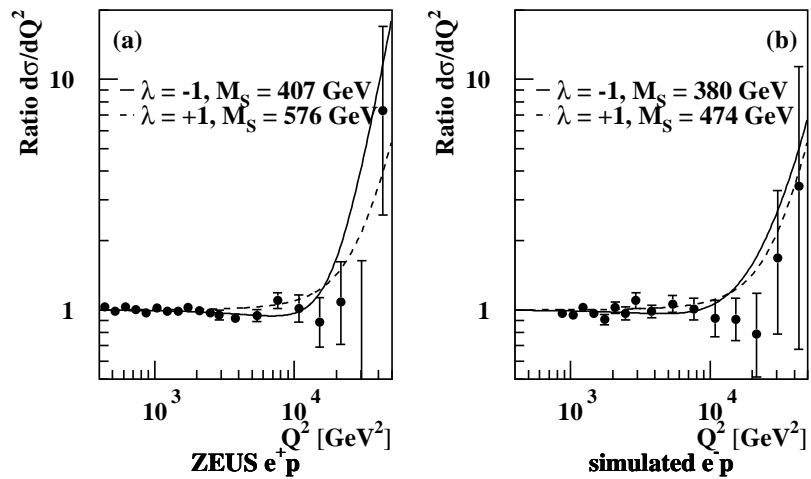, width=5in}
\hfill
\end{minipage}
\caption{\it Fits with a model including graviton exchange 
to HERA NC DIS the cross section $d\sigma/dQ^2$: (a) fit to ZEUS $e^+p$
data; (b) fit to simulated data corresponding to $e^- p$ data recently
taken at HERA.}
\label{fig:disfit}
\end{figure}

Fitting the cross section
expected from the combination of the SM and graviton
exchange to recent $e^+p$ NC DIS data from ZEUS~\cite{ZDIS} 
(similar results are expected from corresponding H1 data~\cite{H1DIS})
using CTEQ4 PDFs
yields $95\%$ CL limits of $M_s > 407$~GeV ($\lambda = -1)$ and
$M_s > 576$~GeV ($\lambda = +1)$ in agreement with expectations
based on preliminary data~\cite{DIStheo}. The results are illustrated in 
figure~\ref{fig:disfit}(a) as the ratio of fitted cross section $d\sigma/dQ^2$ 
to that expected from the SM.


It was further investigated whether the recent HERA $e^- p$ NC DIS
data~\cite{HERAelec} can provide additional information on the mass-scale 
of extra dimensions. For this purpose $e^- p$ NC DIS data were simulated
based on the uncertainty expected from the luminosity of the existing data
sample. Fits similar to above were performed as shown in 
figure~\ref{fig:disfit}(b) yielding $M_s > 380$~GeV ($\lambda = -1)$ and
$M_s > 474$~GeV ($\lambda = +1)$. Thus no stricter limits than already
obtained from the $e^+p$ data should be expected.

\begin{figure}
\begin{minipage}[b]{8in}
\epsfig{file=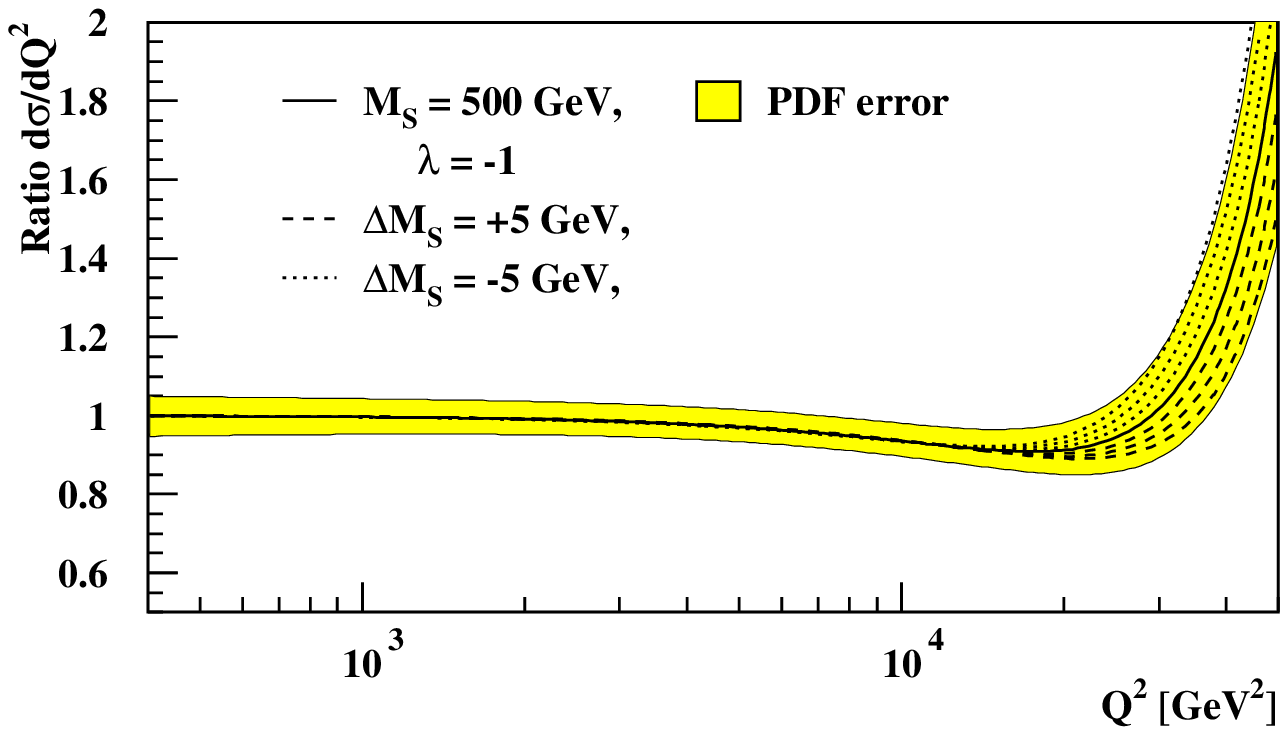 ,width=3.3in}
\epsfig{file=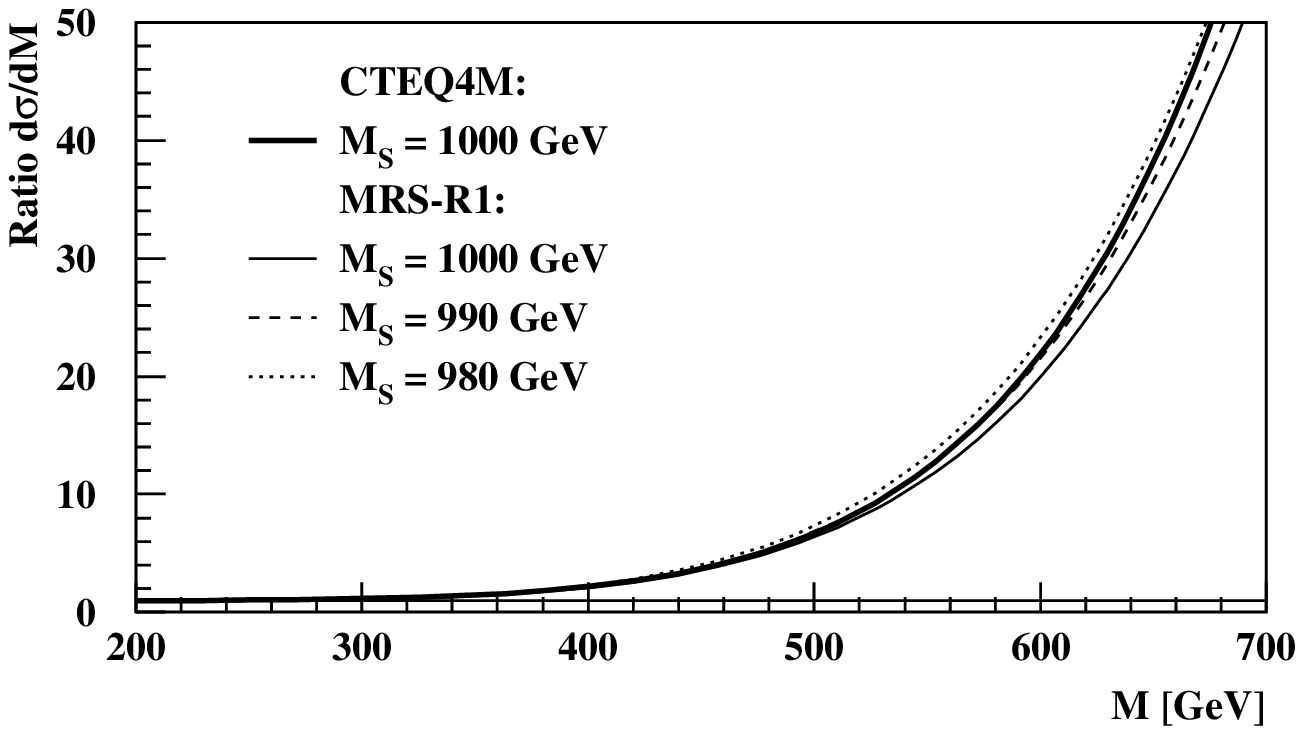,width=3.3in} \hfill
\end{minipage}
\caption{\it
Effect of PDF uncertainties on limits in $M_s$ obtained from fits
to the NC DIS cross section $d\sigma/dQ^2$
(left) and
to the Drell-Yan cross section $d\sigma/dM$ (right).
}
\label{fig:dispdf}
\end{figure}

The predicted cross-sections for process at the TEVATRON and HERA, are
sensitive to uncertainties in the parton distributions functions (PDFs)
of the proton. We 
first estimate the uncertainties in $M_s$ arising from PDF uncertainties in
fits to HERA DIS data. For this purpose
results are used from a NLO QCD fit~\cite{ZDIS}
to measurements of proton structure functions and
quark asymmetries from collider and fixed target 
experiments.
The fit propagates statistical and correlated systematic errors from each 
experiment to corresponding errors in the PDFs which are used to
determine uncertainties in the cross section $d\sigma/dQ^2$, including
contributions from graviton exchange. The result is shown as ratio
of $d\sigma/dQ^2$(SM+graviton) for $M_s = 500$~GeV and
$\lambda = -1$ to $d\sigma/dQ^2$ (SM) 
in figure~\ref{fig:dispdf} (left). 
The band shows the uncertainty
in the ratio $d\sigma/dQ^2$ (SM+graviton)/(SM) arising from PDF uncertainties.
The latter was compared to the variation in the ratio as $M_s$ changes, for
nominal PDFs. These are shown by the dashed and dotted lines, where incremental
changes in $M_s$ of 5~GeV are made. 
This procedure shows that only small errors in 
$M_s$, of approximately 15~GeV, arising from PDF uncertainties
should be expected.

Similar effects from PDF uncertainties are expected for fits to
TEVATRON data. 
To check the effect of PDF uncertainties on this limit the
Drell-Yan cross section $d\sigma/dM$ 
($M$ being the hard scale, ie the $e^+ e^-$ mass)
is determined in leading-order QCD with
two different PDF sets\footnote{The PDF uncertainties from
the QCD fit described above were only available for hard scales
corresponding to $M<300$~GeV, so below the range sensitive to
graviton exchange and could thus not be used here.}
including contributions from graviton exchange, in
figure~\ref{fig:dispdf}
(right).
This analysis indicates that uncertainties in the limits on $M_s$
resulting from PDF uncertainties should be expected to be of order 
10 to 20~GeV.

\subsection{Randall--Sundrum in 
{$e^{+}e^{-} \rightarrow \mu^{+}\mu^{-}$} at LEP II}
So far, we have been referring to models with more
than one extra dimensions and with a factorisable metric.
One can instead envisage a case where a large mass
hierarchy may be generated by an exponential ``warped''
factor of a small compactification radius, $r_c$,
in a case of a 5-dimensional non-factorisable geometry
\cite{RaSun}. It turns out that
a field with a fundamental mass parameter $m_0$
on the visible world 
appears to have a physical mass $m=e^{-kr_c\pi}m_0$,
where $k$ is a scale of
order the Planck scale, relating the 5-dimensional 
Planck scale $M$ to the 
cosmological constant.
The interaction Lagrangian in the 4-dimensional effective theory 
indicates that, while the
zero mode couples with the usual
4-dimensional strength, the 
massive KK states are relatively 
unsuppressed. Thus, unlike the previous
case of more than one factorisable extra dimension, 
now (i)
the individual modes are heavier (${\cal O}(TeV)$).
(ii) the individual modes couple
with weak interaction strength thus may
be produced on resonance.
(iii) as one increases the centre of mass
energy, one may hope to probe a multi-resonance
effect.

For instance, for the first mode, the mass, $m_1$ and the width, $\Gamma_1$, 
of the resonance are given by $m_1 = \Lambda_\pi x_1 ( k / \mpl)$ and 
$\Gamma_1 = \rho m_1 x_1^2 (k / \mpl)^2$
where $x_1$ is the first non-zero root of 
the $J_1$ Bessel function and $\rho$ is 
a constant which depends on the number of decay channels.
Moreover, by making the substitution
 ${\lambda\over M_s^4} \to {i^2\over 8\Lambda_\pi^2}\sum_{n=1}^\infty
{1\over s - m_n^2 - i s \Gamma_n / m_n }$ in the formulas 
obtained for $n$ factorisable extra-dimensions,
one can proceed to calculate any process. Clearly,
as $\frac{k}{\overline{M}_{Pl}}$ grows,
the resonant peaks are substituted by
a contact-interaction behaviour.

\begin{figure}[t]
\begin{center}
\mbox{\epsfig{file=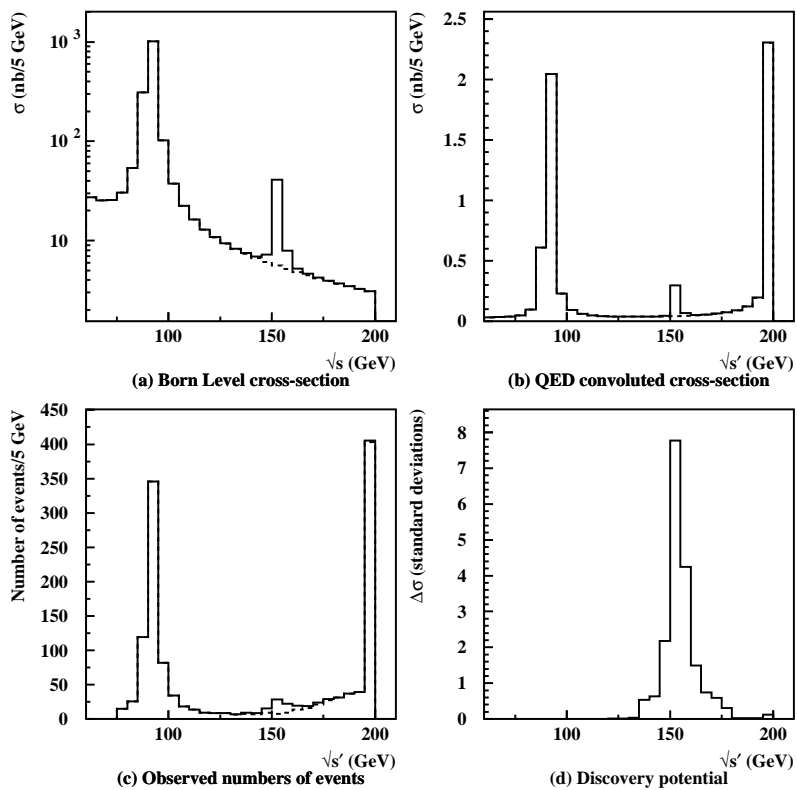,width=0.75\textwidth}}
\end{center}
\caption{\it Predictions for the Randall--Sundrum model with 
$\Lambda_\pi = 800$ GeV and $k/\mpl = 0.05$. The width of the 
Randall--Sundrum resonance with mass of $\sim 150 GeV$ is 1 GeV.
A centre--of--mass energy of 200 GeV and a luminosity of
200 $pb^{-1}$ have been assumed.
In figures (a)-(c) the solid line is the prediction of the Randall--Sundrum 
Model, the dashed line is the prediction of the Standard Model.}
\label{fig:gravrs}
\end{figure}

The possibility of finding a Randall-Sundrum resonance with a mass as low
as 100-200 GeV in $e^{+}e^{-} \rightarrow \mu^{+}\mu^{-}$ at LEP II has
been investigated. It would be possible to
hunt for such a resonance by examining the distribution of the number
of muon events observed as a function of the invariant mass, 
$\sqrt{s^{\prime}}$, of the pair of muons, taking advantage of initial
state radiation which provides access to invariant masses below the 
centre--of--mass energy of the LEP collision energy, $\sqrt{s}$.

Born--level predictions for the cross--section, $\sigma_{0}(s)$,
of the Randall--Sundrum model 
with $\Lambda_\pi = 800$ GeV and $k/\mpl = 0.05$ are shown in 
figure~\ref{fig:gravrs}. The mass of the first resonance is 
approximately 150 GeV. In principle the parameter $\rho$ which determines 
the width can be calculated. For the studies presented here 
$\rho$ was chosen so that the width of the first resonance was 1 GeV. 
The QED convoluted cross--section as a function of $\sqrt{s^{\prime}}$ is 
given by $
 \sigma(s^{\prime}) = R(s^{\prime}) \sigma_{0}(s=s^{\prime}) $.
The radiator function, $R$, was computed for bins of
$s^{\prime}$ for the by computing the Born--level
cross--section and $\sigma(s^{\prime})$ in the Standard Model. This was then
applied to the predictions including the Randall--Sundrum resonance. The QED 
convoluted cross-section for a centre--of--mass energy of 200~GeV is shown 
in figure~\ref{fig:gravrs}(b). The predicted numbers of events 
figure~\ref{fig:gravrs}(c), for a luminosity 200~$pb^{-1}$.
The $\sqrt{s^{\prime}}$ distribution has been smeared to take into account
the experimental resolution, which was obtained from a simulation
of the DELPHI detector.

The difference between the the Randall--Sundrum Model and the Standard Model,
$\Delta\sigma$, is shown in figure~\ref{fig:gravrs}(d) in terms of the
number of statistical standard deviation on the expected numbers of events.
Even taking into account the resolution on $\sqrt{s^{\prime}}$, it is clear 
that a Randall--Sundrum resonance with the parameters given above would be 
observable at LEP II given 200~$pb^{-1}$ at $\sqrt{s}=200$~GeV. In reality
each of the LEP experiments have this much data collected at 
centre--of-mass energies between 192 and 202~GeV. The spread of energies 
should not significantly change the ability to observe such a resonance,
or place limits in the ($\Lambda_\pi , k/\mpl$) plane.
A fit could include all centre of mass energies and all other 
final states in $e^{+}e^{-}$ collisions sensitive to the presence
of a Randall--Sundrum mode.

\newcommand{\ptevt}{\mbox{$p_{t}^{\mathrm{miss}}$}}
\newcommand{\chz}{\mbox{$\tilde{\chi}^0_1$}}
\newcommand{\chpm}{\mbox{$\tilde{\chi}^\pm_1$}}
\newcommand{\cpair}{\mbox{$\tilde{\chi}^+_1\tilde{\chi}^-_1$}}
\newcommand{\stau}{\tilde{\tau}}
\newcommand{\lpair}{\mbox{$\tilde{\ell}^+\tilde{\ell}^-$}}
\newcommand{\llpair}{\mbox{$\tilde{\ell}_L^+\tilde{\ell}_L^-$}}

\newcommand{\erpair}{\mbox{$\tilde{e}_R^+\tilde{e}_R^-$}}
\newcommand{\rpair}{\mbox{$\tilde{\ell}_R^+\tilde{\ell}_R^-$}}
\newcommand{\roots}     {\protect\sqrt{s}}
\newcommand{\stwo}     {\mbox{$\protect\sqrt{s}$~=~200~GeV}}
\newcommand{\seight}     {\mbox{$\protect\sqrt{s}$~=~183~GeV}}
\newcommand{\snine}      {\mbox{$\protect\sqrt{s}$~=~189~GeV}}
\newcommand{\epair}{\mbox{${\mathrm e}^+{\mathrm e}^-$}}
\newcommand{\dm}{\mbox{$\Delta m$}}
\newcommand{\eell}{\mbox{\epair\lpair}}
\newcommand{\epm}{\mbox{${\mathrm e}^\pm$}}
\newcommand{\mupm}{\mbox{$\mu^\pm$}}
\newcommand{\spp}{\mbox{$p$/\Ebeam}}
\newcommand{\Ebeam}{\mbox{$E_{\mathrm{beam}}$}}
\newcommand{\stevt}{\mbox{$p_{t}^{\mathrm{miss}}$/\Ebeam}}
\newcommand{\wpair}{\mbox{$\mathrm{W}^+\mathrm{W}^-$}}
\newcommand{\lept}{\mbox{$\ell^-$}}
\newcommand{\dW}{\mbox{W$^-\rightarrow\lept\nbar$}}
\newcommand{\nbar}{\mbox{$\overline{\nu}$}}
\newcommand{\llnunu}{\mbox{$\ell^+\nu\,\ell^-\nbar$}}
\newcommand{\ee}{{\mathrm e}^+ {\mathrm e}^-}
\newcommand{\sele}{\tilde{e}}
\newcommand{\sll}{\tilde{e}}
\newcommand{\sell}{\tilde{\ell}}
\newcommand{\snu}{\mbox{$\tilde{\nu}$}}
\newcommand{\smu}{\tilde{\mu}}
\newcommand{\chp}{\tilde{\chi}^+_1}
\newcommand{\ellell}{\ell^+ \ell^-}
\newcommand {\slepton}       {\tilde{\ell}}
\newcommand {\neutralino}    {\tilde{\chi }^{0}_{1}}
\newcommand {\chargino}      {\tilde{\chi }^{\pm}_{1}}
\newcommand{\msmu}{\mbox{$m_{\smu}$}}
\newcommand{\mchz}{\mbox{$m_{\tilde{\chi}^0_1}$}}
\newcommand{\ipb}{\mbox{pb$^{-1}$}}
\newcommand{\IL}{\mbox{$\cal L$}}
\newcommand{\nt}{\tilde{\chi}^0}
\newcommand{\Ecm}{\mbox{$E_{\mathrm{cm}}$}}
\newcommand{\smp}{Standard Model processes}
\newcommand{\wenu}{\mbox{W$^-$e$^+ \nu$}}
\newcommand{\chienu}{\mbox{$\tilde{\chi}^-_1$e$^+ \tilde{\nu}$}}
\newcommand{\chiee}{\mbox{$\tilde{\chi}^0_1\tilde{e}^+e^-$}}
\newcommand{\msnu}{\mbox{$m_{\tilde{\nu}}$}}
\newcommand{\mch}{\mbox{$m_{\tilde{\chi}^\pm_1}$}}
\newcommand{\mel}{\mbox{$m_{\tilde{e}_L}$}}
\newcommand{\mer}{\mbox{$m_{\tilde{e}_R}$}}
\newcommand{\mslept}{\mbox{$m_{\tilde{\ell}}$}}

\newcommand{\pfour}{\mbox{$\begin{array}{c} p \\ \sim \end{array}$}}

\section
{Some Alternative Tests of Standard Supersymmetry with
Events Containing Isolated Leptons and Missing 
{\boldmath $p_{t}$} at LEP2 \label{sec:terry}}

\author{D~Hutchcroft, J~Kalinowski, R~McNulty, G~Wilson, T~Wyatt}

\begin{abstract}
A number of potential new physics processes can give rise to
events containing isolated charged leptons and missing $p_{t}$ at LEP2.
Most attention in this field has been focussed on the pair production
of equal mass particles, which leads to events containing two leptons
of roughly equal momenta.
In this report we discuss potential new physics processes with the
following experimental signatures:
(i)~events containing two leptons
of unequal momenta; (ii)~events containing a single visible lepton and
no other activity in the detector.
\end{abstract}
In the Standard Model (SM), low multiplicity events containing 
 charged leptons and 
significant missing transverse momentum, \ptevt , arise from the final state
 \llnunu.
 The most important SM process contributing to this final state 
is \wpair\ production in which both W's
decay leptonically: \dW\ (with $\ell = \mathrm{e}, \mu, \tau $),
thus producing events containing an ``acoplanar''\footnote{The acoplanarity angle is defined  as 180$^{\circ}$ minus the angle
between the two lepton candidates in the plane transverse to the 
beam direction.} 
pair of observed leptons.
The SM subprocess leading to the final state \wenu\ tends to produce
events containing a single observed lepton, since the e$^+$ has a
high probability to be scattered at a small angle to the beam
direction and thus escape detection. 

\begin{figure}
\epsfxsize=\textwidth 
 \epsffile[0 0 580 600]{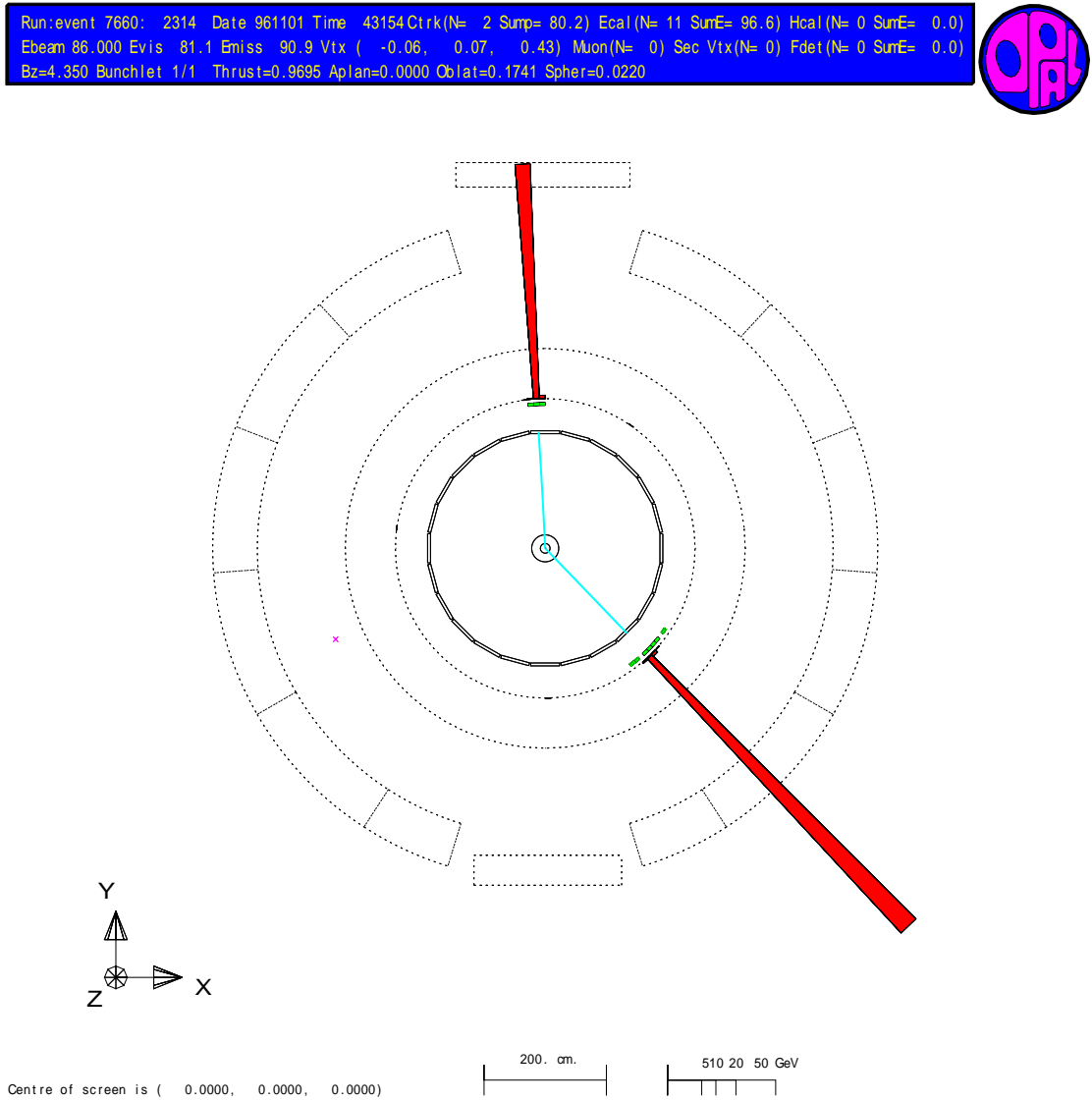}
 \caption{An acoplanar di-lepton candidate selected by OPAL at 172~GeV.
} 
\label{fig-epic6}
\end{figure}

Events containing 
charged leptons and  \ptevt\ are also an experimental
signature for the production of new particles that decay
to a charged lepton accompanied by one or more
invisible particles.
For example, acoplanar di-lepton events are a signal for 
the pair production of new particles such as: 
\begin{description}
\item[charged scalar leptons (sleptons):]
$\sell^\pm \rightarrow  {\ell^\pm} \nt_1$,
where $\sell^\pm$ may be a selectron ($\sele$), smuon ($\smu$) or stau
($\stau$), $\ell^\pm$ is the corresponding charged lepton and $\nt_1$ 
is the lightest neutralino.
\item[charged Higgs bosons:] $\mathrm{H}^{\pm} \rightarrow \tau^\pm \nu_\tau$.
\item[charginos:] $\chpm \rightarrow \ell^\pm \snu$ (``2-body'' decays)
\ \   or \ \ 
$\chpm \rightarrow \ell^\pm \nu \chz$ (``3-body'' decays).
\end{description}
A typical candidate event is  shown in figure~\ref{fig-epic6}.

The LEP detectors provide hermetic detection for showering and minimum
ionising particles, typically down to an angle of around 0.04~rad with respect
to the beam direction.
This means that the potential background from SM 
processes such as  \eell , which have four charged leptons in
the final state (of which only two are observed in the detector),
can be reduced to a low level.
Such potential backgrounds do, however, mean that the
 scaled missing transverse momentum of selected events, \stevt , has
to be required to exceed around 0.04.

A general search for the anomalous production of events of this type
can be made by comparing the number and general properties of the
selected data with the expectations from the SM\@.
However, because of the very large SM cross-section of around
2~pb, such a search is sensitive only to fairly large
deviations from the SM expectations.
When searching for a particular new particle the sensitivity can be
increased by  considering an event as a potential
candidate only if the properties of the observed event are
consistent with expectations for the particular new physics signal
under consideration.

An important property of the selected events that allows new
physics sources to be distinguished from the SM \llnunu\ 
final states is the momentum of the observed leptons.
The  SM \llnunu\ from \wpair\ are characterised by the 
production of two leptons,
both with \spp\ around 0.5.
In the SM \eell\ events both observed leptons tend to 
have low momentum. 
In the new physics signal events the momentum distribution of the expected 
leptons varies strongly
as a function of the mass difference, \dm , between the parent particle (e.g.,
selectron) and the invisible daughter particle (e.g., lightest neutralino),
and, to a lesser extent, $m$, the mass of the parent particle.
When performing a search at a
particular point in $m$ and \dm , the SM background can be minimised
by considering an event as a potential
candidate only if the
momenta of the observed leptons are
consistent with expectations.

The results of the lepton identification and angular distributions may
also help to reduce the SM background in some searches. 
In SM \llnunu\ events from \wpair , equal numbers 
of \epm , \mupm\ and $\tau^\pm$ are produced and there is no
correlation between the flavours of the two charged leptons in the
event.
Some new physics sources of acoplanar lepton pair events, such as
slepton pair production, would produce events in which the two leptons
have the same flavour.
The charge-signed angular distribution of the leptons in the SM events
shows a strong peak in the forward direction due to the dominance of 
the neutrino exchange amplitude and the V-A nature of W decay.
This is in contrast to the expectation, for example, in
 smuon, stau 
and charged Higgs production, in which the angular distribution is 
forward-backward symmetric and peaked towards 
$\cos{\theta}=0$, due to the scalar nature of these particles.  

There is a risk in this approach that the increased sensitivity in the
particular individual search channels considered
may be obtained at the cost of a
lack of generality of the overall search.
In order to avoid the danger that a new physics baby might be thrown
out with the SM bathwater, it is important to ensure that the widest
possible range of experimental signatures from potential new physics
sources is searched for.

Searches for new physics in the acoplanar di-lepton channel including the
data up to \snine\ have been
published by OPAL~\cite{opal} and ALEPH~\cite{aleph}.
Similar searches including the
data up to \seight\ have been
published by L3~\cite{l3} and DELPHI~\cite{delphi}.
These analyses tend to focus primarily on the pair production of equal
mass particles such as  charged scalar leptons (\llpair , \rpair ), or 
leptonically decaying  charged Higgs 
bosons and charginos.
In this case, the two observed leptons are
expected to have the same momentum spectrum, so that one  searches
for events containing two high (low) momentum leptons in the case of high (low)
\dm.

\begin{figure}[htbp]
\vspace{1.5cm}
\epsfxsize=\textwidth 
\epsffile[-100 150 580 600]{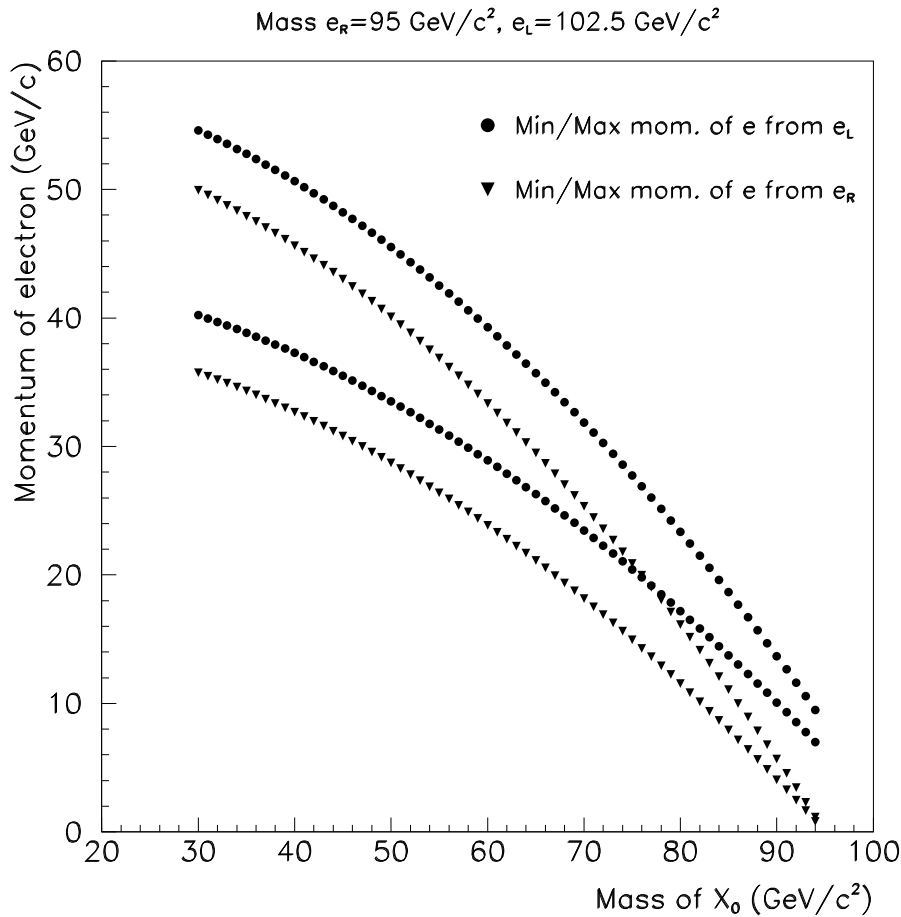}
 \caption{In \lrpair\ production: the kinematically
allowed ranges of the momenta of the two observed electrons as a
function of  \mchz , for the specific choice of
$\mer = 95$~GeV, $\mel = 102.5$~GeV and \stwo. 
} 
\label{fig-momentum}
\end{figure}

A possible  source of acoplanar lepton pair events with unequal
momentum leptons is the associated production of left- and
right-chiral selectrons (\lrpair ), since these particles, in general,
have different masses.
For example,  figure~\ref{fig-momentum} shows, for two-body decays
$\sell^\pm \rightarrow  {\ell^\pm} \nt_1$, the kinematically
allowed ranges of the momenta of the two observed electrons as a
function of the lightest neutralino mass, \mchz , for the specific choice of
$\mer = 95$~GeV, $\mel = 102.5$~GeV and \stwo.  
It can be seen that for low  \mchz\ (and thus high \dm )  
the momentum distributions
of the two electrons overlap substantially, but that as  \mchz\
increases (and thus \dm\ becomes small) the momentum distributions
become quite separated.

\begin{figure}[htbp]
\epsfxsize=\textwidth 
\epsffile{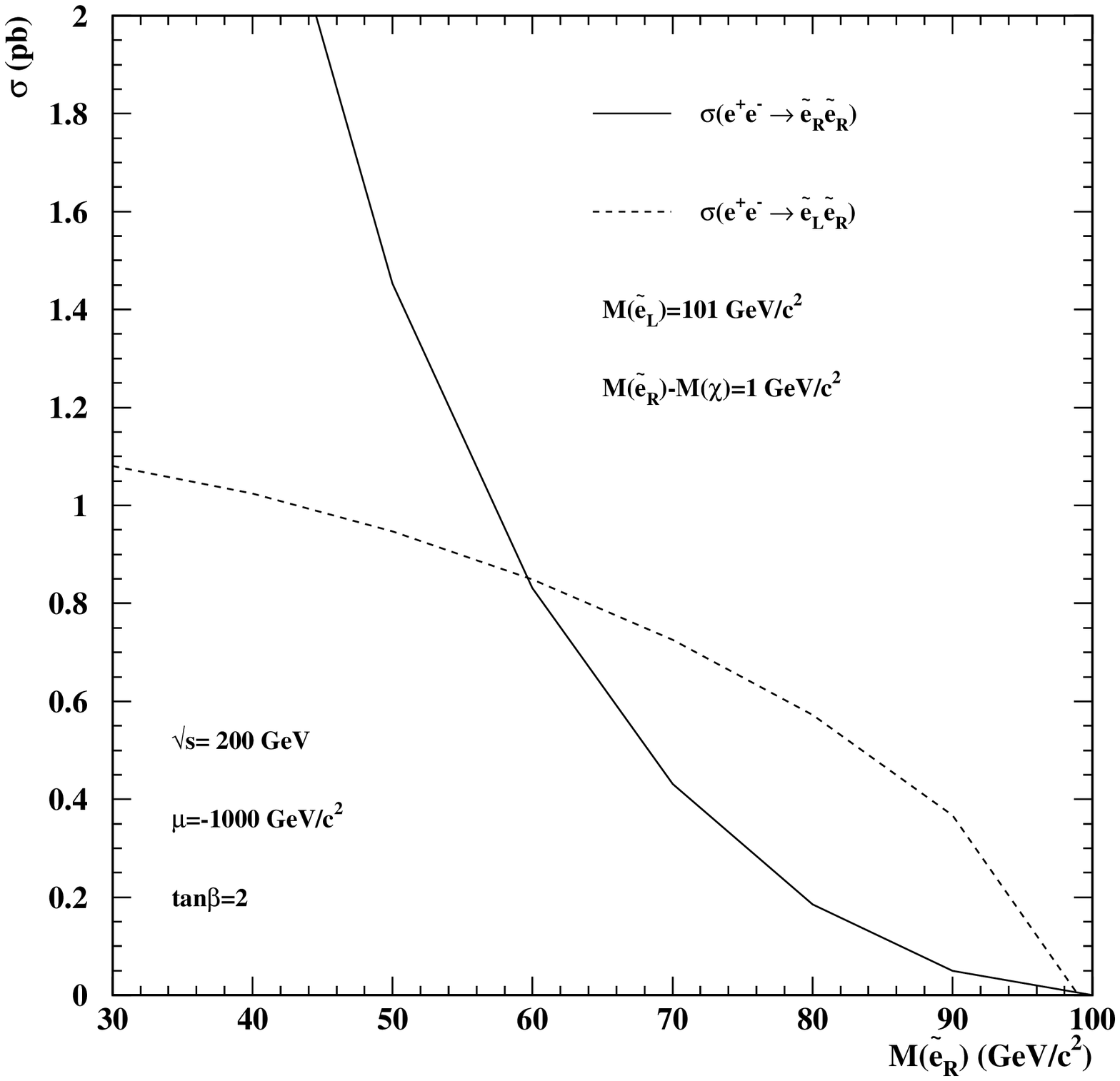}
 \caption{The
cross-sections for  \lrpair\ and \erpair\ as a function of \mer , for
the specific choices $\dm = \mer - \mchz = 1$~GeV ,
$\mel = 101$~GeV and \stwo.
} 
\label{fig-cross}
\end{figure}

Another feature of \lrpair\ production that makes it potentially
interesting is that, because  \lrpair\ results from the t-channel exchange of
a \chz , the expected production cross-section depends on $\beta/s$.
This may be contrasted with the $\beta^3/s$ dependence of
the cross-section for s-channel production of  \llpair\ and \rpair.
Near to the kinematic limit the cross-section for  \lrpair\ may be an
order of magnitude higher than the pair production cross-section for
the lightest selectron.
This is illustrated in figure~\ref{fig-cross}, in which we compare the
cross-sections~\cite{mssmlib} for  \lrpair\ and \erpair\ as a function of \mer.
The cross-sections are shown  for
the specific choices $\dm = \mer - \mchz = 1$~GeV ,
$\mel = 101$~GeV and \stwo.
However, the general features of the plot --- that $\sigma_{\lrpair}$
is around
 100--500~fb and is about an order of magnitude larger than
$\sigma_{\erpair}$ --- are true for a fairly large range of  \mel,
\mer\ and \mchz.

A feasibility study for a search at the example point $\mchz =90$~GeV, 
$\mer = 95$~GeV, $\mel = 102.5$~GeV and \stwo , has been performed 
using SM and selectron Monte Carlo events~\cite{opalmc} processed with a full
simulation of the OPAL experiment.
From the sample of events that pass a general selection of
acoplanar di-lepton events, the lepton identification was required to
be consistent with an electron pair
and the lepton momenta were required to be in the ranges:
$3 < p_1 \mathrm{(GeV)} < 7$; $9 < p_2 \mathrm{(GeV)} < 17$.
(These are
significantly broader than the kinematically allowed ranges
from figure~\ref{fig-momentum} in order to allow for the effects of
detector resolution.) 
A selection efficiency of around 65\% was achieved with a SM expected
background of 8~fb.
With an integrated luminosity of 500~pb$^{-1}$ per experiment
collected at LEP2, such searches are clearly feasible and should be performed.

How to organise such a search does present some problems, however.
In the more standard search for pair production of equal mass
particles there are two unknown masses, e.g., \mslept\ and \mchz.
Signal Monte Carlo events have to be generated, event selection cuts
or multivariate discriminants have to be optimised, and limits have to be
calculated, at each point in a finely spaced grid that covers 
the whole of the kinematically allowed region of this
2-D parameter space. 
This is time consuming, but achievable.
A search for the associated production of unequal mass particles
involves three unknown masses, e.g., \mel , \mer\ and \mchz.
Further work is needed to determine how best to perform the 
experimental search and present limits in this 3-D  parameter space. 

The associated production of \lrpair\ clearly motivates the search for
events containing two electrons of unequal momentum.
However, this is no reason to limit the experimental search to
electron pair events.
In addition to grounds of experimental generality, specific new
physics models predict the possibility of observing acoplanar lepton
pairs of unequal momentum with arbitrary lepton flavour.
For example,~\cite{inv} describes the scenario of \wpair\ production
in which one W decays normally and the other decays via
$W^\pm\rightarrow\chz\chpm$ followed by
$\chpm \rightarrow \ell^\pm \snu$.
If the mass difference between  \chpm\ and \snu\ is less than about
2~GeV the direct searches for \cpair\ followed by
$\chpm \rightarrow \ell^\pm \snu$, such as~\cite{opal}, are
insensitive because the events contain two very soft leptons with
insufficient \ptevt\ to be selected as acoplanar di-lepton candidates.
In contrast, the \wpair\ events considered above have a large \ptevt\
from the normally decaying W.
The soft lepton from the $W^\pm\rightarrow\chz\chpm$ decay is visible
down to a $p_{t}$ of 50--100~MeV.

It is interesting to search also for the anomalous production of
events containing a single observed lepton.
This has been done by the LEP experiments, e.g., in the context of
their selection of ``single W'' events (\wenu\ final state)~\cite{w}.  
An example of a potential new physics source of such events is the
final state \chienu , with the e$^+$ scattered at a small angle to the beam
direction and thus unobserved. 
An additional interest in this process is provided by the fact that,
whereas the pair production of charginos is clearly limited to 
$\mch < \Ebeam$, the
final state \chienu\ is kinematically possible for $\mch > \Ebeam$.
Unfortunately, the expected cross-section is quite small.
For the specific example:
$\mch = 100$~GeV, $\msnu = 45$~GeV and \stwo , the expected
cross-section is about 20~fb~\cite{epa}.
A feasibility study using  Monte Carlo events~\cite{opalmc} processed 
with a full
simulation of the OPAL experiment suggests that a selection efficiency
of about 60\% can be achieved for such events by requiring a single
lepton, significant \ptevt\ and no other activity in the event.
However the predicted SM background is around 200~fb.
Although the lepton momentum may give some additional discrimination, it 
looks difficult to achieve the sensitivity required to observe the
expected cross-section.
Another potential source of
events containing a single observed lepton is the final state \chiee ,
although the expected cross-section is even smaller than for \chienu.

\ack
JK was partially supported by KBN Grant number 2P03B 030 14.

\section{Implications of LEP Precision Electroweak Data for Higgs Searches
Beyond the Standard Model\label{higgsphen}}
\author{B~C~Allanach, J~J~van~der~Bij, G~G~Ross, M~Spira}

\begin{abstract}
We briefly review precision electroweak fits, focussing upon their
implications for the standard model Higgs mass. We review attempts to extend
the analysis beyond the Standard Model in order to obtain information upon
Higgs masses in a general scenario.
\end{abstract}

\begin{figure}
\begin{center}
\leavevmode
\hbox{\epsfxsize=10cm
\epsffile{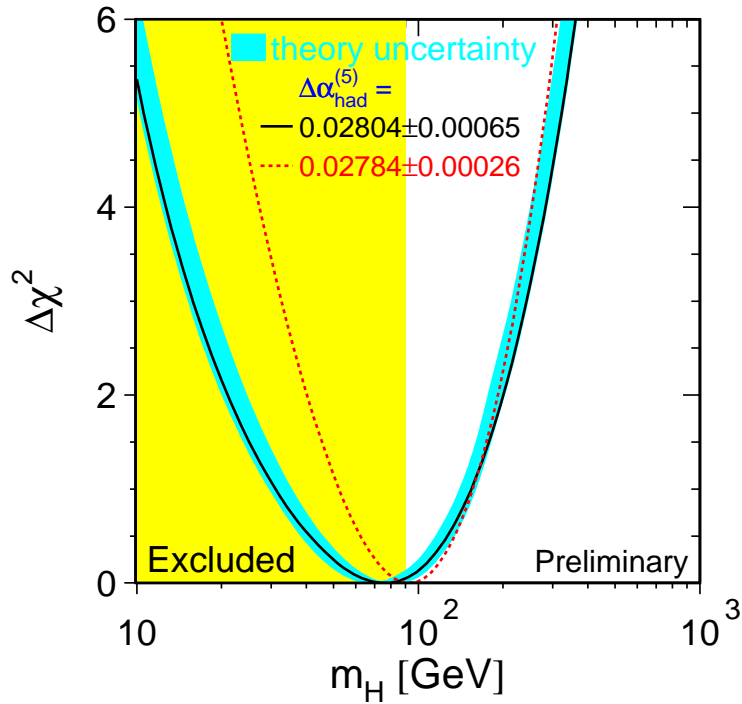}}
\end{center}
\caption{LEP Electroweak Working Group fit to Higgs
mass~\protect\cite{LEPEWWG}. The light shaded area is excluded by the direct
Higgs search. The dotted and full lines show two different values for the
hadronic part of the extraction of the fine structure constant $\Delta
\alpha_{had}^{(5)}$.}
\label{fig:ewwg}
\end{figure}

Figure~\ref{fig:ewwg} displays the implications of the combined LEP
Electroweak Working Group fit to the minimal Standard Model for the mass of
the Higgs boson. From the figure, one can extract 
\begin{equation}
m_{h^0} < 230\mbox{~GeV at 95\% C.L.} \label{higgs_pred}
\end{equation}
even accounting for the theoretical uncertainty in its determination.
The figure shows that the value of $m_{h^0}$ most favoured by the fit is
already excluded by the direct searches at LEP, favouring imminent discovery
within the context of the Standard Model. It is tempting to
infer from the fit that any model beyond the Standard Model must have something
that behaves just like a Higgs boson with mass less than 230 GeV, providing
the LHC, for example, with complete coverage in its Higgs search.
We now provide brief reviews of recent literature which critically examine this
inference.

A number of authors~\cite{bagfalk,cn4} have used effective Lagrangians to
describe low energy effects of beyond the standard model physics. Assuming the
Standard Model with Higgs $\phi$, one can add the effective Lagrangian
pieces~\cite{cn4} 
\begin{equation}
-\,{a \over 2!\,\Lambda^2} \left\{ [D_\mu,D_\nu] \phi \right\}^\dagger[D^\mu,D^\nu]
\phi
+{\tilde{b}\, \kappa^2 \over 2!\, \Lambda^2} (\phi^\dagger \stackrel{\leftrightarrow}{D^\mu} \phi) 
(\phi^\dagger \stackrel{\leftrightarrow}{D_\mu} \phi)~,
\label{operators}
\end{equation}
where $a$ and $\tilde{b}$ are expected to be of order one. 
$\Lambda$
represents the mass scale associated with new physics and $\kappa$ is a
measure of the size of its dimensionless couplings (of order 4$\pi$ for a
strongly coupled theory). The terms in Equation~\ref{operators} then
parameterise the effect of the new physics upon the Higgs. They lead to
corrections to the Peskin-Takeuchi $S$ and $T$ parameters~\cite{PT}
\begin{equation}
\Delta S = {4 \pi a  v^2\over  \Lambda^2},\hspace{1cm}
\& \hspace{1cm} \Delta T = {{\tilde b}\kappa v^2 \over \alpha\Lambda^2}
\label{corrections}
\end{equation}
which are extracted from electroweak fits and strongly constrain physics
beyond the SM\@.
Without the operators in Equation~\ref{operators}, $\Delta S=\Delta T=0$ and
one retains the prediction in Equation~\ref{higgs_pred}.
When the additional operators are included, the authors of
reference~\cite{cn4} conclude that satisfactory
electroweak fits are obtained if
\begin{equation}
m_H < 500 \mbox{~GeV}, \qquad \Lambda < 10 \mbox{~TeV},
\end{equation}
without unnatural magnitudes of the parameters $a,{\tilde b},\kappa,\Lambda$.

Another approach~\cite{bagfalk} abandons the Higgs completely and asks the
question: can the electroweak data be explained by the SM without a Higgs but
with some unspecified (other) new physics. The parameter $\Lambda$ then
defines the scale of the physics responsible for the electroweak symmetry
breaking. 
Gauged chiral Lagrangians provide a model independent description of the
effect of the electroweak symmetry breaking physics upon low energy phenomena.
The Lagrangian is constructed from the Goldstone bosons $w^a$ coming from the
electroweak symmetry breaking.  The $w^a$
appear in the group element $\Sigma =
\exp(2i w^a \tau^a/v)$, where $\tau^a$
are Pauli matrices, normalised to $1/2$, and $v =
246$ GeV is the scale of the symmetry breaking.
The gauge bosons appear
through their field strengths, $W_{\mu\nu} =
W^a_{\mu\nu} \tau^a$ and $B_{\mu\nu} = B^3_{\mu
\nu} \tau^3$, as well as through the covariant
derivative, $D_\mu \Sigma = \partial_\mu \Sigma
+ i g W^a_\mu \tau^a \Sigma - i g' \Sigma B^3_\mu
\tau^3$.
The gauged chiral Lagrangian is built from these
objects.  It can be organised in a derivative
expansion,
\begin{equation}
L = L^{(2)} + L^{(4)} + \ldots,
\end{equation}
where
\begin{eqnarray}
L^{(2)} &=& \ \ {v^2\over 4}\, \Tr D_\mu \Sigma
D_\mu \Sigma^\dagger
   + {{g'}^2 v^2\over 16 \pi^2}\, b_1\, (\Tr
T\, \Sigma^\dagger D_\mu \Sigma )^2 \nonumber\\
&& \mbox{}+ {gg'\over 16 \pi^2}\,a_1\,
\Tr B_{\mu\nu} \Sigma^\dagger W_{\mu\nu} \Sigma\, \label{chiralL}
\end{eqnarray}
and $T = \Sigma^\dagger \tau^3 \Sigma$.  
$a_1$, $b_1$ are the dimensionless couplings associated with the new physics
and have been normalised so they would be naturally of order 1 for a strongly
interacting sector at $\Lambda \sim 3$ TeV. From equation~\ref{chiralL}, the
authors of~\cite{bagfalk} obtain
\begin{eqnarray}
S &=& -\frac{a_1}{\pi} + \frac{1}{6 \pi} \log \left( \frac{\Lambda}{M_Z}
\right), \nonumber \\
T &=& \frac{b_1}{\pi \cos^2 \theta_W} - \frac{3}{8 \pi \cos^2 \theta_W} \log
\left( \frac{\Lambda}{M_Z} \label{vindaloo}
\right).
\end{eqnarray}
When incorporated into a fit of electroweak precision observables, the above
scheme provides acceptable fits without unnatural cancellations between $a_1$
and $b_1$ and the second terms in $S$ and $T$ for 
\begin{equation}
\Lambda \leq 3 \mbox{~TeV}.
\end{equation}

Some comments about this last result are in order. The main concern about the
result is that the mechanism of electroweak symmetry breaking would be hidden
from the LHC\@. 
However, if the scale of the new physics were of order 3 TeV,
the LHC might still see some signals of strongly interacting
$W$'s, for example longitudinal $W$ pair production~\cite{ATLASTDR}. 
It remains to be seen whether a model can be built which gives $a_1$, $b_1$
and $\Lambda$ of the correct values to fit the electroweak data.
For example the most naive technicolour theories predicted the
wrong sign for $a_1$ compared to the fit and were consequently ruled
out~\cite{PT}.
The model then has to 
simultaneously {\em not}\/ generate four-fermion effective interactions which
are excluded by current data. 
The above analysis does not include these fermion
interactions. 

In the SM with Higgs, $m_H$ replaces $\Lambda$ in equation~\ref{vindaloo}.
The coefficient in front of the logarithm is the
same in both cases. Since we do not know $m_H$ (or $\Lambda$), 
$S$ and $T$ are not uniquely predicted.
However, the Higgs-mass or $\Lambda$ independent combination
\begin{equation}
V\equiv \frac{8}{3} T \cos^2 \theta_W  + 6 S = 0
\end{equation}
is a firm prediction of the standard model.
With the precise measurement of $M_W$, a second Higgs mass
independent prediction can be made based on the $U$ parameter.
We think it would  be useful, in order to test whether the data
are in agreement with the standard model {\em independent of the mechanism of
electroweak symmetry breaking}, that two-dimensional plots
in $U-V$ space be made,
particularly because the fit to the SM is only moderately good.

\ack
We would like to thank J Forshaw and G Weiglein for helpful discussions.

\section{The stealthy type of Higgs models\label{sec:stealth}}
\author{J J van der Bij}

\begin{abstract}
We briefly review the effects of singlet scalars
on the Higgs sector. 
\end{abstract}

\subsection{Introduction}

Understanding of the electroweak symmetry breaking mechanism is one of the main
tasks in particle physics. The establishment of the structure of the Higgs sector would 
be a break-through in our knowledge about matter. So it is important to 
think about alternatives to the Standard Model Higgs sector.
Most alternatives give rise to some effects at low energy, that can be measured at
LEP and  are therefore already constrained.
However the simplest possible
extension, by scalar  singlets, does not give rise to extra radiative corrections 
at the one-loop
level and is therefore indistinguishable from the Standard Model as far as 
precision measurements at LEP1 are concerned.
While leaving the gauge-sector of the Standard Model unchanged singlets can have important
effects within the Higgs sector of the model. For example strong interactions can
be present. These effects can significantly change the Higgs signal at future colliders.
Singlets change the Higgs signal in two ways, mixing and invisible decay, which can appear
separately or in combination.

\subsection{Mixing}

A pure mixing model for singlets was analysed in ref.~\cite{hill}.
This model is the simplest possible extension of the Standard Model,
containing only two extra parameters. The Lagrangian of the Higgs sector
is given by:
\bea
\label{hillmodel}
{\cal L} &=& -1/2\,(D_{\mu} \Phi)^{\dagger}(D_{\mu}\Phi) 
-1/2\,(\partial_{\mu} X)^2 -\lambda_1/8(\Phi^{\dagger}\Phi - f_1^2)^2
\nonumber \\
&&-\lambda_2/8 (2 f_2 X - \Phi^{\dagger}\Phi)^2 \nonumber
\eea
where $\Phi$ is the standard Higgs doublet and X a real scalar singlet.
After spontaneous symmetry breaking and diagonalisation of the mass
matrix one finds two Higgs with different masses and each having a reduced
coupling $g_i$ to matter : $g_1 = g_{SM} cos(\theta)$, $ g_2 = g_{SM} sin (\theta)$.
The branching ratio of decay products is the same as for the standard model with the
same mass.
This model will therefore give rise to two Higgs peaks at the LHC, each with reduced
significance. In the mass range where the Higgs can only be studied by rare 
decays this could marginalise the Higgs signal. The situation is however worse. One can consider
not just one X-field, but many \cite{krasnikov}. In this case the Higgs signal can be spread
out over a large energy range, thereby hiding the Higgs signal at the LHC\@.
However at a linear $e^+ e^-$-collider one can use the process $e^+ e^- \rightarrow Z H$
to study this process.

\subsection{Invisible decay}

To check the influence of a hidden sector we will study the coupling
of a Higgs boson to an O(N) symmetric set of scalars~\cite{binoth}.
 The effect of the extra scalars is practically
the presence of a possibly large invisible decay width of the Higgs particle.
When the coupling is large enough the Higgs resonance can become
wide even for a light Higgs boson.

The scalar sector of the model consists of the usual Higgs sector coupled 
to a real N--component vector $\vec\varphi$ of scalar fields, denoted by 
Phions in the following. The Lagrangian density is given by,
\bea
\label{definition}
 {\cal L}  &=&
 - D_{\mu}\Phi^+ D_{\mu}\Phi -\lambda (\Phi^+\Phi - v^2/2)^2
   - 1/2\,\partial_{\mu} \vec\varphi \partial^{\mu}\vec\varphi
     -1/2 \, m^2 \,\vec\varphi^2 \nonumber \\
     &&- \kappa/(8N) \, (\vec\varphi^2 )^2
    -\omega/(2\sqrt{N})\, \, \vec\varphi^2 \,\Phi^+\Phi \nonumber
\eea
where $\phi$ is the standard Higgs doublet. 
Couplings to fermions and vector bosons are the same as in the Standard Model.
The ordinary
Higgs field acquires the vacuum expectation value $v/\sqrt{2}$. For positive $\omega$
the $\vec\varphi$--field acquires no vacuum expectation
value. After spontaneous
symmetry breaking one is left with the ordinary Higgs boson,
coupled to the Phions into which it decays. Also the Phions
receive an induced mass from the spontaneous symmetry breaking which is suppressed
by a factor $1/\sqrt{N}$.
If the factor N is taken
to be large,  the model can be analysed with $1/N$--expansion techniques.
By taking this limit the Phion mass is suppressed, whereas
the decay width of the Higgs boson is not. Because the Higgs width
is now depending on the Higgs Phion coupling its value is arbitrary. 
Therefore the main effect of the presence of the Phions is to give
a possibly large invisible decay rate to the Higgs boson. The 
invisible decay width is given by 
\bea 
\Gamma_H =\frac {\omega^2 v^2}{32 \pi M_H} = 
\frac {\omega^2 (\sin\theta_W\cos\theta_W M_Z)^2)}{32 \pi^2 \alpha_{em} M_H}.\nonumber 
\eea
The model is different
from Majoron models \cite{valle}, since the width is not necessarily small.
The model is similar to the technicolor--like model of ref.~\cite{chivukula}.

It is clear that looking for an invisibly decaying wide Higgs resonance is essentially
hopeless at the LHC\@. One should therefore study the signal at a linear $e^+ e^-$-collider.
A typical exclusion plot is given in figure 1. from ref.~\cite{binoth2}.

\subsection{The general case}

In the general case there will be both mixing and invisible decay.
This can be arranged i.e.\ by spontaneously breaking the O(N) symmetry in the model above
or by allowing $X^3, X^4$ interactions in the first model. 
A model of this type was presented in ref.~\cite{bjorken}. The general picture
consists therefore of a Higgs sector that consists of an arbitrary number of
mass peaks, with an arbitrary invisible width. The analysis of this general
situation is not significantly different from the special cases studied above.
The general conclusion is that the LHC might very well be unable to establish
a Higgs sector of this type. However an $e^+ e^-$-collider  
 will be able to study such a Higgs sector 
using the process $e^+ e^- \rightarrow Z H$~\cite{binoth,binoth2,gunion}. 
This can be done in a clean way using
the decay of the Z boson to leptons if a high luminosity is provided. 

\begin{figure}[htb]
\vspace{0.1cm}
\centerline{\epsfig{figure=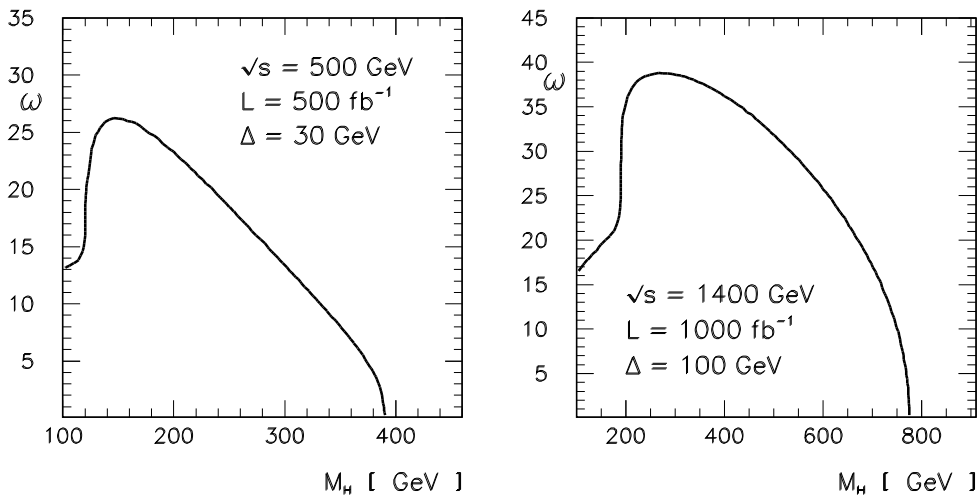,height=6.5cm,angle=0}}
\caption{\it Exclusion limits at a LC at an energy of 500 (1400) GeV and
luminosity 500 (1000) $fb^{-1}$, respectively. }
\label{exclu1}
\end{figure}

\ack
This work was supported by the ARC-Program, the DFG-Forschergruppe Quantenfeldtheorie,
Computeralgebra und Monte Carlo Simulation and by the NATO-grant 
CRG 970113.

%


\newcommand{\gsim}{\lower.7ex\hbox{$\;\stackrel{\textstyle>}
{\sim}\;$}}


\def\citer{\@ifnextchar [{\@tempswatrue\@citexr}{\@tempswafalse\@citexr[]}}
 
\def\@citexr[#1]#2{\if@filesw\immediate\write\@auxout{\string\citation{#2}}\fi
  \def\@citea{}\@cite{\@for\@citeb:=#2\do
    {\@citea\def\@citea{--\penalty\@m}\@ifundefined
       {b@\@citeb}{{\bf ?}\@warning
       {Citation `\@citeb' on page \thepage \space undefined}}%
\hbox{\csname b@\@citeb\endcsname}}}{#1}}

\renewcommand{\textfraction}{0}
\renewcommand{\floatpagefraction}{1.0}

\def\refeq#1{\mbox{eq.~(\ref{#1})}}
\def\refeqs#1{\mbox{eqs.~(\ref{#1})}}
\def\reffi#1{\mbox{Fig.~\ref{#1}}}
\def\reffis#1{\mbox{Figs.~\ref{#1}}}
\def\refta#1{\mbox{Tab.~\ref{#1}}}
\def\reftas#1{\mbox{Tabs.~\ref{#1}}}
\def\refse#1{\mbox{Sect.~\ref{#1}}}
\def\refses#1{\mbox{Sects.~\ref{#1}}}
\def\refapp#1{\mbox{App.~\ref{#1}}}
\def\citere#1{\mbox{Ref.~\cite{#1}}}
\def\citeres#1{\mbox{Refs.~\cite{#1}}}

\newcommand{\mst}{m_{\tilde{t}}}
\newcommand{\delmst}{\Delta\mst}
\newcommand{\mstr}{m_{\tilde{t}_R}}
\newcommand{\mstl}{m_{\tilde{t}_L}}
\newcommand{\mste}{m_{\tilde{t}_1}}
\newcommand{\mstz}{m_{\tilde{t}_2}}
\newcommand{\msti}{m_{\tilde{t}_i}}
\newcommand{\mstip}{m_{\tilde{t}_{i'}}}
\newcommand{\mstj}{m_{\tilde{t}_j}}
\newcommand{\msb}{m_{\tilde{b}}}
\newcommand{\msbr}{m_{\tilde{b}_R}}
\newcommand{\msbl}{m_{\tilde{b}_L}}
\newcommand{\msbe}{m_{\tilde{b}_1}}
\newcommand{\msbz}{m_{\tilde{b}_2}}
\newcommand{\msbi}{m_{\tilde{b}_i}}
\newcommand{\msbj}{m_{\tilde{b}_j}}
\newcommand{\MstL}{M_{\tilde{t}_L}}
\newcommand{\MstR}{M_{\tilde{t}_R}}
\newcommand{\MsbL}{M_{\tilde{b}_L}}
\newcommand{\MsbR}{M_{\tilde{b}_R}}
\newcommand{\MsqL}{M_{\tilde{q}_L}}
\newcommand{\MsqR}{M_{\tilde{q}_R}}
\newcommand{\Mtlr}{M_{t}^{LR}}
\newcommand{\Mtlrz}{\KL M_{t}^{LR}\KR^2}
\newcommand{\Mtlrd}{\KL M_{t}^{LR}\KR^3}
\newcommand{\Mtlrv}{\KL M_{t}^{LR}\KR^4}
\newcommand{\Mtlrf}{\KL M_{t}^{LR}\KR^5}
\newcommand{\Mtlrse}{\KL M_{t}^{LR}\KR^6}
\newcommand{\Mtlrsi}{\KL M_{t}^{LR}\KR^7}
\newcommand{\Mtlra}{\KL M_{t}^{LR}\KR^8}
\newcommand{\Mblr}{M_{b}^{LR}}
\newcommand{\Mblrz}{\KL M_{b}^{LR}\KR^2}
\newcommand{\Mqlr}{M_{q}^{LR}}
\newcommand{\At}{A_t}
\newcommand{\Ab}{A_b}
\newcommand{\Xt}{X_t}

\newcommand{\mf}{m_f}
\newcommand{\mfp}{m_{f'}}
\newcommand{\mfi}{m_{f_i}}
\newcommand{\msfr}{m_{\tilde{f}_R}}
\newcommand{\msfl}{m_{\tilde{f}_L}}
\newcommand{\msf}{m_{\tilde{f}}}
\newcommand{\msfi}{m_{\tilde{f}_i}}
\newcommand{\msfe}{m_{\tilde{f}_1}}
\newcommand{\msfz}{m_{\tilde{f}_2}}

\newcommand{\Ms}{M_S}
\newcommand{\msusy}{M_{\mathrm{SUSY}}}
\newcommand{\msq}{m_{\tilde{q}}}
\newcommand{\msqi}{m_{\tilde{q}_i}}
\newcommand{\msqip}{m_{\tilde{q}_{i'}}}
\newcommand{\msqj}{m_{\tilde{q}_j}}
\newcommand{\msqe}{m_{\tilde{q}_1}}
\newcommand{\msqz}{m_{\tilde{q}_2}}

\newcommand{\sfn}{\tilde{f}}
\newcommand{\sfl}{\tilde{f}_L}
\newcommand{\sfr}{\tilde{f}_R}
\newcommand{\sfe}{\tilde{f}_1}
\newcommand{\sfz}{\tilde{f}_2}
\newcommand{\sfi}{\tilde{f}_i}
\newcommand{\sfj}{\tilde{f}_j}
\newcommand{\sfez}{\tilde{f}_{12}}
\newcommand{\sfze}{\tilde{f}_{21}}

\newcommand{\Pe}{\phi_1}
\newcommand{\Pz}{\phi_2}
\newcommand{\Pez}{\phi_{1,2}}
\newcommand{\PePz}{\phi_1\phi_2}
\newcommand{\mpe}{m_{\Pe}}
\newcommand{\mpz}{m_{\Pz}}
\newcommand{\mpez}{m_{\PePz}}

\newcommand{\smallz}{{\scriptscriptstyle Z}} 
\newcommand{\smallD}{{\scriptscriptstyle D}} %
\newcommand{\smallQ}{{\scriptscriptstyle Q}} %
\newcommand{\smallU}{{\scriptscriptstyle U}} %

\newcommand{\SU}{\mathrm {SUSY}}
\newcommand{\SM}{\mathrm {SM}}
\newcommand{\msbar}{$\overline{\rm{MS}}$}
\newcommand{\oa}{{\cal O}(\alpha)}
\newcommand{\oas}{{\cal O}(\alpha_s)}
\newcommand{\oaas}{{\cal O}(\alpha\alpha_s)}
\newcommand{\ogmzmts}{{\cal O}(\gf^2\mt^6)}
\def\order#1{${\cal O}(#1)$}
\def\cL{{\cal L}}
\newcommand{\cp}{{\cal CP}}
\newcommand{\wz}{\sqrt{2}}
\newcommand{\edz}{\frac{1}{2}}
\def\ed#1{\frac{1}{#1}}
\newcommand{\twol}{two-loop}
\newcommand{\onel}{one-loop}
\newcommand{\mma}{{\em Mathematica}}
\newcommand{\tc}{{\em TwoCalc}}
\newcommand{\fa}{{\em FeynArts}}
\newcommand{\fh}{{\em FeynHiggs}}
\newcommand{\fhf}{{\em FeynHiggsFast}}
\newcommand{\rp}{$\rho\,$-parameter}

\newcommand{\MW}{M_W}
\newcommand{\MZ}{M_Z}
\newcommand{\MA}{M_A}
\newcommand{\mh}{m_h}
\newcommand{\mhmax}{m_h^{\rm max}}
\newcommand{\mH}{m_H}
\newcommand{\Mh}{M_h}
\newcommand{\MH}{M_H}

\newcommand{\sudxsuzxue}{SU(3)_C \otimes SU(2)_L \otimes U(1)_Y}
\newcommand{\suzxue}{SU(2)_L \otimes U(1)_Y}

\newcommand{\dr}{\De\rho}
\newcommand{\mt}{m_{t}}
\newcommand{\mtexp}{\mt^{\rm exp}}
\newcommand{\mtms}{\overline{m}_t}
\newcommand{\mq}{m_{q}}
\newcommand{\mb}{m_{b}}
\newcommand{\gl}{\tilde{g}}
\newcommand{\Mgl}{m_{\tilde{g}}}
\newcommand{\mgl}{m_{\tilde{g}}}
\newcommand{\sq}{\tilde{q}}
\newcommand{\sqi}{\tilde{q}_i}
\newcommand{\sqj}{\tilde{q}_j}
\newcommand{\sql}{\tilde{q}_L}
\newcommand{\sqr}{\tilde{q}_R}
\newcommand{\sqe}{\tilde{q}_1}
\newcommand{\sqz}{\tilde{q}_2}
\newcommand{\sqez}{\tilde{q}_{12}}
\newcommand{\Stop}{\tilde{t}}
\newcommand{\StopL}{\tilde{t}_L}
\newcommand{\StopR}{\tilde{t}_R}
\newcommand{\Stope}{\tilde{t}_1}
\newcommand{\Stopz}{\tilde{t}_2}
\newcommand{\Stopi}{\tilde{t}_i}
\newcommand{\Stopj}{\tilde{t}_j}
\newcommand{\Sbot}{\tilde{b}}
\newcommand{\SbotL}{\tilde{b}_L}
\newcommand{\SbotR}{\tilde{b}_R}
\newcommand{\Sbote}{\tilde{b}_1}
\newcommand{\Sbotz}{\tilde{b}_2}
\newcommand{\Sboti}{\tilde{b}_i}
\newcommand{\Sbotj}{\tilde{b}_j}
\newcommand{\dst}{\Delta_{\tilde{t}}}
\newcommand{\tst}{\theta_{\tilde{t}}}
\newcommand{\tsb}{\theta_{\tilde{b}}}
\newcommand{\tsf}{\theta\kern-.20em_{\tilde{f}}}
\newcommand{\tsfp}{\theta\kern-.20em_{\tilde{f}\prime}}
\newcommand{\tsq}{\theta\kern-.15em_{\tilde{q}}}
\newcommand{\sw}{s_W}
\newcommand{\cw}{c_W}
\newcommand{\sweff}{\sin^2\theta_{\mathrm{eff}}}

\newcommand{\sintt}{\sin\tst}
\newcommand{\sinQtt}{\sin^2\tst}
\newcommand{\sinZtt}{\sin 2\tst}
\newcommand{\sinQZtt}{\sin^2 2\tst}
\newcommand{\sintb}{\sin\tsb}
\newcommand{\sinQtb}{\sin^2\tsb}
\newcommand{\sinZtb}{\sin 2\tsb}
\newcommand{\sintf}{\sin\tsf}
\newcommand{\sintfp}{\sin\tsfp}
\newcommand{\sinQtf}{\sin^2\tsf}
\newcommand{\sinZtf}{\sin 2\tsf}
\newcommand{\sintq}{\sin\tsq}
\newcommand{\sinQtq}{\sin^2\tsq}
\newcommand{\sinZtq}{\sin 2\tsq}

\newcommand{\costt}{\cos\tst}
\newcommand{\cosQtt}{\cos^2\tst}
\newcommand{\cosZtt}{\cos 2\tst}
\newcommand{\costb}{\cos\tsb}
\newcommand{\cosQtb}{\cos^2\tsb}
\newcommand{\cosZtb}{\cos 2\tsb}
\newcommand{\costf}{\cos\tsf}
\newcommand{\costfp}{\cos\tsfp}
\newcommand{\cosQtf}{\cos^2\tsf}
\newcommand{\cosZtf}{\cos 2\tsf}
\newcommand{\costq}{\cos\tsq}
\newcommand{\cosQtq}{\cos^2\tsq}
\newcommand{\cosZtq}{\cos 2\tsq}

\newcommand{\stt}{s_{\tilde{t}}}
\newcommand{\stb}{s_{\tilde{b}}}
\newcommand{\stf}{s_{\tilde{f}}}
\newcommand{\stq}{s_{\tilde{q}}}
\newcommand{\ctt}{c_{\tilde{t}}}
\newcommand{\ctb}{c_{\tilde{b}}}
\newcommand{\ctf}{c_{\tilde{f}}}
\newcommand{\ctq}{c_{\tilde{q}}}

\newcommand{\KL}{\left(}
\newcommand{\KR}{\right)}
\newcommand{\KKL}{\left[}
\newcommand{\KKR}{\right]}
\newcommand{\KKKL}{\left\{}
\newcommand{\KKKR}{\right\}}
\newcommand{\BL}{\lbrack}
\newcommand{\BR}{\rbrack}

\newcommand{\VL}{\left( \begin{array}{c}}
\newcommand{\VR}{\end{array} \right)}
\newcommand{\ML}{\left( \begin{array}{cc}}
\newcommand{\MLd}{\left( \begin{array}{ccc}}
\newcommand{\MLv}{\left( \begin{array}{cccc}}
\newcommand{\MR}{\end{array} \right)}
\newcommand{\dd}{\partial}
\newcommand{\dmu}{\partial_{\mu}}
\newcommand{\dmo}{\partial^{\mu}}
\newcommand{\mn}{{\mu\nu}}
\newcommand{\hc}{\mbox {h.c.}}
\newcommand{\re}{\mbox {Re}\,}
\newcommand{\OP}{\omega_+}
\newcommand{\OM}{\omega_-}
\newcommand{\Tb}{\tan \beta\hspace{1mm}}
\newcommand{\tb}{\tan \beta}
\newcommand{\TQb}{\tan^2 \beta\hspace{1mm}}
\newcommand{\CTb}{\cot \beta\hspace{1mm}}
\newcommand{\CTQb}{\cot^2 \beta\hspace{1mm}}
\newcommand{\Sb}{\sin \beta\hspace{1mm}}
\newcommand{\sbe}{\sin \beta}
\newcommand{\SQb}{\sin^2\beta\hspace{1mm}}
\newcommand{\SDb}{\sin^3\beta\hspace{1mm}}
\newcommand{\SVb}{\sin^4\beta\hspace{1mm}}
\newcommand{\Cb}{\cos \beta\hspace{1mm}}
\newcommand{\CQb}{\cos^2\beta\hspace{1mm}}
\newcommand{\CDb}{\cos^3\beta\hspace{1mm}}
\newcommand{\CVb}{\cos^4\beta\hspace{1mm}}
\newcommand{\Sa}{\sin \alpha\hspace{1mm}}
\newcommand{\SQa}{\sin^2\alpha\hspace{1mm}}
\newcommand{\Ca}{\cos \alpha\hspace{1mm}}
\newcommand{\CQa}{\cos^2\alpha\hspace{1mm}}
\newcommand{\Ta}{\tan\al}
\newcommand{\CTa}{\cot\al}
\newcommand{\Sab}{\sin (\alpha + \beta)\hspace{1mm}}
\newcommand{\Cab}{\cos (\alpha + \beta)\hspace{1mm}}
\newcommand{\Sba}{\sin (\beta - \alpha)\hspace{1mm}}
\newcommand{\Cba}{\cos (\beta - \alpha)\hspace{1mm}}
\newcommand{\SZa}{\sin 2\alpha\hspace{1mm}}
\newcommand{\SQZa}{\sin^2 2\alpha\hspace{1mm}}
\newcommand{\CZa}{\cos 2\alpha\hspace{1mm}}
\newcommand{\CQZa}{\cos^2 2\alpha\hspace{1mm}}
\newcommand{\SZb}{\sin 2\beta\hspace{1mm}}
\newcommand{\SQZb}{\sin^2 2\beta\hspace{1mm}}
\newcommand{\CZb}{\cos 2\beta\hspace{1mm}}
\newcommand{\CQZb}{\cos^2 2\beta\hspace{1mm}}
\newcommand{\CDZb}{\cos^3 2\beta\hspace{1mm}}
\newcommand{\CVZb}{\cos^4 2\beta\hspace{1mm}}
\newcommand{\CQba}{\cos^2 (\beta - \alpha)\hspace{1mm}}
\newcommand{\SQba}{\sin^2 (\beta - \alpha)\hspace{1mm}}
\newcommand{\SDba}{\sin^3 (\beta - \alpha)\hspace{1mm}}
\newcommand{\CQab}{\cos^2 (\alpha + \beta)\hspace{1mm}}
\newcommand{\SQab}{\sin^2 (\alpha + \beta)\hspace{1mm}}
\newcommand{\SDab}{\sin^3 (\alpha + \beta)\hspace{1mm}}
\newcommand{\SDea}{\sin \De\al}
\newcommand{\CDea}{\cos \De\al}
\newcommand{\CQDea}{\cos^2\De\al}
\newcommand{\TDea}{\tan \De\al}
\newcommand{\TQDea}{\tan^2\De\al}

\newcommand{\tev}{\,\, \mathrm{TeV}}
\newcommand{\gev}{\,\, \mathrm{GeV}}
\newcommand{\mev}{\,\, \mathrm{MeV}}
\def\Gcs{\mathrm{GeV}}

\newcommand{\BC}{\begin{center}}
\newcommand{\EC}{\end{center}}
\newcommand{\BE}{\begin{equation}}
\newcommand{\EE}{\end{equation}}
\newcommand{\BEA}{\begin{eqnarray}}
\newcommand{\BEAnn}{\begin{eqnarray*}}
\newcommand{\EEA}{\end{eqnarray}}
\newcommand{\EEAnn}{\end{eqnarray*}}
\newcommand{\non}{\nonumber}
\newcommand{\id}{{\rm 1\kern-.12em
\rule{0.3pt}{1.5ex}\raisebox{0.0ex}{\rule{0.1em}{0.3pt}}}}
\newcommand{\lesim}
{\;\raisebox{-.3em}{$\stackrel{\displaystyle <}{\sim}$}\;}

\newcommand{\cf}{C_F}
\newcommand{\gf}{G_F}
\newcommand{\nf}{N_f}
\def\al{\alpha}
\def\aeff{\al_{\rm eff}}
\def\aeffapprox{\al_{\rm eff}{\rm (approx)}}
\newcommand{\Saeff}{\sin\aeff}
\newcommand{\Caeff}{\cos\aeff}
\def\als{\alpha_s}
\def\altl{\alpha_{\rm 2-loop}}
\def\gs{g_s}
\def\be{\beta}
\def\ben{\be_0}
\def\bee{\be_1}
\def\bez{\be_2}
\def\ga{\gamma}
\def\gan{\ga_0}
\def\gae{\ga_1}
\def\gaz{\ga_2}
\def\de{\delta}
\def\deqed{\de\Ga_\ga}
\def\eps{\varepsilon}
\def\la{\lambda}
\def\si{\sigma}
\def\ie{i\epsilon}
\def\om{\omega}
\def\Ga{\Gamma}
\def\Gh{\Ga_h}
\def\GH{\Ga_H}
\def\Gz{\Ga_0}
\def\Ge{\Ga_1}
\def\Geeff{\Ga_{1,\rm eff}}
\def\Geg{\Ga_{1,g}}
\def\Gga{\Ga_\ga}
\def\Gg{\Ga_g}
\def\Ggl{\Ga_{\gl}}
\def\Gtot{\Ga_{\rm tot}}
\def\De{\Delta}
\def\La{\Lambda}
\def\Laqcd{\La_{\rm QCD}}
\def\Si{\Sigma}
\def\Sip{\Sigma'}
\def\Sie{\Sigma^{(1)}}
\def\Siz{\Sigma^{(2)}}
\def\Sii{\Sigma^{(i)}}
\def\Siep{\Sigma'^{(1)}}
\def\Sizp{\Sigma'^{(2)}}
\def\Siip{\Sigma'^{(i)}}

\def\hSi{\hat{\Sigma}}
\def\hSip{\hat{\Sigma}'}
\def\hSie{\hat{\Sigma}^{(1)}}
\def\hSiz{\hat{\Sigma}^{(2)}}
\def\hSiep{\hat{\Sigma}^{(1)}'}
\def\hSizp{\hat{\Sigma}^{(2)}'}
\def\hSiH{\hSi_{H}}
\def\hSih{\hSi_{h}}
\def\hSihH{\hSi_{hH}}
\def\hSipH{\hSip_{H}}
\def\hSiph{\hSip_{h}}
\def\hSiphH{\hSip_{hH}}
\def\hSieh{\hSie_{h}}
\def\hSieH{\hSie_{H}}
\def\hSiehH{\hSie_{hH}}
\def\hSizh{\hSiz_{h}}
\def\hSizH{\hSiz_{H}}
\def\hSizhH{\hSiz_{hH}}
\newcommand{\htad}{\hat{t}}
\newcommand{\tade}{t^{(1)}}
\newcommand{\tadz}{t^{(2)}}

\newcommand{\Ph}{\mathswitchr h}
\newcommand{\PH}{\mathswitchr H}

\newcommand{\deMZe}{\de \MZ^{2\,(1)}}
\newcommand{\deMZz}{\de \MZ^{2\,(2)}}
\newcommand{\deMZi}{\de \MZ^{2\,(i)}}
\newcommand{\deMAe}{\de \MA^{2\,(1)}}
\newcommand{\deMAz}{\de \MA^{2\,(2)}}
\newcommand{\deMAi}{\de \MA^{2\,(i)}}
\newcommand{\deZHie}{\de Z_{H_i}^{(1)}}
\newcommand{\deZHiz}{\de Z_{H_i}^{(2)}}
\newcommand{\deZHee}{\de Z_{H_1}^{(1)}}
\newcommand{\deZHez}{\de Z_{H_1}^{(2)}}
\newcommand{\deZHze}{\de Z_{H_2}^{(1)}}
\newcommand{\deZHzz}{\de Z_{H_2}^{(2)}}
\newcommand{\deTbe}{\de\Tb^{(1)}}
\newcommand{\deTbz}{\de\Tb^{(2)}}
\newcommand{\detie}{\de t_i^{(1)}}
\newcommand{\detiz}{\de t_i^{(2)}}
\newcommand{\detee}{\de t_1^{(1)}}
\newcommand{\detez}{\de t_1^{(2)}}
\newcommand{\detei}{\de t_1^{(i)}}
\newcommand{\detze}{\de t_2^{(1)}}
\newcommand{\detzz}{\de t_2^{(2)}}
\newcommand{\detzi}{\de t_2^{(i)}}
\newcommand{\deVie}{\de V_i^{(1)}}
\newcommand{\deViz}{\de V_i^{(2)}}
\newcommand{\deVee}{\de V_{\Pe}^{(1)}}
\newcommand{\deVez}{\de V_{\Pe}^{(2)}}
\newcommand{\deVei}{\de V_{\Pe}^{(i)}}
\newcommand{\deVze}{\de V_{\Pz}^{(1)}}
\newcommand{\deVzz}{\de V_{\Pz}^{(2)}}
\newcommand{\deVzi}{\de V_{\Pz}^{(i)}}
\newcommand{\deVeze}{\de V_{\PePz}^{(1)}}
\newcommand{\deVezz}{\de V_{\PePz}^{(2)}}
\newcommand{\deVezi}{\de V_{\PePz}^{(i)}}

\newcommand{\SLASH}[2]{\makebox[#2ex][l]{$#1$}/}
\newcommand{\Dslash}{\SLASH{D}{.5}\,}
\newcommand{\dslash}{\SLASH{\dd}{.15}}
\newcommand{\kslash}{\SLASH{k}{.15}}
\newcommand{\pslash}{\SLASH{p}{.2}}
\newcommand{\qslash}{\SLASH{q}{.08}}

\newcommand{\sqmsmt}{\sqrt{\Ms^2-\mt^2}}
\newcommand{\sqmsmtbar}{\sqrt{\Ms^2-\mtms^2}}
\newcommand{\lmtms}{\KL\frac{\mt^2}{\Ms^2}\KR}
\newcommand{\lmtmsms}{\KL\frac{\mtms^2}{\Ms^2}\KR}
\newcommand{\lmsmt}{\KL\frac{\Ms^2 - \mt^2}{\mt^2}\KR}

\newcommand{\lc}{linear collider}
\newcommand{\epem}{$e^+e^-$}
\newcommand{\ifb}{$\mbox{fb}^{-1}$}
\newcommand{\iab}{$\mbox{ab}^{-1}$}
\newcommand{\gvf}{g_V^f}
\newcommand{\gaf}{g_A^f}

\newcommand{\hff}{h \to f\bar{f}}
\newcommand{\Hff}{H \to f\bar{f}}
\newcommand{\hbb}{h \to b\bar{b}}
\newcommand{\Hbb}{H \to b\bar{b}}
\newcommand{\htautau}{h \to \tau^+\tau^-}
\newcommand{\hcc}{h \to c\bar{c}}
\newcommand{\rf}{R_f}
\newcommand{\cff}{C_f^{(u)}}
\newcommand{\cfpfp}{C_f^{(d)}}
\newcommand{\Zh}{Z_h}
\newcommand{\wZh}{\sqrt{\Zh}}
\newcommand{\ZH}{Z_H}
\newcommand{\wZH}{\sqrt{\ZH}}
\newcommand{\ZhH}{Z_{hH}}
\newcommand{\ZHh}{Z_{Hh}}

\newcommand{\mpar}[1]{{\marginpar{\hbadness10000%
                      \sloppy\hfuzz10pt\boldmath\bf#1}}%
                      \typeout{marginpar: #1}\ignorespaces}
\def\mua{\marginpar[\boldmath\hfil$\uparrow$]%
                   {\boldmath$\uparrow$\hfil}%
                    \typeout{marginpar: $\uparrow$}\ignorespaces}
\def\mda{\marginpar[\boldmath\hfil$\downarrow$]%
                   {\boldmath$\downarrow$\hfil}%
                    \typeout{marginpar: $\downarrow$}\ignorespaces}
\def\mla{\marginpar[\boldmath\hfil$\rightarrow$]%
                   {\boldmath$\leftarrow $\hfil}%
                    \typeout{marginpar: $\leftrightarrow$}\ignorespaces}

\section{Upper limit on 
$\mh$ in the MSSM and M-SUGRA vs.\ prospective reach of LEP \label{sec:msugra}}

\author{A~Dedes, S~Heinemeyer, P~Teixeira-Dias and G~Weiglein}

\begin{abstract}
The upper limit on the lightest $\cp$-even Higgs boson mass, $\mh$,
is analysed within the MSSM as a function of $\tb$ for fixed $\mt$ and
$\msusy$. The impact of recent diagrammatic \twol\ results on this limit
is investigated. We compare the MSSM theoretical upper bound on $\mh$
with the lower bound obtained from experimental searches at LEP. We
estimate that with the LEP data taken until the end of 1999, the region
$\mh<108.2~\Gcs$ can be excluded at the 95\% confidence level. This
corresponds to an excluded region $0.6\lesim\tb\lesim 1.9$ within the MSSM
for $\mt = 174.3 \gev$ and $\msusy \leq 1$~TeV. 
The final exclusion sensitivity after the end of
LEP, in the year 2000, is also briefly discussed. Finally, we determine
the upper limit on $\mh$ within the Minimal Supergravity (M-SUGRA)
scenario up to the \twol\ level, consistent with radiative
electroweak symmetry breaking. We find an upper bound of $\mh \approx
127 \gev$ for $\mt = 174.3 \gev$ in this scenario, which is slightly
below the bound in the unconstrained MSSM.
\end{abstract}



\subsection{Introduction}

Within the MSSM the masses of the $\cp$-even neutral Higgs bosons are
calculable in terms of the other MSSM parameters. The mass of the
lightest Higgs boson, $\mh$, has been of particular interest, as
it is bounded to be smaller than the $Z$~boson mass at the tree level. 
The \onel\ results~\cite{mhiggs1l,mhiggsf1l,mhiggsf1ldab,pierce} 
for $\mh$ have been supplemented in the
last years with the leading \twol\ corrections, performed in the
renormalisation group (RG)
approach~\cite{mhiggsRG1,mhiggsRG2}, in the effective
potential approach~\cite{mhiggsEP} and most recently in
the Feynman-diagrammatic (FD)
approach~\cite{mhiggsletter,mhiggslong}. 
The \twol\ corrections have turned out to be sizeable. They can
change the \onel\ results by up to 20\%.

Experimental searches at LEP now exclude a light MSSM Higgs boson
with a mass below 
$\sim$90~GeV~\cite{lepc-aleph,lepc-delphi,lepc-l3,lepc-opal}.
In the low $\tb$ region, in which the limit is the same as for the 
Standard Model Higgs boson, a mass limit of even $\mh \gsim 106 \gev$ has
been obtained~\cite{lepc-aleph,lepc-delphi,lepc-l3,lepc-opal}.
Combining this experimental bound with the theoretical upper limit 
on $\mh$ as a function of $\tb$ within the MSSM, it is possible to derive
constraints on $\tb$. In this paper we investigate, for which MSSM
parameters the maximal $\mh$ values are obtained and discuss in this
context the impact of the new FD two-loop result. Resulting constraints
on $\tb$ are analysed on the basis of the present LEP data and of the
prospective final exclusion limit of LEP.

The Minimal Supergravity (M-SUGRA) scenario provides a relatively
simple and constrained version of the MSSM. 
In this paper we explore, how the maximum possible values for $\mh$ change
compared to the general MSSM, if one restricts to the M-SUGRA framework. 
As an additional constraint we impose that the condition of radiative 
electroweak symmetry breaking (REWSB)~\cite{REWSB} should be fulfilled. 


\subsection{The upper bound on $\mh$ in the MSSM} 
\label{section:mssm}

The most important radiative corrections to $\mh$ arise from the top and
scalar top sector of the MSSM, with the input parameters $\mt$, $\msusy$
and $\Xt$. Here we assume the soft SUSY breaking parameters in the
diagonal entries of the scalar top mixing matrix to be equal for simplicity,
$\msusy = \MstL = \MstR$. This has been shown to yield upper values for 
$\mh$ which comprise also the case where $\MstL \neq \MstR$,
if $\msusy$ is identified with the heavier one of $\MstL$, 
$\MstR$~\cite{mhiggslong}. For the off-diagonal entry of the mixing
matrix we use the convention 
\BE
\label{eq:xt}
\mt \Xt = \mt (A_t - \mu \cot\beta).
\EE
Note that the sign convention used for $\mu$ here is the opposite of the
one used in \citere{sakis}.


Since the predicted value of $\mh$ depends
sensitively on the precise numerical value of $\mt$, it has become
customary to discuss the constraints on $\tb$ within a so-called
``benchmark'' scenario (see \citere{lephiggs183} and references therein), 
in which $\mt$ is kept fixed at the value $\mt = 175 \gev$ and in which 
furthermore a large value of $\msusy$ is chosen,
$\msusy = 1 \tev$, giving rise to large values of $\mh(\tb)$. 
In \citere{tbexcl} it has recently been analysed how the values chosen
for the other SUSY parameters in the benchmark scenario should be modified
in order to obtain the maximal values of $\mh(\tb)$ for given $\mt$ and
$\msusy$. The corresponding scenario ($\mhmax$ scenario) is defined 
as~\cite{tbexcl,bench}
\BEA
&& \mt = \mt^{\mathrm{exp}}~( = 174.3 \gev ), \quad \msusy = 1 \tev \non \\
&& \mu = -200 \gev, \; M_2 = 200 \gev, \; \MA = 1 \tev, 
   \; \mgl = 0.8 \, \msusy ({\rm FD}) \non \\
&& \Xt = 2\, \msusy ({\rm FD})\;\; {\rm or}\;\; 
   \Xt = \sqrt{2}\, \msusy ({\rm RG}) ,
\label{benchmarkdef}
\EEA
where the parameters are chosen such that the chargino masses are beyond
the reach of LEP2 and that the lightest $\cp$-even Higgs boson does not
dominantly decay invisibly into neutralinos. In \refeq{benchmarkdef}
$\mu$ is the Higgs mixing parameter, $M_2$ denotes the
soft SUSY breaking parameter in the gaugino sector, and $\MA$ is the
$\cp$-odd Higgs boson mass. The gluino mass, $\mgl$, can only be
specified as a free parameter 
in the FD result (program {\tt FeynHiggs}~\cite{feynhiggs}).
The effect of varying $\mgl$ on $\mh$ is up to $\pm 2
\gev$~\cite{mhiggslong}. Within the RG result 
(program {\tt subhpole}~\cite{mhiggsRG1}) $\mgl$ is fixed to $\mgl = \msusy$. 
Compared to the maximal values for $\mh$ (obtained for $\mgl \approx
0.8\,\msusy$) this leads to a reduction of the Higgs boson mass by up to
$0.5 \gev$. Different values of $\Xt$ are specified in
\refeq{benchmarkdef} for the results of the FD and the RG calculation,
since within the two approaches the maximal values for $\mh$ are
obtained for different values of $\Xt$. This fact is partly due to the
different renormalisation schemes used in the two approaches~\cite{bse}.

The maximal values for $\mh$ as a function of $\tb$ within the
$\mhmax$ scenario are higher by about 5~GeV than in the
previous benchmark scenario. The constraints on $\tb$ derived within the
$\mhmax$ scenario are thus more conservative than the ones based on the
previous scenario.

The investigation of the constraints on $\tb$ that can be obtained from the 
experimental search limits on $\mh$ has so far been based on the 
results for $\mh$ obtained within the RG approach~\cite{mhiggsRG1}.
The recently obtained FD~\cite{mhiggsletter,mhiggslong} result differs 
from the RG result by a more complete treatment of the one-loop
contributions~\cite{mhiggsf1ldab} and in particular by genuine
non-logarithmic two-loop terms that go beyond the leading logarithmic 
two-loop contributions contained in the RG result~\cite{bse,mhiggslle}. 
Comparing the FD result (program {\tt FeynHiggs}) with
the RG result (program {\tt subhpole}) we find that the
maximal value for $\mh$ as a function of $\tb$ within the FD result is
higher by up to 4~GeV.

In \reffi{fig:rgv} we show both the effect of modifying the previous
benchmark scenario to the $\mhmax$ scenario and the impact of the new FD
two-loop result on the prediction for $\mh$. The maximal value for the 
Higgs boson mass is plotted as a function of $\tb$ for $\mt = 174.3$~GeV
and $\msusy = 1$~TeV. The dashed curve displays the benchmark scenario,
used up to now by the LEP collaborations~\cite{lephiggs183}. The dotted
curve shows the $\mhmax$ scenario. Both curves are based on the RG
result (program {\tt subhpole}). The solid curve corresponds to the FD
result (program {\tt FeynHiggs}) in the $\mhmax$ scenario. The increase 
in the maximal value for $\mh$ by about $4$~GeV from the new FD result 
and by further 5~GeV if the benchmark scenario is replaced by the 
$\mhmax$ scenario has a
significant effect on exclusion limits for $\tb$ derived from the
Higgs boson search. Combining both effects, which of course have a very
different origin, the maximal Higgs boson masses are increased by almost
$10 \gev$ compared to the previous benchmark scenario.

\begin{figure}[ht!]
\begin{center}
\mbox{\psfig{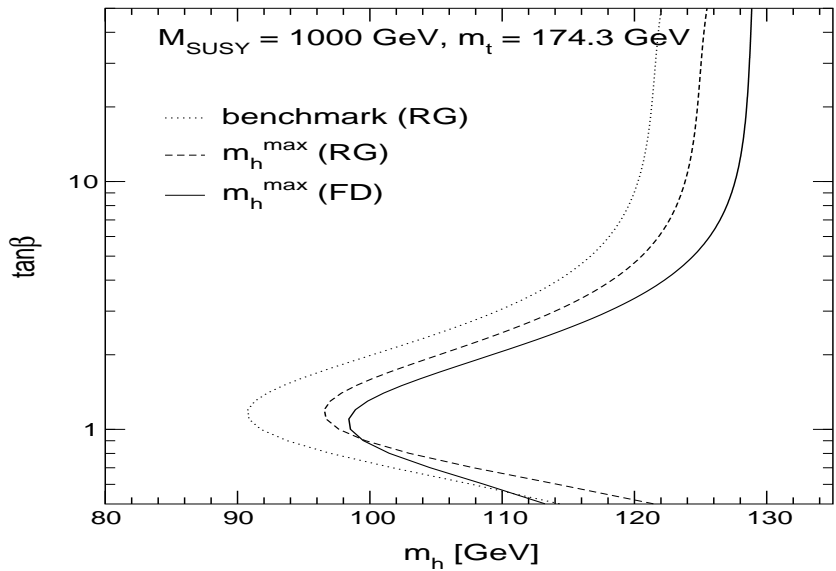}}
\end{center}
\caption[]{The upper bound on $\mh$ is shown as a function of $\tb$ for given $\mt$
and $\msusy$.
The dashed curve displays the previous benchmark scenario. 
The dotted curve shows the RG result for the $\mhmax$ scenario, 
while the solid curve represents the FD result for the $\mhmax$ scenario.
}
\label{fig:rgv}
\end{figure}

{}From the FD result we find the upper bound of $\mh \lesim 129$~GeV in
the region of large $\tb$ within the MSSM for $\mt = 174.3 \gev$ and
$\msusy = 1$~TeV. Higher values for $\mh$ are
obtained if the experimental uncertainty in $\mt$ of currently 
$\De\mt = 5.1 \gev$ is taken into account and higher values are allowed 
for the top quark mass. As a rule of thumb, increasing $\mt$ by 1~GeV
roughly translates into an upward shift of $\mh$ of 1~GeV. An
increase of $\msusy$ from 1~TeV to 2~TeV enhances $\mh$ by about 2~GeV
in the large $\tb$ region. As an extreme case, choosing $\mt =
184.5$~GeV, i.e.\ two standard deviations above the current experimental
central value, and using $\msusy = 2$~TeV leads to an upper bound on 
$\mh$ of $\mh \lesim 141$~GeV within the MSSM.




\subsection{The prospective upper $\mh$ reach of LEP}
%
%
The four LEP experiments are very actively searching for the Higgs
boson.  Results presented recently by the LEP collaborations revealed
no evidence of a SM Higgs boson signal in the data collected in 1999
at centre-of-mass energies of approximately 192, 196, 200 and 202
GeV\cite{lepc-aleph,lepc-delphi,lepc-l3,lepc-opal}. From the negative results of their searches ALEPH, DELPHI and L3 
have therefore individually excluded a SM Higgs boson lighter than
$\sim$101--106 $\Gcs$ (at the 95\% confidence 
level)~\cite{lepc-aleph,lepc-delphi,lepc-l3}.

Here we will present the expected exclusion reach of LEP assuming all
the data taken by the four experiments in 1999 is combined. The
ultimate exclusion reach of LEP -- assuming no signal were found in
the data to be collected in the year 2000 -- will also be estimated
for several hypothetical scenarios of luminosity and centre-of-mass
energy. These results are then confronted with the theoretical MSSM
upper limit on $\mh(\tb)$ presented in section~\ref{section:mssm},
in order to establish to what extent the LEP data can probe the
low $\tb$ region.  We recall that models in which b-$\tau$ Yukawa
coupling unification at the GUT scale is imposed favour low $\tb$
values, $\tb \approx 2$, which can 
severely be constrained experimentally by searches at LEP. Alternatively, 
such models can favour $\tb \approx 40$, a region which however can only 
be partly covered at LEP.

 All experimental exclusion limits quoted in this section are
 implicitly meant at the 95\% confidence level (CL).

%
%
It has been proposed \cite{safari} that the LEP-combined expected 95\% CL
lower bound on $\mh$, $\mh^{95}$, for a data set consisting of data
accumulated at given centre-of-mass energies can be estimated by
solving the equation
\BE  
 n(\mh^{95}) = (\sigma_0 \mathcal{L}_{eq})^{\alpha}, \label{eqn:predictor}
\EE
 where $n(\mh^{95})$ is the number of signal events produced at the
 95\% CL limit. The equivalent luminosity, $\mathcal{L}_{eq}$, is
 the luminosity that one would have to accumulate at the highest
 centre-of-mass energy in the data set in order to have the same
 sensitivity as in the real data set, where the data is split between
 several different $\sqrt{s}$ values. For a SM Higgs boson signal, the
 parameters $\sigma_{0}$ and $\alpha$ are $\sim$38 pb and $\sim$0.4,
 respectively \cite{safari}. (These parameter values are obtained from
 a fit to the actual LEP-combined expected limits from
 $\sqrt{s}=161$ GeV up to $\sqrt{s}=188.6$ GeV
 \cite{lephiggs172,lephiggs183,lephiggs189}.) The predicted $\mh$
 limits obtained with this method are expected to approximate the more
 accurate combinations done by the LEP Higgs Working Group, with
 an uncertainty of the order of $\pm$ 0.3 $\Gcs$.

 Solving \refeq{eqn:predictor} for the existing LEP data with
 183 GeV $\lesim \sqrt{s} \lesim 202 $ GeV (Table~\ref{tab:lepdata})
 results in a predicted combined exclusion of $\mh < 108.2~\Gcs$ for
 the SM Higgs boson (see Figure~\ref{fig:predictor}a).

\begin{table}[htbp]
\caption{{\footnotesize Summary of the total LEP data luminosity accumulated
since 1997. The luminosities for the data taken in 1999 ($\sqrt{s}\ge$
191.6 GeV) are the (still preliminary) values
quoted by the four LEP experiments at the LEPC open session
\cite{lepc-aleph,lepc-delphi,lepc-l3,lepc-opal}.}
\label{tab:lepdata} 
}
\begin{center}
\begin{tabular}{|c|cccccc|}
\hline
 $\sqrt{s}$ (GeV)          & 182.7 & 188.6 & 191.6 & 195.5 & 199.5 & 201.6 \\
\hline
 $\mathcal{L}$ (pb$^{-1}$) & 220.0  & 682.7 & 113.9 & 316.4 & 327.8 & 148.1 \\
\hline
\end{tabular}
\end{center}
\end{table}

 Based on the current LEP operational experience, it is believed that
 in the year 2000 stable running is possible up to $\sqrt{s}=206$
 GeV\cite{lepc-lep}. Figure~\ref{fig:predictor}b demonstrates the
 impact of additional data collected at $\sqrt{s}=206$ GeV
 on the exclusion. For instance, if no evidence of a signal were
 found in the data, collecting 500 (1000) pb$^{-1}$ at this
 centre-of-mass energy would increase the $\mh$ limit to 113.0 (114.1)
 $\Gcs$. Figure~\ref{fig:predictor}c shows the degradation in the
 sensitivity to a Higgs boson signal if the data in the year 2000 were
 accumulated at $\sqrt{s}=205$ GeV instead: in this case the
 luminosity required to exclude up to $\mh=113~\Gcs$ would be
 840~pb$^{-1}$.

\begin{figure}[htbp]
\begin{tabular}{rlr}
{\epsfig{file=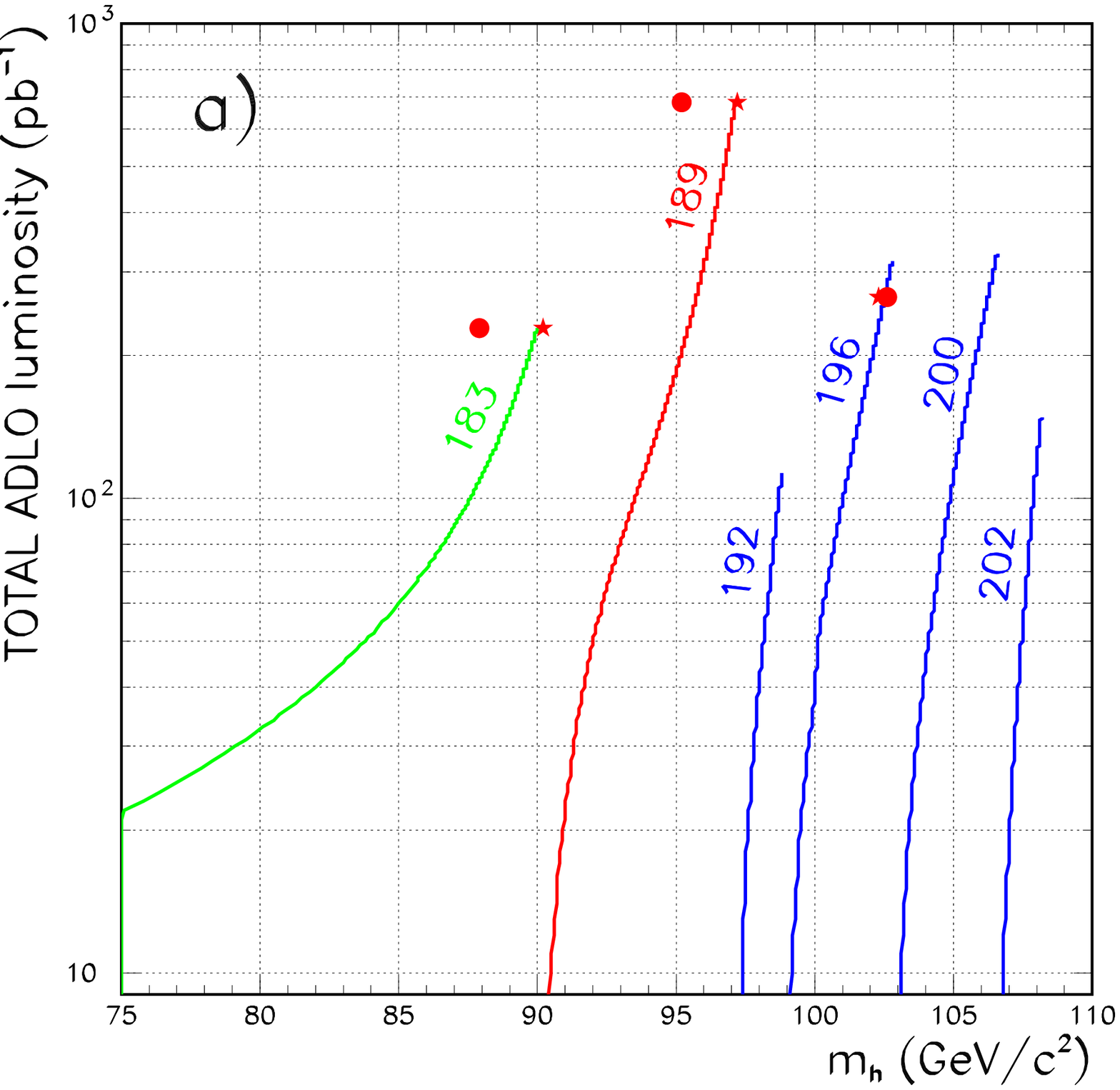,width=7.9cm}}
{\epsfig{file=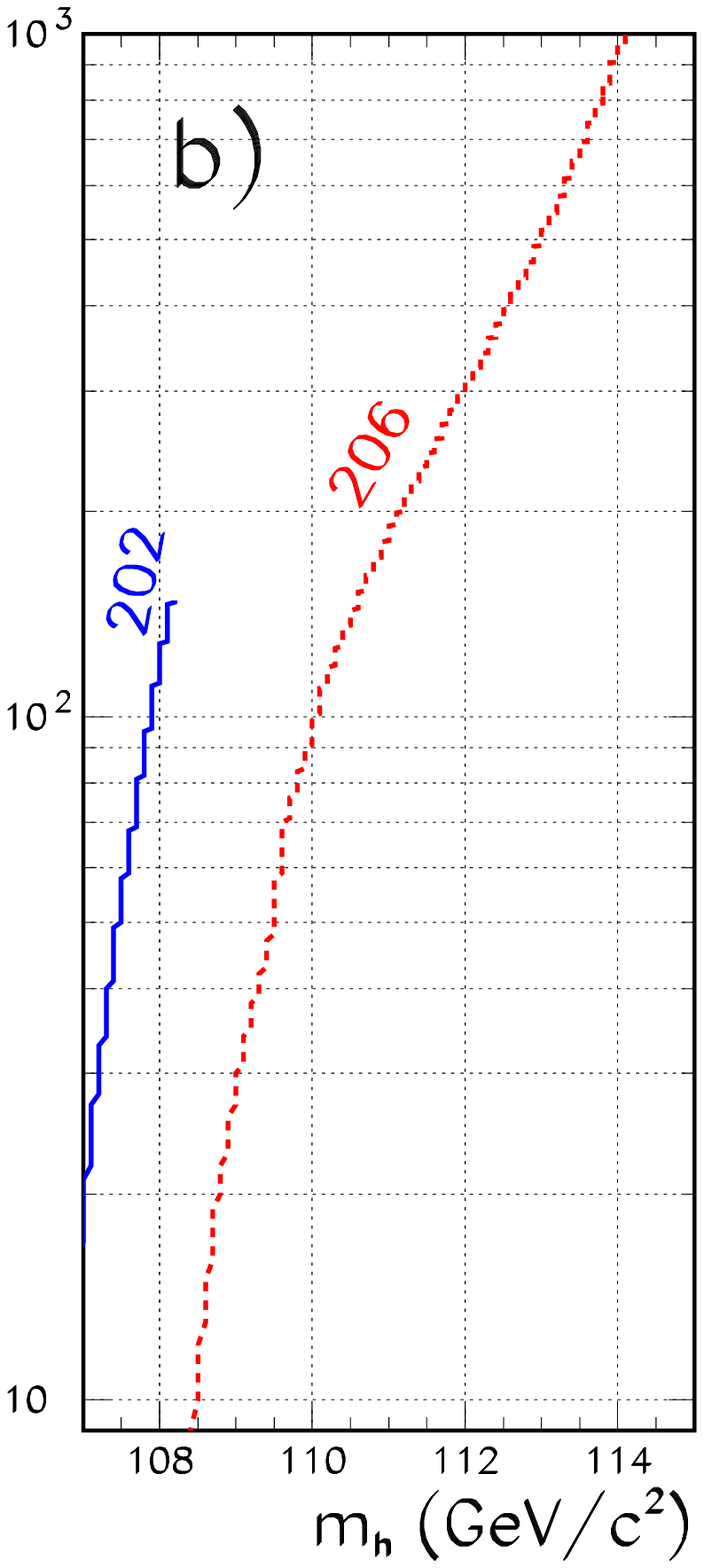,width=3.7cm}}
{\epsfig{file=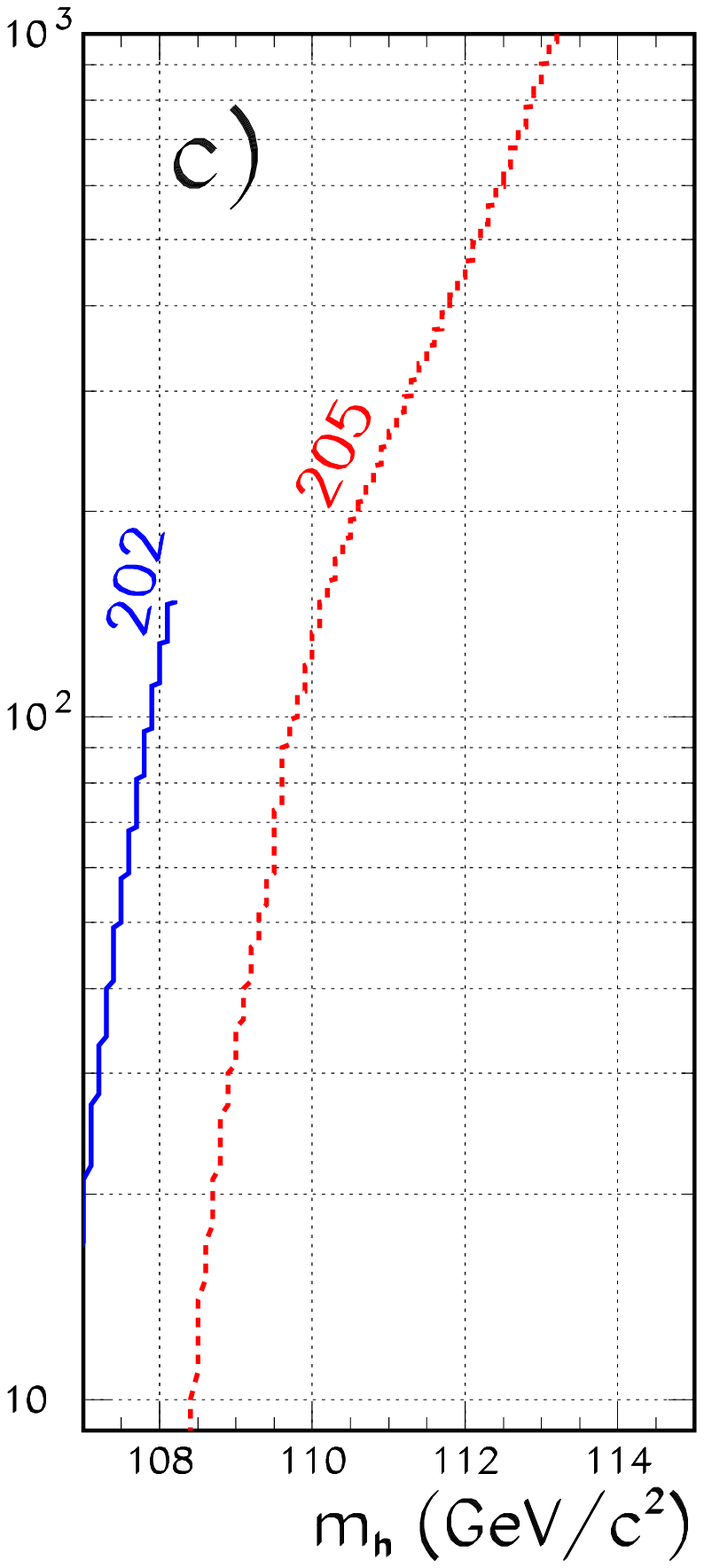,width=3.7cm}}
\end{tabular}
\caption{{\footnotesize Predictions of the expected combined
ALEPH+DELPHI+L3+OPAL 95\% CL $\mh$ exclusion; a) obtained from the
data taken until the end of 1999 (solid lines). For comparison the
expected (stars) and observed (dots) combined LEP limits obtained from
actual data combinations\cite{lephiggs183,lephiggs189,lephiggs196} are
also shown. The effect of adding to this data set new data at
b) $\sqrt{s}=$206 GeV or c) 205 GeV is indicated by the dashed line.
\label{fig:predictor}}
}
\end{figure}

\begin{table}[htbp]
\caption{{\footnotesize Predictions of the sensitivity of the four LEP
experiments combined, for several hypothetical data sets. The table
shows the expected excluded SM Higgs boson mass ($\mh^{95}$, in GeV) as well
as the corresponding excluded $\tb$ region in the $\mhmax$ benchmark
scenario (with $\mt = 174.3$~GeV, $\msusy = 1$~TeV),
when new data at the indicated $\sqrt{s}$ is combined with the
existing data set (Table~\ref{tab:lepdata}). The luminosities indicated are for the 4 LEP experiments 
combined. The results shown are valid only if no signal were found in
the data. (Note that, as it is not foreseen at the moment that it will
be possible to run LEP at
$\sqrt{s}>206$ GeV, scenario 8 is probably unrealistic.) }
\label{tab:lepreach} 
}
\begin{center}
\begin{tabular}{|c||cccc||cc|}
\hline
 $\sqrt{s}$ (GeV)             & 204. & 205. & 206. & 208. & $\mh^{95}$ & $\tb^{95}$ \\
\hline \hline
 1) $\mathcal{L}$ (pb$^{-1}$) &  -   &  -   & 100. &  -   & 110.0 & 0.6 -- 2.1 \\
\hline
 2) $\mathcal{L}$ (pb$^{-1}$) &  -   &  -   & 500. &  -   & 113.0 & 0.5 -- 2.4 \\
\hline
 3) $\mathcal{L}$ (pb$^{-1}$) &  -   &  -   & 1000.&  -   & 114.1 & 0.5 -- 2.5 \\
\hline
 4) $\mathcal{L}$ (pb$^{-1}$) &  -   & 120. &  -   &  -   & 110.0 & 0.6 -- 2.1 \\
\hline
 5) $\mathcal{L}$ (pb$^{-1}$) &  -   & 840. &  -   &  -   & 113.0 & 0.5 -- 2.4 \\
\hline
 6) $\mathcal{L}$ (pb$^{-1}$) & 100. & 100. & 400. &  -   & 113.1 & 0.5 -- 2.4 \\
\hline
 7) $\mathcal{L}$ (pb$^{-1}$) & 150. & 300. & 300. &  -   & 113.3 & 0.5 -- 2.4 \\
\hline
 8) $\mathcal{L}$ (pb$^{-1}$) & 150. & 300. & 300. & 280. & 115.0 & 0.5 -- 2.6 \\
\hline
\end{tabular}
\end{center}
\end{table}

 In Table~\ref{tab:lepreach} the expected SM Higgs boson limit is
 shown for several possible LEP running scenarios in the year
 2000. Taking into account that the {\sl experimental} \/MSSM $\mh$
 exclusion in the range $0.5 \lesim \tb \lesim 3$ is {\sl (i)}
 \/essentially independent of $\tb$ and {\sl (ii)} \/equal in value to the
 SM $\mh$ exclusion (see e.g. \cite{lephiggs189,lephiggs196}), 
 $\mh^{95}$ can be
 converted into an excluded $\tb$ range in the $\mhmax$ benchmark
 scenario described in Section~\ref{section:mssm}.
 This is done by intersecting
 the experimental exclusion and the solid curve in Figure~\ref{fig:rgv}. Using
 the LEP data taken until the end of 1999 (for 
 which $\mh^{95}=108.2~\Gcs$) one can already expect to exclude 
 $0.6 \lesim \tb \lesim 1.9$ 
 within the MSSM for $\mt = 174.3$~GeV and $\msusy = 1$~TeV. 
 Note that in determining the excluded $\tb$ regions in
 Table~\ref{tab:lepreach} the theoretical uncertainty from unknown
 higher-order corrections has been neglected.
 As can be seen from Table~\ref{tab:lepreach}, several
 plausible scenarios for adding new data at higher energies can extend
 the exclusion to $\mh\lesim 113~\Gcs$ ($0.5 \lesim \tb \lesim 2.4$).


\subsection{The upper limit on $\mh$ in the M-SUGRA scenario}

The M-SUGRA scenario is described by four independent parameters and a
sign,
namely the common squark mass $M_0$, the common gaugino mass $M_{1/2}$, 
the common trilinear coupling $A_0$, $\tb$ and the sign of $\mu$.
The universal parameters are fixed at the GUT scale, where we assumed
unification of the gauge couplings. Then they are run down
to the electroweak scale with the help of renormalisation group
equations~\cite{ross,nath,faraggi,castano,barger,kolda,sakis,pierce}.
The condition of REWSB 
puts an upper bound on $M_0$ of about $M_0 \lesim $ 5 TeV
(depending on the values of the other four parameters). 


In order to obtain a precise prediction for $\mh$ within the M-SUGRA
scenario, we employ the complete \twol\ RG running with appropriate
thresholds (both logarithmic and finite for the gauge couplings
and using the so called $\theta$-function approximation for 
the masses~\cite{sakis}) including full \onel\ minimisation
conditions for the effective potential, in order to
extract all the parameters of the M-SUGRA scenario at the EW
scale. This method has been combined with the presently most precise
result of $\mh$ based on a Feynman-diagrammatic
calculation~\cite{mhiggsletter,mhiggslong}.
This has been carried out by combining the codes of two
programs namely, 
{\tt SUITY}~\cite{sakis2} and {\tt FeynHiggs}~\cite{feynhiggs}.

\begin{figure}
\centerline{\psfig{figure=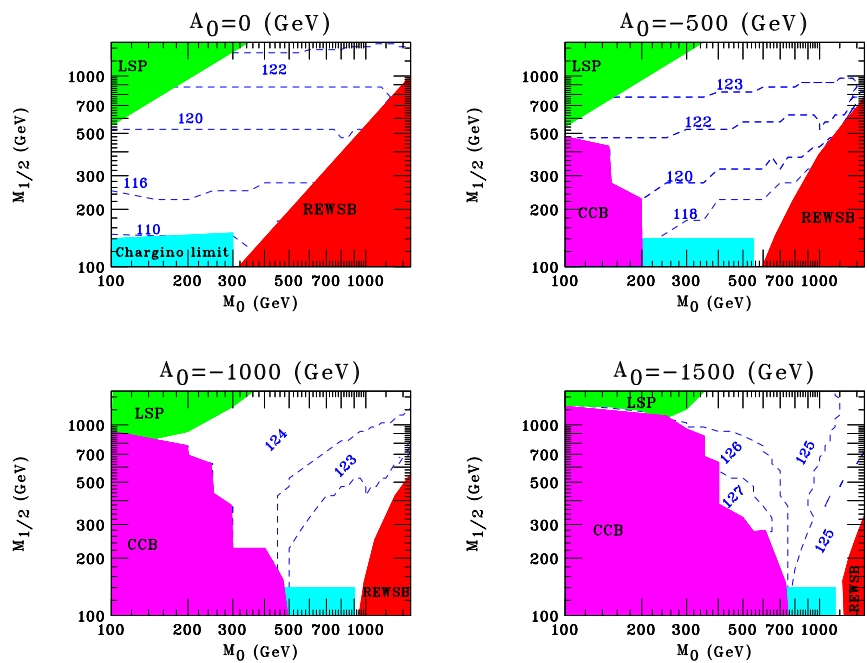,height=4.5in,angle=90}}
\caption{In the $M_0-M_{1/2}$-plane the contour lines of $\mh$ are shown
for four values of $A_0$. The numbers refer to $\mh$ in the
respective region within $\pm 0.5 \gev$. The regions that are excluded
by REWSB, the CCB or LSP conditions, or by direct chargino search are
also indicated.
}
\label{fig:msugra}
\end{figure}

In order to investigate the upper limit on the Higgs boson mass in the
M-SUGRA scenario, we keep $\tb$ fixed at a large value, $\tb = 30$.
Concerning the sign of the Higgs mixing parameter, $\mu$, we find larger
$\mh$ values (compatible with the constraints discussed below) for 
negative $\mu$ (in the convention of~\refeq{eq:xt}).
In the following we analysed the upper limit on $\mh$ as a function of 
the other M-SUGRA parameters, $M_0$, $M_{1/2}$ and $A_0$. 
Our results are displayed in~\reffi{fig:msugra} for four values of
$A_0$: $A_0 = 0, -500, -1000, -1500 \gev$. We show contour lines of
$\mh$ in the $M_0-M_{1/2}$-plane. The numbers inside the plots indicate the
lightest Higgs boson mass in the respective area within $\pm 0.5 \gev$.
The upper bound on the lightest $\cp$-even Higgs boson mass 
is found to be at most 127 GeV. This upper limit is reached for 
$M_0 \approx 500 \gev$, $M_{1/2} \approx 400 \gev$ and $A_0 = -1500 \gev$.
Concerning the analysis the following should be noted:
\begin{itemize}
\item 
We have chosen the current experimental central value for
the top quark mass, $\mt = 174.3$ GeV. As mentioned above, increasing 
$\mt$ by 1 GeV results in an increase of $\mh$ of approximately $1 \gev$.
%
%
%
%
\item 
The M-SUGRA parameters are taken to be real, no SUSY $\cp$-violating
phases are assumed.
\item  
We have chosen negative values for the trilinear 
coupling, because $\mh$ turns out to be increased
by going from positive to negative values of $A_0$. 
$|A_0|$ is restricted from
above by the condition that no negative squares of squark masses and no
charge or colour breaking minima appear.
%
%
%
%
\item
The regions in the $M_0-M_{1/2}$-plane that are excluded for the
following reasons are also indicated:
\begin{itemize}
\item
REWSB: parameter sets that do not fulfil the REWSB condition.
\item
CCB: regions where charge or colour breaking minima occur
or negative squared squark masses are obtained at the EW scale.
\item
LSP: sets where the lightest neutralino is not the LSP. Mostly there the
lightest scalar tau becomes the LSP.
\item
Chargino limit: parameter sets which correspond to a chargino mass
that is already excluded by direct searches.
\end{itemize}
\item 
We do not take into account the $b\rightarrow s\gamma$
constraint as the authors of~\citere{pierce2,deboer} do. 
This could reduce the upper limit but still
the experimental and theoretical uncertainties of this constraint
are quite large.
%
%
%
\end{itemize}
%
%

\subsection{Conclusions}

We have analysed the upper bound on $\mh$ within the MSSM. Using the
Feynman-diagrammatic result for $\mh$, which contains new
genuine two-loop corrections, leads to an increase of $\mh$ of up to 
$4 \gev$ compared to the previous result obtained by renormalisation
group methods. We have furthermore investigated the MSSM parameters for
which the maximal $\mh$ values are obtained and have compared the 
$\mhmax$ scenario with the previous benchmark scenario. For $\mt =
174.3$~GeV and $\msusy = 1$~TeV we find $\mh \lesim 129$~GeV as upper
bound in the MSSM.
In case that no evidence of a Higgs signal is found before the
end of running in 2000, experimental searches for the Higgs boson at
LEP can ultimately be reasonably expected to exclude $\mh\lesim
113~\Gcs$. In the context of the $\mhmax$ benchmark scenario
(with $\mt = 174.3$~GeV, $\msusy = 1$~TeV) this rules out the interval 
$0.5\lesim\tb\lesim 2.4$ at the 95\% confidence level within the MSSM.
Within the M-SUGRA scenario, the upper bound on $\mh$ is
found to be $\mh \lesim 127 \gev$ for $\mt = 174.3 \gev$. 
This upper limit is reached for the M-SUGRA parameters $M_0 \approx 500
\gev$, $M_{1/2} \approx 400 \gev$ and $A_0 = -1500 \gev$. The upper
bound within the M-SUGRA scenario 
is lower by 2 and 4~GeV than the bound obtained in the general
MSSM for $\msusy = 1 \tev$ and $\msusy = 2 \tev$, respectively.


\ack
A.D.\ acknowledges financial support from the Marie Curie Research
Training Grant ERB-FMBI-CT98-3438. A.D.\ would also like to thank Ben
Allanach for useful discussions. P.T.D.\ would like to thank Jennifer
Kile for providing the Standard Model Higgs boson production
cross-sections. G.W.\ thanks C.E.M.~Wagner for useful discussions.


\section{An Update of the program HDECAY\label{hdecay}}

\author{A~Djouadi, J~Kalinowski and M~Spira}

\begin{abstract}
{\small 
The program HDECAY determines the decay widths and branching
ratios of the Higgs bosons within the Standard Model and its minimal
supersymmetric extension, including the dominant higher-order
corrections. New theoretical developments are briefly discussed and
the new ingredients incorporated in the program are summarised.}
\end{abstract}

The search strategies for Higgs bosons at LEP, Tevatron, LHC and
future $e^+e^-$ linear colliders (LC) exploit various Higgs boson
decay channels. The strategies depend not only on the experimental
setup (hadron versus lepton colliders) but also on the theoretical
scenarios: the Standard Model (SM) or some of its extensions such as
the Minimal Supersymmetric Standard Model (MSSM). It is of vital
importance to have reliable predictions for the branching ratios of
the Higgs boson decays for these theoretical models.

The current version of the program HDECAY \cite{hdecay} can be used to
calculate Higgs boson partial decay widths and branching ratios within
the SM and the MSSM and includes:  

\noindent
-- All decay channels that are kinematically allowed and which have
branching ratios larger than $10^{-4}$, {\it y compris}\/ the loop
mediated, the three body decay modes and in the MSSM the cascade and
the supersymmetric decay channels \cite{review}. \\
-- All relevant higher-order QCD corrections to the decays into quark
pairs and to the loop mediated decays into gluons are incorporated in
a complete form \cite{QCD}; the small leading electroweak corrections
are also included. \\
-- Double off--shell decays of the CP--even Higgs bosons into massive
gauge bosons which then decay into four massless fermions, and all
important below--threshold three--body decays \cite{below}. \\
-- In the MSSM, the complete radiative corrections in the effective
potential approach with full mixing in the stop/sbottom sectors; it
uses the renormalisation group improved values of the Higgs masses and
couplings and the relevant next--to--leading--order corrections are
implemented \cite{SUBH}. \\
-- In the MSSM, all the decays into SUSY particles (neutralinos,
charginos, sleptons and squarks including mixing in the stop, sbottom
and stau sectors) when they are kinematically allowed \cite{SUSY}. The
SUSY particles are also included in the loop mediated $\gamma \gamma$
and $gg$ decay channels.

The source code of the program, written in FORTRAN, has been tested on
computers running under different operating systems. The program
provides a very flexible and convenient usage, fitting to all options
of phenomenological relevance. The basic input parameters, fermion and
gauge boson masses and their total widths, coupling constants and, in
the MSSM, soft SUSY-breaking parameters can be chosen from an input
file. In this file several flags allow switching on/off or changing
some options [{\it e.g.}\/ choosing a particular Higgs boson,
including/excluding the multi-body or SUSY decays, or
including/excluding specific higher-order QCD corrections].

Since the release of the original version of the program several bugs
have been fixed and a number of improvements and new theoretical
calculations have been implemented. The following points have
recently been done:

\noindent 
-- Link to the {\tt FeynHiggsFast} routine for Higgs masses and couplings 
\cite{fhf}.  \\
-- Link to the {\tt SUSPECT} routine for RG evolution of SUGRA parameters
\cite{suspect}.  \\
-- Implementation of Higgs boson decays to  gravitino + gaugino \cite{gg}. \\
-- Inclusion of  gluino loops in Higgs decays to $b\bar{b}$ \cite{hbb}. \\ 
-- Inclusion of QCD corrections  in Higgs decays to squarks \cite{hsqsq}.  \\
-- Determination and inclusion of the RG improved two-loop contributions to
   the MSSM Higgs  self-interactions. 

The logbook of all modifications and the most recent version of the
program can be found on the web page {\tt
http://www.desy.de/$\sim$spira/prog}.

\ack JK has been supported in
parts by the KBN grant No.\ 2 P03B 030 14 and the Foundation for
Polish-German Collaboration grant No.\ 3310/97/LN. We thank Peter
Zerwas for continuous interest and support.


\subsection*{References}
\begin{thebibliography}{99}
\bibitem{focuspoints}
J.L Feng \etal, \PR {\bf D61} (2000) 075005, hep-ph/9908309
\bibitem{ATLASTDR2}
ATLAS Collaboration, Detector and Physics Performance TDR, Volume II,
Technical Report CERN/LHCC 99-15, (1999) CERN
\bibitem{hbaeretc}
H Baer \etal \PR {\bf D53} (1996) 6241
\bibitem{measures}
See for example R. Barbieri and A. Strumia, \PL {\bf B433} (1998) 63;
B. de Carlos and J.A. Casas, \PL {\bf B320} (1993) 320
\bibitem{ISASUSY}
H. Baer \etal hep-ph/9810440
\bibitem{HERWIG}
{\small HERWIG6.1} collaboration,
http://home.cern.ch/$\sim$seymour/herwig/herwig61.html, Cavendish-HEP-99/03
\bibitem{usnext}
B.C. Allanach \etal {\em work in progress}
\end{thebibliography}

\subsection*{References}
\begin{thebibliography}{99}

\bibitem{Haber}H.R. Haber and G.L.Kane,
{\em Phys. Rep.}{\bf 117}, 75 (1984).
\bibitem{Chikovani96}E.Chikovani, V.Kartvelishvili, R.Shanidze and G.Shaw,
{\em Phys. Rev.} {\bf D53}, 6653 (1996).
\bibitem{Keung}W.Y. Keung and A. Khare, 
{\em Phys. Rev.} {\bf D29}, 2657 (1984);
J.H. K\"uhn and S. Ono, 
{\em Phys. Lett.} {\bf B142}, 436 (1984);
T. Goldman and H.E. Haber, {\em Physica} {\bf {15D}}, 181 (1985). 
\bibitem{Chikovani89}E.G. Chikovani, V.G. Kartvelishvili and A.V. Tkabladze, 
{\em Z. Phys.}{\bf C43}, 509 (1989); {\em Sov. J. Nucl.
Phys.}{\bf 51} 546, (1990). 

\end{thebibliography}

\section*{References}
\begin{thebibliography}{9}

\bibitem{Dim}
N. Arkani-Hamed, S. Dimopoulos and G. Dvali, 
Phys. Rev. D59 (1999) 086004.

\bibitem{GRW}
G.F. Giudice, R. Rattazzi and J.D. Wells,
Nucl. Phys. B544 (1999) 3.

\bibitem{gravlim}
L3 Collaboration, M Acciarri {\em{et al.}}, Phys. Lett. {{B464}} (1999) 135;
P. Mathews {\em{et al.}}, hep-ph/9904232 (1999);
A.K. Gupta {\em{et al.}}, hep-ph/9904234 (1999).

\bibitem{RaSun}
L. Randall and R. Sundrum, Phys. Rev. 
Lett. 83 (1999) 3370.

\bibitem{warped}
H. Davoudiasl, J.L. Hewett and
T.G. Rizzo,  hep-ph/9909255.


\bibitem{DIStheo} P.~Mathews, S.~Raychaudhuri and K.~Sridhar,
Phys. Lett. B455 (1999) 115.

\bibitem{ZDIS} ZEUS Collaboration, J.~Breitweg et al., Eur. Phys. Jnl. {\bf
C11} (1999) 427.

\bibitem{H1DIS} H1 Collaboration, C.~Adloff et al., DESY 99-107 (1999), hep-ex/9908059.

\bibitem{HERAelec} H1 Collaboration, Abstract No 157b, 
presented at HEP99, 
Tampere, Finland, July 1999; ZEUS Collaboration, Abstract No 549, ibid.

\bibitem{DYdata} CDF Collaboration, F.~Abe et al., Phys. Rev. {D59}
(1999) 052002; 
D0 Collaboration, B.~Abbott et al., Phys. Rev. Lett. {82} (1999) 4769.


\end{thebibliography}

\section*{References}
\begin{thebibliography}{99}
\bibitem{opal}
OPAL Collaboration, G.~Abbiendi \etal  , 
hep-ex/9909552, CERN-EP/99-122
\bibitem{aleph}
ALEPH Collaboration, R.~Barate  \etal , Phys. Lett. {\bf B469} (1999) 303.
\bibitem{l3}
L3 Collaboration, M. Acciarri  \etal , Eur.~Phys.~J.~{\bf C4} (1998) 207.
\bibitem{delphi}
DELPHI Collaboration, P.~Abreu \etal , Eur.~Phys.~J. {\bf C6} (1999) 385.
\bibitem{mssmlib} We calculated these cross-sections using the program {\small
MSMLIB} from Gerado Ganis (private communication).
\bibitem{opalmc}
The selectron samples were generated with 
{\sc Susygen}: S.~Katsanevas and S.~Melachroinos,
in {\it Physics at LEP2},
edited by G.~Altarelli, T.~Sj\"{o}strand and
F.~Zwirner, CERN 96-01, Vol.~2 (1996) p.~216.
S.~Katsanevas and P.~Morawitz, Comp. Phys. Comm. {\bf 112} (1998)
227. \\
The most important SM samples were generated with:
{\sc Koralz}~4.0:
S.~Jadach, B.F.L.~Ward, Z.~W\c{a}s, Comp.~Phys.~Comm. {\bf 79} (1994) 503,
and the generator of J.A.M.~Vermaseren, Nucl.~Phys. {\bf B229} (1983) 347.\\
All samples were processed with the full
simulation program of the OPAL experiment:
J.~Allison \etal , Nucl.~Instr.~Meth.~{\bf A317} (1992) 47.
\bibitem{inv}
J. Kalinowski, Acta Phys. Polon.~{\bf B28} (1997) 1437;
J.~Kalinowski and P.~M.~Zerwas, Phys.~Lett.~{\bf B400} (1997) 112.
\bibitem{w}
L3 Collaboration, M. Acciarri  \etal , Phys.~Lett.~{\bf B436} (1998) 417. \\
ALEPH Collaboration, R.~Barate  \etal ,
Physics Letters Phys.~Lett.~{\bf B462} (1999) 389.   
\bibitem{epa}
We calculated this result by using the effective photon approximation 
and the results on photon-electron scattering in S.~Hesselbach and
   H.~Fraas, Phys.~Rev.~{\bf D55} (1997) 1343. 
\end{thebibliography}

\subsection*{References}
\begin{thebibliography}{99}
\bibitem{LEPC}
LEP C collaboration meeting, CERN, Nov 1999
\bibitem{LEPEWWG}
LEP electroweak working group, 
see http://www.cern.ch/LEPEWWG/plots/
\bibitem{bagfalk}
J.A. Bagger, A.F. Falk and M. Swartz, hep-ph/9908327
\bibitem{cn4}
R.S. Chivukula and N. Evans, \PL {\bf B464} (1999) 244
\bibitem{PT}
M.E. Peskin and T. Takeuchi, \PRL {\bf 65} (1990) 964
\bibitem{ATLASTDR}
ATLAS Collaboration, Detector and Physics Performance TDR, Volume II,
Technical Report CERN/LHCC 99-15, (1999) CERN.
\end{thebibliography}

\section*{References}
\begin{thebibliography}{99}
\bibitem{hill}
A.~Hill, J.~J.~van~der~Bij, Phys. Rev. D36, 3463 (1987).
\bibitem{krasnikov}
N.~V.~Krasnikov, Mod. Phys. Lett. A13, 893 (1998).
\bibitem{binoth}
T.~Binoth, J.~J.~van~der~Bij, Z. Phys. C75, 17 (1997) and references therein.
\bibitem{valle}
J.~Valle et al. LEP2 Higgs Report, CERN 96-01, 350 (1996).
\bibitem{chivukula}
R.~S.~Chivukula, M.~Golden, Phys. Lett. B267, 233 (1991).
\bibitem{binoth2}
T.~Binoth, J.~J.~van~der~Bij, contribution to the Linear Collider
Workshop, Sitges 1999.
\bibitem{bjorken}
J.~D.~Bjorken, Int. J. Mod. Phys. A7, 4221 (1992).
\bibitem{gunion}
J.~R.~Espinosa, J.~F.~Gunion, Phys. Rev. Lett. 82, 1084 (1999).
\end{thebibliography}

\section*{References}

\begin{thebibliography}{99}

\bibitem{mhiggs1l} H.~Haber and R.~Hempfling,
                   {\em Phys. Rev. Lett.} {\bf 66} (1991) 1815;
                   J.~Ellis, G.~Ridolfi and F.~Zwirner,
                   {\em Phys. Lett.} {\bf B 257} (1991) 83; 
                   {\em Phys. Lett.} {\bf B 262} (1991) 477.

\bibitem{mhiggsf1l} P.~Chankowski, S.~Pokorski and J.~Rosiek,
                    {\em Nucl. Phys.} {\bf B 423} (1994) 437.
\bibitem{mhiggsf1ldab} A.~Dabelstein, 
                    {\em Nucl. Phys.} {\bf B 456} (1995) 25,
                    hep-ph/9503443;
                    {\em Z. Phys.} {\bf C 67} (1995) 495,
                    hep-ph/9409375.

\bibitem{pierce} J.~Bagger, K.~Matchev, D.~Pierce and R.~Zhang,
                    {\em Nucl. Phys.} {\bf B 491} (1997) 3,
                    hep-ph/9606211.

\bibitem{mhiggsRG1}  M.~Carena, J.~Espinosa, M.~Quir\'os and C.~Wagner,
                      {\em Phys. Lett.} {\bf B 355} (1995) 209,
                      hep-ph/9504316;
                      M.~Carena, M.~Quir\'os and C.~Wagner,
                      {\em Nucl. Phys.} {\bf B 461} (1996) 407,
                      hep-ph/9508343.

\bibitem{mhiggsRG2} H.~Haber, R.~Hempfling and A.~Hoang,
                    {\em Z. Phys.} {\bf C 75} (1997) 539,
                    hep-ph/9609331.

\bibitem{mhiggsEP} R.~Hempfling and A.~Hoang,
                   {\em Phys. Lett.} {\bf B 331} (1994) 99,
                   hep-ph/9401219;
                   R.-J.~Zhang, 
                   {\em Phys. Lett.} {\bf B 447} (1999) 89,
                   hep-ph/9808299.

\bibitem{mhiggsletter} S.~Heinemeyer, W.~Hollik and G.~Weiglein,
                    {\em Phys. Rev.} {\bf D 58} (1998) 091701,
                    hep-ph/9803277;
                    {\em Phys. Lett.} {\bf B 440} (1998) 296,
                    hep-ph/9807423.

\bibitem{mhiggslong} S.~Heinemeyer, W.~Hollik and G.~Weiglein,
                     {\em Eur. Phys. Jour.} {\bf C 9} (1999) 343, 
                     hep-ph/9812472.

\bibitem{lepc-aleph} A.~Blondel, ALEPH Collaboration, 
	talk given at the LEPC meeting, November 9, 1999.

\bibitem{lepc-delphi} J.~Marco, DELPHI Collaboration, 
	talk given at the LEPC meeting, November 9, 1999.

\bibitem{lepc-l3} G.~Rahal-Callot, L3 Collaboration, 
	talk given at the LEPC meeting, November 9, 1999.

\bibitem{lepc-opal} P.~Ward, OPAL Collaboration, 
	talk given at the LEPC meeting, November 9, 1999.

\bibitem{REWSB}
L.E.~Iba\~nez  and G.G.~Ross, {\em Phys. Lett.}
{\bf 110} (1982) 215; \\
K.~Inoue, A.~Kakuto, H.~Komatsu and S.~Takeshita,
{\em Progr. Theor. Phys.} {\bf 68} (1982) 927; {\it ibidem}, {\bf 71} (1984)
96;\\
J.~Ellis, D.V.~Nanopoulos and K.~Tamvakis,
{\em Phys. Lett.} {\bf B121} (1983) 123;\\
L.E.~Iba\~nez , {\em Nucl. Phys.} {\bf B218} (1983) 514;\\
L.~Alvarez-Gaum\'e, J.~Polchinski and M.~Wise, {\em Nucl. Phys.} {\bf B221}
(1983) 495;\\
J.~Ellis, J.S.~Hagelin, D.V.~Nanopoulos and K.~Tamvakis, {\em Phys. Lett.}
{\bf B125} (1983) 275;\\
L.~Alvarez-Gaum\'e, M.~Claudson and M.~Wise, {\em Nucl. Phys.} {\bf B207}
(1982) 96.

\bibitem{sakis}
A.~Dedes, A.B.~Lahanas and K.~Tamvakis,
{\em Phys.\ Rev.} {\bf D53}, 3793 (1996)
hep-ph/9504239;

\bibitem{lephiggs183} The LEP working group for Higgs boson searches,
                     CERN-EP/99-060.

\bibitem{tbexcl} S.~Heinemeyer, W.~Hollik and G.~Weiglein,
                 DESY 99-120, hep-ph/9909540.

\bibitem{bench} M.~Carena, S.~Heinemeyer, C.~Wagner and G.~Weiglein,
                hep-ph/9912223.

\bibitem{feynhiggs} S.~Heinemeyer, W.~Hollik and G.~Weiglein,
                    Comp. Phys. Comm. 124 (2000) 76.

\bibitem{bse} M.~Carena, H.~Haber, S.~Heinemeyer, W.~Hollik, C.~Wagner
              and G.~Weiglein,
	hep-ph/0001002.

\bibitem{mhiggslle} S.~Heinemeyer, W.~Hollik and G.~Weiglein,
                    {\em Phys. Lett.} {\bf B 455} (1999) 179,
                    hep-ph/9903404; hep-ph/9910283.

\bibitem{safari} P.~Janot, {\sl How should we organize the Higgs safari?}, 
	published in the Proceedings of the 9th Chamonix SPS \& LEP Performance 
	Workshop, January 1999 (J.~Poole, Ed.); CERN-SL-99-007-DI.

\bibitem{lephiggs172} The LEP working group for Higgs boson searches,
                     CERN-EP/98-046.


\bibitem{lephiggs189} The LEP working group for Higgs boson searches, paper 
  \#6\_49, submitted to the the International Europhysics Conference on
  High Energy Physics, July 1999, Tampere, Finland; ALEPH 99-081,
  DELPHI 99-142, L3 note 2442, OPAL TN-614.

\bibitem{lepc-lep} A.~Butterworth,
	talk given at the LEPC meeting, November 9, 1999.

\bibitem{lephiggs196} P.~McNamara, {\sl Combined LEP Higgs search results up to 
      $\sqrt{s}=$196 GeV}, talk given at the LEPC meeting, September
      7, 1999.

\bibitem{ross}
G.G.~Ross and R.G.~Roberts,
{\em Nucl.\ Phys.} {\bf B377}, 571 (1992).

\bibitem{nath}
P.~Nath and R.~Arnowitt, {\em Phys. \ Lett.} {\bf B287}, 89 (1992).

\bibitem{faraggi}
A.~Faraggi and B.~Grinstein,  
{\em Nucl.\ Phys.} {\bf B422}, 3 (1994).

\bibitem{castano}
D.J.~Castano, E.J.~Piard and P.~Ramond,
{\em Phys.\ Rev.} {\bf D49}, 4882 (1994)
hep-ph/9308335.


\bibitem{barger}
V.~Barger, M.S.~Berger and P.~Ohmann,
{\em Phys.\ Rev.} {\bf D49}, 4908 (1994)
hep-ph/9311269.


\bibitem{kolda}
G.L.~Kane, C.~Kolda, L.~Roszkowski and J.D.~Wells,
{\em Phys.\ Rev.} {\bf D49}, 6173 (1994)
hep-ph/9312272.


\bibitem{sakis2}
A.~Dedes, A.~B.~Lahanas, V.~Spanos and K.~Tamvakis,
``{\tt SUITY:} A program for the minimal {\tt SU}pergrav{\tt ITY}
spectrum", {\em in preparation}.


\bibitem{pierce2}
K.~Matchev and D.~Pierce, {\em Phys.\ Lett.} {\bf B445}, 331 (1999). 

\bibitem{deboer}
W.~de Boer, H.J.~Grimm, A.V.~Gladyshev, D.I.~Kazakov,
{\em Phys. Lett.} {\bf B438}, 281 (1998);\\
W.~de Boer, talk given at the ECFA/DESY Linear Collider Workshop, 
Obernai, October 1999. 

\end{thebibliography}

\subsection*{References}
\begin{thebibliography}{99}
%
\bibitem{hdecay} A.~Djouadi, J.~Kalinowski and M.~Spira, Comput.\ Phys.\ 
Commun.\ {\bf 108} (1998) 56. 
%
\bibitem{review} M.~Spira, Fortschr.\ Phys.\ {\bf 46} (1998) 203.
%
\bibitem{QCD} A.\ Djouadi, M.\ Spira and P.M.\ Zerwas, Z.~Phys. 
{\bf C70} (1996) 427.
%
\bibitem{below} A.~Djouadi, J.~Kalinowski and P.M.~Zerwas, 
Z.~Phys.~{\bf C70} (1996) 435.
%
\bibitem{SUBH}
M.~Carena, M.~Quiros and C.E.M.~Wagner, Nucl.~Phys.~{\bf B461} (1996) 407;
H.E.\ Haber, R.\ Hempfling and A.H.\ Hoang, Z.\ Phys.\ {\bf C75} (1997) 539;
S.\ Heinemeyer, W.\ Hollik and G.\ Weiglein, Phys.\ Rev.\ {\bf D58}
(1998) 091701.
%
\bibitem{SUSY} A.~Djouadi, J.~Kalinowski and P.M.~Zerwas, Z.~Phys.~{\bf C57} 
(1993) 569; A.~Djouadi, P.~Janot, J.~Kalinowski and P.M.~Zerwas,
Phys.~Lett.~{\bf B376} (1996) 220;
A.~Djouadi, J.~Kalinowski, P.~Ohmann and P.M.~Zerwas, Z.\ Phys.\ {\bf C74}
(1997) 93.
%
\bibitem{fhf} S.\ Heinemeyer, W.\ Hollik and G.\ Weiglein,
Comp. Phys. Comm. {\bf 124} (2000) 76.
%
\bibitem{suspect} A.\ Djouadi, J.L.\ Kneur and G.\ Moultaka, 
in hep-ph/9901246. 
%
\bibitem{gg} A.\ Djouadi and M.\ Drees, Phys.\ Lett.\ {\bf B407} (1997) 243.
%

\bibitem{hbb} A.\ Dabelstein, Nucl.\ Phys.\ {\bf B456} (1995) 25;
R.A.\ Jim\'enez and J.\ Sol\`a, Phys.\ Lett.\ {\bf B389} (1996) 53;
J.A.\ Coarasa, R.A.\ Jim\'enez and J.\ Sol\`a, Phys.\ Lett.\ {\bf B389}
(1996) 312.
%

\bibitem{hsqsq} A.\ Bartl, H.\ Eberl, K.\ Hidaka, T.\ Kon, W.\ Majerotto
and Y.\ Yamada, Phys.\ Lett.\ {\bf B402} (1997) 303;
A.\ Arhrib, A.\ Djouadi, W.\ Hollik and C.\ J\"unger, Phys.\ Rev.\ {\bf D57}
(1998) 5860.
%
\end{thebibliography}
\end{document}